\documentclass[%
 reprint,
superscriptaddress,
 amsmath,amssymb,
 aps,
]{revtex4-2}

\usepackage{graphicx}
\usepackage{dcolumn}
\usepackage{bm}
\usepackage{braket}
\usepackage{color}
\usepackage{hhline}
\usepackage{appendix}
\usepackage{soul}

\begin{document}

\title{Demonstrating a long-coherence dual-rail erasure qubit using tunable transmons}

\author{H. Levine}
\affiliation{AWS Center for Quantum Computing, Pasadena, CA 91125, USA}

\author{A. Haim}
\affiliation{AWS Center for Quantum Computing, Pasadena, CA 91125, USA}
\affiliation{Institute of Applied Physics, The Hebrew University of Jerusalem, Jerusalem 91904, Givat Ram, Israel}

\author{J. S.C.~Hung}
\affiliation{AWS Center for Quantum Computing, Pasadena, CA 91125, USA}

\author{N. Alidoust}
\affiliation{AWS Center for Quantum Computing, Pasadena, CA 91125, USA}

\author{M. Kalaee}
\affiliation{AWS Center for Quantum Computing, Pasadena, CA 91125, USA}

\author{L. DeLorenzo}
\affiliation{AWS Center for Quantum Computing, Pasadena, CA 91125, USA}

\author{E.~A.~Wollack}
\affiliation{AWS Center for Quantum Computing, Pasadena, CA 91125, USA}

\author{P.~Arrangoiz-Arriola}
\affiliation{AWS Center for Quantum Computing, Pasadena, CA 91125, USA}

\author{A.~Khalajhedayati}
\affiliation{AWS Center for Quantum Computing, Pasadena, CA 91125, USA}

\author{R. Sanil}
\affiliation{AWS Center for Quantum Computing, Pasadena, CA 91125, USA}

% ...

\date{\today}

\begin{abstract}
Quantum error correction with erasure qubits promises significant advantages over standard error correction due to favorable thresholds for erasure errors. To realize this advantage in practice requires a qubit for which nearly all errors are such erasure errors, and the ability to  check for erasure errors without dephasing the qubit.
We demonstrate that a “dual-rail qubit” consisting of a pair of resonantly coupled transmons can form a highly coherent erasure qubit, where transmon $T_1$ errors are converted into erasure errors and residual dephasing is strongly suppressed, leading to millisecond-scale coherence within the qubit subspace.
We show that single-qubit gates are limited primarily by erasure errors, with erasure probability $p_\text{erasure} = 2.19(2)\times 10^{-3}$ per gate while the residual errors are $\sim 40$ times lower.
We further demonstrate mid-circuit detection of erasure errors while introducing $< 0.1\%$ dephasing error per check. Finally, we show that the suppression of transmon noise allows this dual-rail qubit to preserve high coherence over a broad tunable operating range, offering an improved capacity to avoid 
frequency collisions. This work establishes transmon-based dual-rail qubits as an attractive building block for hardware-efficient quantum error correction.

\end{abstract}

\maketitle

\section{Introduction}

The path towards useful quantum error correction requires  qubits which have physical error rates well below the error correction threshold. While many physical qubit platforms have reached error rates below the  surface code threshold \cite{google2023suppressing,clark2021high,ryan2021realization,zajac2021spectator,evered2023high}, a major frontier challenge is to continue driving down error rates to reduce the overhead necessary to protect logical qubits. In parallel with this effort, the growing field of hardware-efficient error correction seeks to leverage or engineer the noise properties of physical qubits to more efficiently correct errors, effectively raising thresholds and reducing overhead \cite{guillaud2019repetition, PhysRevX.9.041031, darmawan_practical_2021, cong_hardware-efficient_2022, chamberland_building_2022}. Recent examples include noise-bias engineering and robust qubit encoding in superconducting bosonic modes
\cite{lescanne_exponential_2020, sivak_real-time_2023}.

The recently developed \emph{erasure qubit} offers a compelling path towards more relaxed error correction requirements \cite{wu_erasure_2022, kubica_erasure_2023,DualRailPatent, teoh2023dual}. While most physical qubits exhibit undetectable errors that occur within the qubit subspace, erasure qubits are those in which the dominant error is leakage out of the computational subspace, and for which these leakage errors can be detected in real time.
These \emph{erasure errors} can be corrected more efficiently, leading to higher thresholds and reduced logical error rates \cite{grassl_codes_1997,DualRailPatent, knill_scalable_2004, kubica_erasure_2023, wu_erasure_2022, sahay_high_2023}.
To leverage this advantage, erasure qubits must demonstrate (1) a large erasure noise bias, which is the ratio of erasure errors to residual errors within the qubit subspace,
and (2) the ability to detect erasure errors without introducing additional errors within the subspace. While erasure qubit implementations have been proposed in several architectures including neutral atoms \cite{wu_erasure_2022, sahay_high_2023}, ions~\cite{campbell2020certified, kang_quantum_2023}, and superconducting qubits \cite{kubica_erasure_2023, teoh2023dual}, 
{these ideas have only recently been implemented experimentally using neutral atoms, demonstrating that a large fraction of gate errors can be detected mid-circuit and converted into erasures \cite{ma2023high}, as well as the excision of state preparation and Rydberg decay errors detected as erasures \cite{scholl2023erasure}.
}

In this article, we demonstrate that a superconducting ``dual-rail qubit'' formed by two resonantly coupled transmons meets both requirements for serving as an erasure qubit.
We show that the dual-rail qubit 
converts transmon $T_1$ errors into detectable erasure errors while passively suppressing residual transmon dephasing, enabling a large erasure noise bias
while idling and during single-qubit gates. We then demonstrate high-fidelity mid-circuit erasure detection using an ancilla qubit, and show that this  detection minimally degrades coherence within the dual-rail subspace. Finally, we demonstrate that the dual-rail qubit's passive noise suppression allows it to maintain high coherence while remaining broadly tunable,
offering an extra knob to dodge frequency collisions
and improving the prospects for qubit yield at scale.

\section{Dual-rail qubit}
\begin{figure*}
    \includegraphics[width=\textwidth]{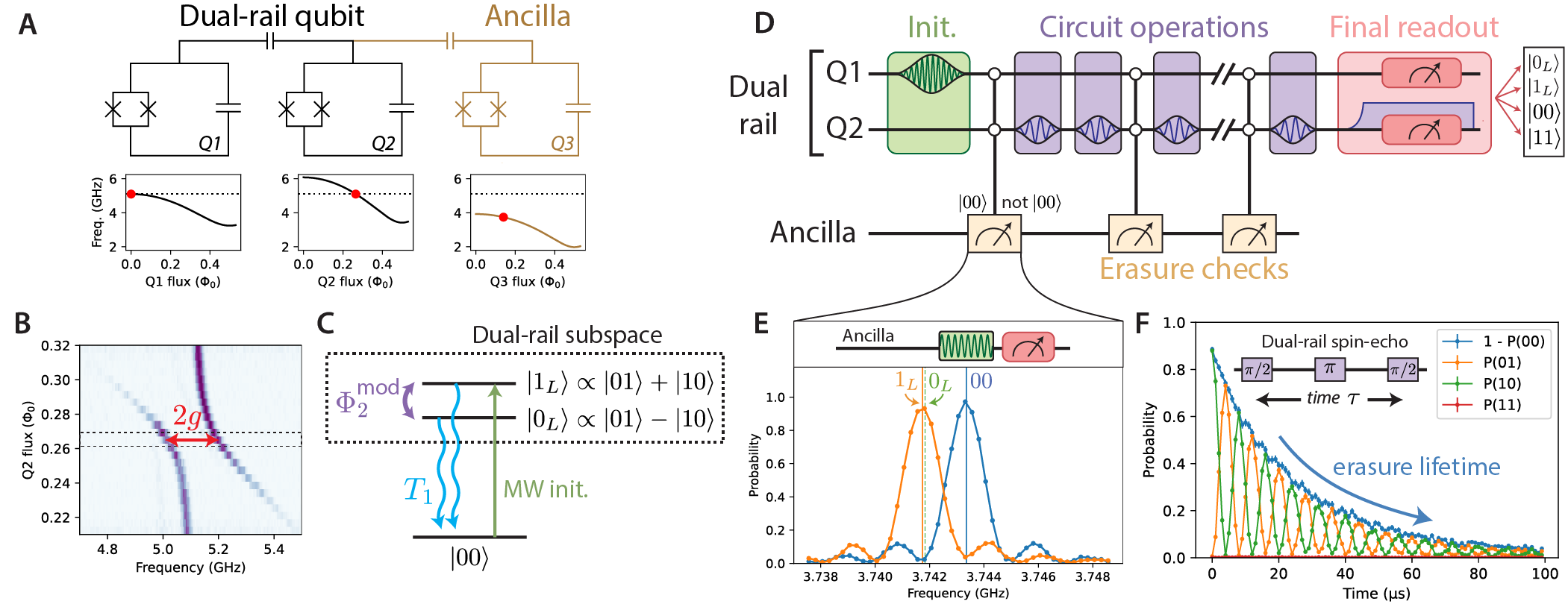}
    \caption{
    \textbf{Dual-rail erasure qubit encoding.}
    (A) Our circuit has three tunable transmons: Q1 and Q2 form the dual-rail qubit, and Q3 is an ancilla qubit used for erasure detection (insets: tuning ranges, with primary operating point marked).
    (B) Microwave spectroscopy of Q1 as Q2 is tuned into resonance, exhibiting an avoided crossing with gap $2g/2\pi=180~$MHz.
    (C) The dual-rail subspace is the symmetric and antisymmetric combination of $\ket{01}$ and $\ket{10}$. 
    Decay of the underlying transmons leads to detectable leakage to $\ket{00}$, constituting an erasure error.
    (D) Circuits begin with a 40~ns microwave pulse to initialize $\ket{1_L}$, 
    followed by single-qubit gates enacted by flux modulation of Q2, and finally the qubits are adiabatically separated and  jointly read-out.
    (E) We perform spectroscopy on the ancilla after initializing the dual-rail pair in $\ket{00}$ (blue), $\ket{1_L}$ (orange), or $\ket{0_L}$ (green, trace not shown). 
    Erasure checks are performed by applying a microwave pulse on the ancilla which is resonant only if the dual-rail pair is in $\ket{00}$ (blue vertical line).
    (F) Spin-echo experiment on the dual-rail qubit where the phase of the final $\pi/2$ is stepped, showing both coherent fringes within the subspace and also decay out of the subspace with a lifetime of $\sim 30~\mu$s.
    Error bars denote $68\%$ confidence  intervals unless otherwise specified.
    }
    \label{fig:Fig1}
\end{figure*}
The dual-rail qubit is defined in the single-excitation manifold of a pair of superconducting modes \cite{chuang_simple_1995, duan_multi-error-correcting_2010, zakka-bajjani_quantum_2011, shim_semiconductor-inspired_2016, campbell_universal_2020,kubica_erasure_2023, teoh2023dual}. We use a pair of resonantly coupled transmons, as first demonstrated in Ref.~\cite{campbell_universal_2020}, where the logical subspace consists of the hybridized symmetric and antisymmetric states $\ket{0_L}=\frac{1}{\sqrt{2}}(\ket{01} - \ket{10})$ and $\ket{1_L} = \frac{1}{\sqrt{2}}(\ket{01}+\ket{10})$. This encoding allows $T_1$ decay of the underlying transmons, which constitutes the fundamental limit to transmon coherence, to be converted to erasure errors in the form of leakage to the $\ket{00}$ state (Fig.~\ref{fig:Fig1}a-c).

Crucially, in addition to converting $T_1$ errors into erasure errors, the resonant coupling between $\ket{01}$ and $\ket{10}$ with strength $g$ acts as a passive decoupling
mechanism which strongly suppresses the impact of low-frequency noise on the underlying transmons \cite{campbell_universal_2020, kubica_erasure_2023}, analogous to continuous dynamical decoupling in driven systems \cite{timoney_quantum_2011, miao_universal_2020, huang_engineering_2021}. 
This is illustrated by the dual-rail energy gap $E_{DR}\approx \sqrt{(2g)^2 + \delta^2}$, where $\delta$ is the frequency difference between the two transmons which inherits frequency noise from each transmon. When $g \gg \delta$, frequency noise is suppressed according to $E_{DR} \approx 2g + \delta^2 / 4g$, allowing the dual-rail energy gap to be orders of magnitude more stable than the underlying transmons (see Appendix~\ref{appendix:CoherenceAnalysis} for a more detailed analysis). This noise suppression enables a large erasure noise bias, in which dephasing errors within the subspace are suppressed and the dominant error mechanism is erasure errors due to the underlying transmon $T_1$.

Our experiments utilize a superconducting quantum circuit with three tunable transmons, two of which (Q1 and Q2) form the dual-rail qubit and are parked on resonance with one another, while the third (Q3) is used as an ancilla qubit
(Fig.~\ref{fig:Fig1}).
Experimental sequences involving dual-rail erasure qubits are illustrated in Fig.~\ref{fig:Fig1}d, with calibration routines presented in Appendices~\ref{Appendix:Calibration},\ref{Appendix:GateCalibration}. Each sequence begins with a microwave pulse on Q1 to initialize the logical $\ket{1_L}$. 
Single-qubit gates within the dual-rail subspace are performed by flux modulation of Q2 at the frequency $2g=2\pi\times180~$MHz, with the phase of the flux modulation defining the axis of the drive field. At the end of the circuit, the dual-rail qubit is measured by adiabatically separating the two transmons with a flux pulse on Q2 such that the two logical states $\ket{0_L}, \ket{1_L}$ map to the pair states $\ket{01}$ and $\ket{10}$, respectively, whereupon the two transmons are jointly read-out \cite{campbell_universal_2020}. Dual-rail qubit operation is illustrated in Fig.~\ref{fig:Fig1}f by a  spin-echo experiment, where the final readout shows both coherent oscillations within the $\ket{0_L}, \ket{1_L}$ subspace as well as leakage to $\ket{00}$ due to transmon $T_1$ decay.

During the experimental sequence, periodic erasure checks may be  performed to identify if the system decayed into $\ket{00}$.
This is done by leveraging a dispersive shift on the ancilla qubit which depends on whether the dual-rail is in $\ket{00}$ or if it remains in the logical subspace (Fig.~\ref{fig:Fig1}e). The erasure check consists of a conditional $\pi-$pulse on the ancilla which is resonant only if the dual-rail is in $\ket{00}$, followed by ancilla state readout.
We note that a similar procedure using a directly coupled readout resonator could also be used for erasure detection, but would require more carefully engineered symmetric coupling strengths to avoid dephasing as discussed in Ref.~\cite{kubica_erasure_2023}.
Such erasure checks are critical tools when using erasure qubits for quantum error correction, and as discussed below, are also necessary for characterizing  the coherence properties of the dual-rail qubit.

\section{Coherence within the dual-rail subspace}

\begin{figure}
    \centering
    \includegraphics{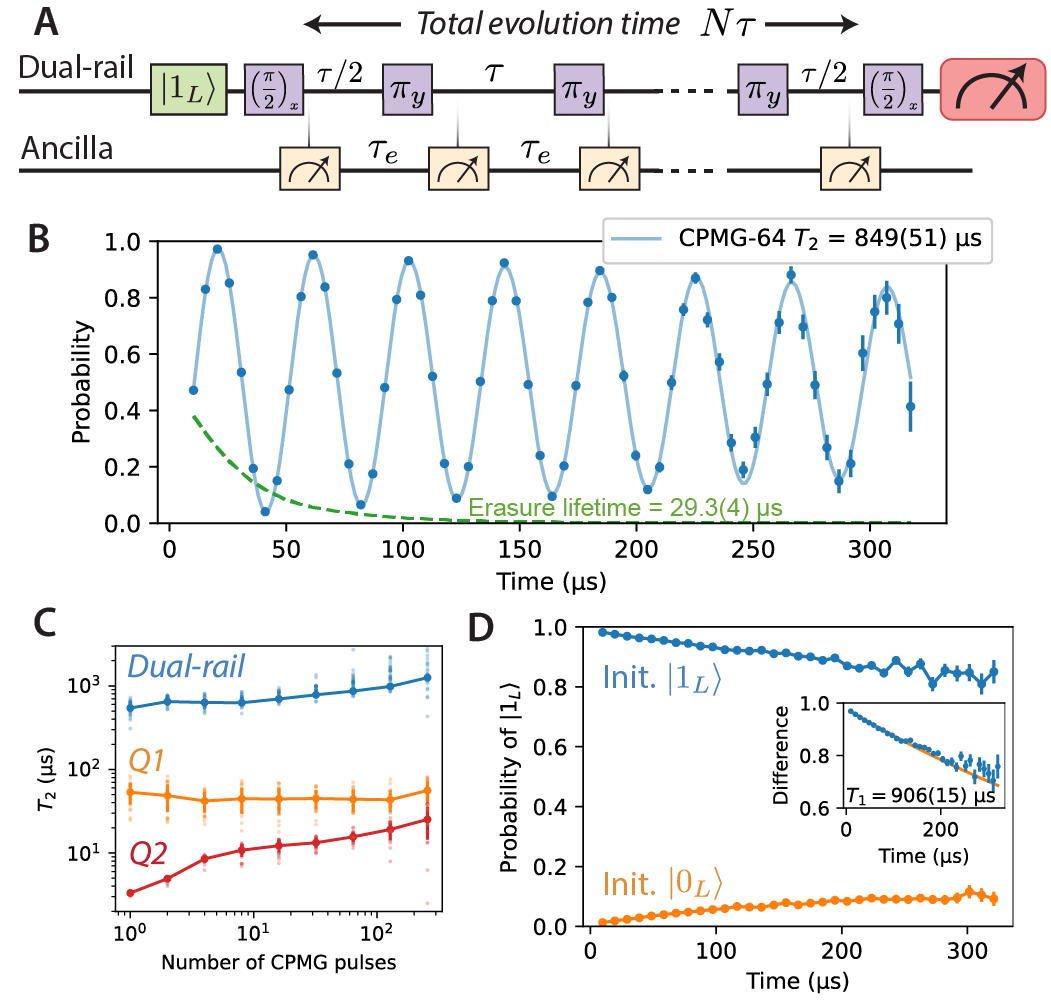}
    \caption{\textbf{Millisecond-scale coherence within the dual-rail subspace.} (A) We measure $T_2$ coherence with a CPMG sequence \cite{gullion_new_1990} consisting of $N$ $\pi$-pulses  with a total evolution time of $N\tau$. In parallel, we perform a sequence of $N_{e} = 11$ erasure checks, evenly spaced by $\tau_e = N\tau / (N_e-1)$.
    Coherence data is postselected against leakage using both the final dual-rail readout and also the mid-circuit erasure checks.
    (B) Example experiment with N=64 $\pi$-pulses, where the final $\pi/2$ pulse phase is stepped to produce fringes. The postselection probability decays with time according to a $29.3(4)~\mu$s erasure lifetime, while the postselected shots are fit to an exponential decay with extrapolated $T_2=849(51)~\mu$s.  330,000 shots were taken per point, but only a small fraction of these survive postselection. 
    (C) CPMG measurements on each underlying transmon, if parked alone at the operating point, as well as on the dual-rail qubit. Error bars are standard deviation of many individual measurements.
    (D) $T_1$ within the dual-rail subspace. We measure the probability of ending in $\ket{1_L}$, postselected against erasure errors, after initializing in either  $\ket{1_L}$  (blue) or $\ket{0_L}$  (orange). The difference between these traces is fit to an exponential decay with fixed offset 0, giving $T_1 = 906(15)~\mu$s.
    The 24 hour time-trace of data comprising these dual-rail $T_1$, $T_2$ plots is shown in Appendix~\ref{appendix:DualRailStability}.
    }
    \label{fig:Fig2}
\end{figure}

A key metric for erasure qubits is the erasure noise bias, which should be $\gg 1$ in order to benefit most significantly from erasure conversion \cite{wu_erasure_2022}. This should hold for errors while idling, requiring that the coherence within the logical subspace is much longer than the erasure lifetime, as well as during gates.
To probe error rates within the subspace, we perform coherence and gate benchmarking experiments and postselect on shots where the system stays in the dual-rail subspace.
We note that postselection is used here only to characterize the non-erasure error rates and would not be needed in a concatenated surface code architecture as described in Ref.~\cite{kubica_erasure_2023}.

We postselect against leakage by relying both on the final readout of the dual-rail pair as well as a sequence of erasure checks performed during the measurement (Fig.~\ref{fig:Fig2}a). These mid-circuit erasure checks are important to catch events in which the system decays into $\ket{00}$ during the circuit but then heats back into the dual-rail subspace; without such checks, this decay/heating process limits the dephasing time that can be measured using only the final readout.
Erasure checks are evenly spaced during the coherence measurement and successfully mitigate the contributions of decay/heating events as long as they are repeated with a spacing shorter than the transmon $T_1$ times (see further discussion in Appendix~\ref{appendix:TransmonHeating}).

We find that while the two transmons Q1 and Q2 individually have $T_2^{\text{CPMG}}$ at the microsecond and tens of microseconds scales, the $T_2^\text{CPMG}$ within the dual-rail subspace (postselected against erasure errors) is $> 500~\mu$s, ranging from $543(23)~\mu$s for one echo pulse to $1.25(8)~$ms for 256 echo pulses (Fig.~\ref{fig:Fig2}b,c). The dual-rail $T_1$
fits to an extrapolated $T_1 = 906(15)~\mu$s (Fig.~\ref{fig:Fig2}d). These millisecond-scale coherence times are far longer than the timescale for decay out of the subspace into $\ket{00}$, which occurs at a characteristic erasure lifetime $T_{eras} \sim 30~\mu$s (Fig.~\ref{fig:Fig2}b) set by the underlying transmon $T_1$ values $15 - 35~\mu$s (Table~\ref{tab:DeviceProperties}). The ratio of these timescales defines the erasure noise bias for idling errors, $ T_2^\text{CPMG} / T_{eras} \gtrsim 20$, confirming that the dominant error on the dual-rail qubit while idling is erasure errors.

Several effects can limit the coherence within the dual-rail subspace at the millisecond level and are discussed in Appendix~\ref{appendix:DualRailDecoherence}. The most dominant contribution is thermal Johnson-Nyquist noise on the flux lines, which is expected to limit the $T_1$ within the dual-rail subspace to $\sim 1$~ms. We therefore hypothesize that this is the limiting contribution for our measured $T_1$; this effect can be mitigated with further cryogenic attenuation of the fast-flux line and will be studied in future work. Other effects, including residual sensitivity to transmon frequency noise, photon fluctuations in readout resonators, and leakage into the two-photon manifold, can all limit dual-rail coherence at the level of several milliseconds. While the measured $T_1$ is attributed to Johnson noise, the $T_2$ may be limited by a combination of these effects and requires further investigation.

We  find that the dual-rail $T_2^*$ phase coherence without dynamical decoupling 
is limited by a separate effect: slow, seconds-scale telegraph noise on the dual-rail frequency (Fig.~\ref{fig:RamseyBeating}).
We correlate this frequency instability with similar telegraph noise  on one of the underlying transmons (Q1) and attribute this to a nearby toggling two-level system (TLS) defect at 4.98 GHz (Appendix~\ref{appendix:TelegraphNoise}).
This behavior is consistent with a dispersive coupling model discussed in Appendix~\ref{appendix:TLS} and may offer a new probe for TLS behavior.
Nonetheless, the switching time is sufficiently slow that this noise is effectively mitigated with dynamical decoupling.

\begin{figure}
    \centering
    \includegraphics[width=3.43in]{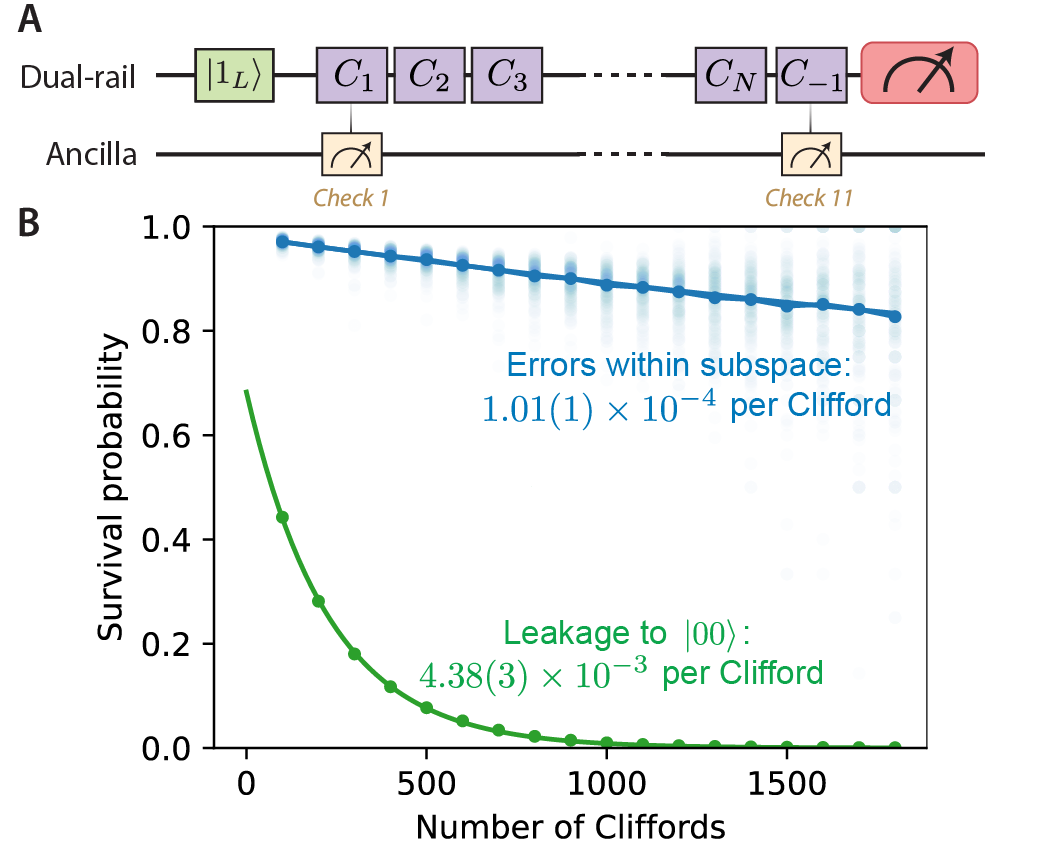}
    \caption{\textbf{Single-qubit randomized benchmarking.} (A) The randomized benchmarking circuit uses random Cliffords, each implemented as two X90 pulses with appropriate phase shifts \cite{mckay_efficient_2017}. In parallel, we perform  11 erasure checks evenly spaced throughout the circuit, regardless of depth. The X90 pulse is a Gaussian-shaped 48~ns flux pulse on Q2 along with a $0.067~$rad phase correction. An example timing diagram is presented in Fig.~\ref{fig:PulseSequence}.
    (B) We plot the postselection probability (green) as a function of circuit depth and fit to an exponential decay to 0 to extract the erasure error per Clifford of $4.38(3)\times 10^{-3}$.
    The postselected survival probability within the dual-rail subspace (blue), averaged over 200 random circuits for each depth, is fit to an exponential decay with offset $1/2$ and gives a residual error rate of $1.01(1) \times 10^{-4}$ per Clifford.
    Individual outcomes for random circuits are plotted with transparency.
    We symmetrize readout errors by alternating measurements in which the ideal outcome is $\ket{0_L}$ or $\ket{1_L}$. 
    The X90 gate errors are half those of the Clifford gate, with erasure error $2.19(2)\times 10^{-3}$ and residual error $5.06(6)\times 10^{-5}$. Similar error rates are computed when postselecting only on mid-circuit erasure checks (Appendix~\ref{appendix:RBSequence}).
    }
    \label{fig:Fig3}
\end{figure}

To validate that long dual-rail coherence times with dynamical decoupling translate to high-fidelity gates, we characterize single-qubit X90 gates on the dual-rail qubit, enacted by 48~ns flux modulation pulses on Q2 (Appendix~\ref{Appendix:GateCalibration}). 
We perform Clifford randomized benchmarking (Fig.~\ref{fig:Fig3}), in parallel with frequent erasure checks, and measure an erasure error probability of $2.19(2) \times 10^{-3}$ per X90, consistent with the duration of the gate relative to the erasure lifetime. Postselected against erasures, we find a low residual error rate of $5.06(6) \times 10^{-5}$ per X90, 
 averaged over a 24 hour period, with the remaining error likely limited by a combination of residual decoherence within the logical subspace and coherent calibration errors. These error rates constitute an erasure noise bias of $43(1)$, and 
show that single-qubit gates preserve the large noise bias achieved while idling.

\section{Mid-circuit erasure detection using an ancilla qubit}
\begin{figure}
    \includegraphics{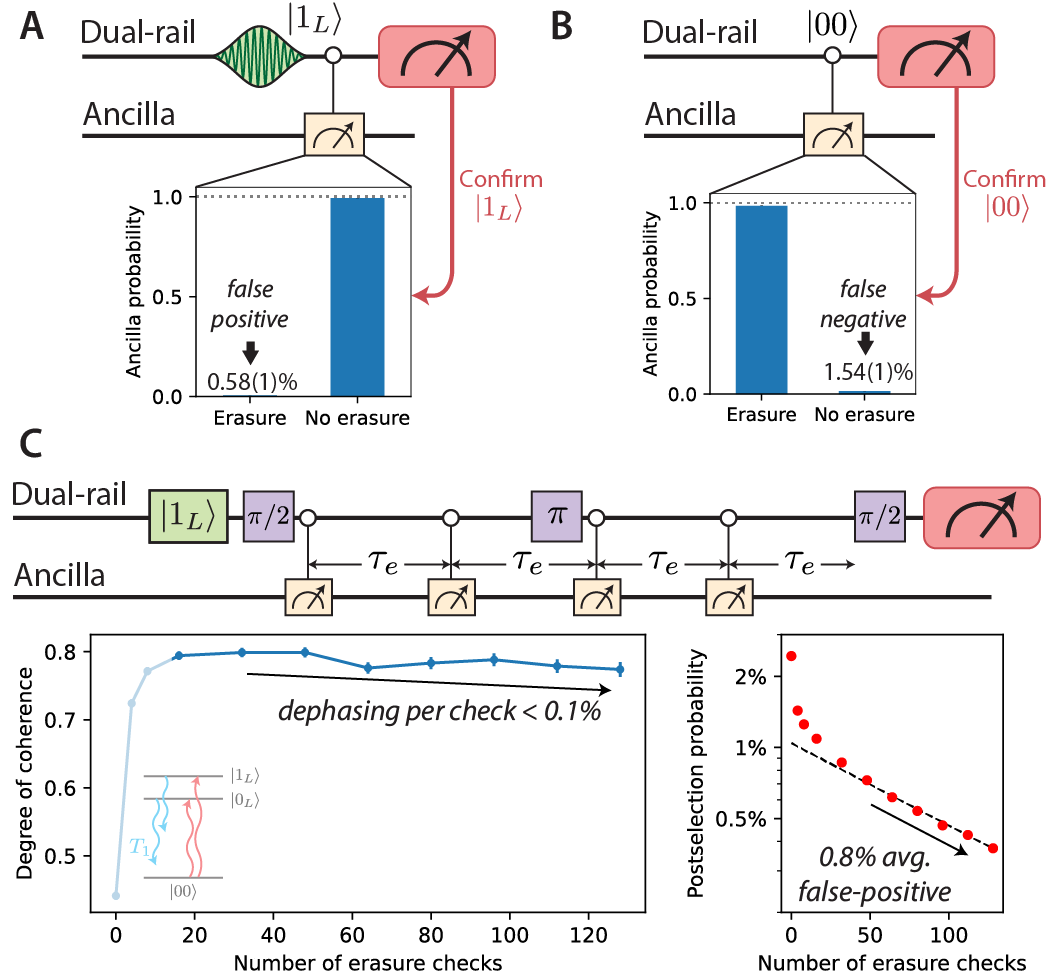}
    \caption{\textbf{Characterizing mid-circuit erasure detection.} (A,B) We characterize false-positive and false-negative errors by initializing a target state, either $\ket{1_L}$ or $\ket{00}$, performing an erasure check, and then analyzing its results post\-selected on a final readout confirming the correct initialization. We ensure initialization of the ancilla in $\ket{0}$ by postselecting on an additional pre-readout before the sequence begins (not shown).
    (C) We characterize erasure-check-induced-dephasing  by inserting a variable number of checks into a spin-echo experiment with fixed total evolution-time  $T=114~\mu$s. The $N$ erasure checks are evenly spaced with separation time $\tau_e = T/N$. 
    We vary the phase of the final $\pi/2$ pulse to measure the remaining degree of coherence, postselected on all erasure checks indicating no erasure. Coherence appears to grow with a small number of checks due to properly catching and postselecting against leakage into $\ket{00}$ which re-heats into the subspace. This effect saturates, after which we observe little dephasing per check at the level of $< 0.1\%$ (see Appendix~\ref{appendix:ErasureCheckDephasing}). The postselection probability (right subplot) decreases with increasing erasure checks, first due to discarding shots with leakage, then saturating at a rate of $\sim 0.8\%$ per check which we interpret as a combination of false-positive assignment errors and check-induced erasure events.
    We note that a positive erasure detection event can with low-probability $0.0293(3)\%$ re-excite the dual-rail qubit (see Appendix~\ref{appendix:Readout}).
    }
    \label{fig:Fig4}
\end{figure}

We next turn to the characterization of mid-circuit erasure detection.
Our erasure check relies on a dispersive shift ($2\pi \times 1.6~$MHz) induced on the ancilla qubit when the dual-rail qubit is in the logical subspace ($\ket{0_L}, \ket{1_L}$) relative to when it is in $\ket{00}$ (Fig.~\ref{fig:Fig1}e, precise shifts given in Appendix~\ref{appendix:DispersiveShifts}) \cite{livingston_experimental_2022}. The erasure check consists of a 540~ns square microwave pulse on the ancilla qubit whose length is chosen to excite the ancilla only if the dual-rail is in $\ket{00}$ and not if the dual-rail is in its logical subspace (Fig.~\ref{fig:Fig1}e). After the microwave pulse, we perform a 340~ns readout of the ancilla transmon (Appendix~\ref{appendix:Readout}).

There are three metrics to evaluate the performance of erasure detection, motivated by the proposed use of erasure checks in an error-correcting code \cite{wu_erasure_2022, kubica_erasure_2023}.
First, \emph{false-positive} errors are those in which no error occurred but an erasure is falsely flagged. Second, \emph{false-negative} errors are those in which an erasure error occurs but is not correctly flagged.  Finally,  \emph{erasure-check-induced-dephasing} is the dephasing on the dual-rail qubit induced by  checking for an erasure error when no such error had occurred. To establish target performance levels for these metrics, we consider a surface code protocol in which an erasure check is performed {alongside each} two-qubit gate to prevent the diffusion of errors. 
Erasure check errors will add to the effective error rates of the gates, {and in instances where the erasure check is slower than the gates an additional error will also be introduced due to the time associated with the erasure check.}
For concreteness, we benchmark against $1\%$ erasure error per gate and $0.1\%$ Pauli error per gate; this pair of error rates is comfortably below the surface code threshold \cite{kubica_erasure_2023}.

False positives are characterized by initializing the dual-rail in $\ket{1_L}$, performing an erasure check, and then reading out the dual-rail  pair (Fig.~\ref{fig:Fig4}a). We postselect on the final pairwise readout showing the correct state was initialized, and measure the probability that the erasure check correctly indicated no error. We find a false-positive rate of $0.58(1)\%$ for $\ket{1_L}$, limited by accidental excitation of the ancilla when the dual-rail has not decayed. We find a higher false-positive rate of $\sim 0.8\%$ for the logical state $\ket{+_L}$ (measured independently in Fig.~\ref{fig:Fig4}c), which we take as an effective average rate that does not distinguish between assignment errors and any mechanism which may induce an erasure error during the check. In a surface code context, this type of error would ``inject" extra erasure errors into the system, and should thus be compared to the $1\%$ erasure error target.

False negatives are characterized  by initializing the dual-rail in $\ket{00}$, performing an erasure check, and then postselecting on a final readout which shows $\ket{00}$ (Fig.~\ref{fig:Fig4}b). We find a false-negative rate of $1.54(1)\%$, limited by $T_1$ decay of the ancilla during readout. In a surface code, the quantity of interest is the probability that an erasure event occurred and was not properly flagged in detection, thus persisting  and possibly propagating to undetectable Pauli errors. The probability of such an event is the chance of an erasure error during the erasure check, $T_{check} / T_{eras} \sim 2.9\%$, multiplied by the false negative rate. The probability of a missed erasure is then $0.04\%$, comfortably below the $0.1\%$ target Pauli level.

Finally, we measure the dephasing induced by each erasure check by performing a spin-echo measurement on the dual-rail and inserting a variable number of erasure checks (Fig.~\ref{fig:Fig4}c). The dual-rail coherence appears to improve and then plateau when inserting a small number of erasure checks, as these checks correctly eliminate shots in which the system decayed to $\ket{00}$ and then heated back into the subspace. As more checks are inserted, remaining phase coherence degrades only minimally, from which we compute a conservative upper bound of $< 0.1\%$ error per erasure check (Appendix~\ref{appendix:ErasureCheckDephasing}).
Such dephasing errors inject undetectable Pauli errors in each check, and thus should be compared to the target $0.1\%$ Pauli error level.

{While the fidelities for these three  metrics are below their respective surface code thresholds, the current erasure check time (880~ns) is slower than two-qubit gate protocols in this architecture of $\sim 200$~ns \cite{campbell_universal_2020,kubica_erasure_2023}, and would contribute a larger erasure error of $T_{check} / T_{eras} \sim 2.9\%$ per gate.
Faster erasure checks can be achieved through larger dispersive couplings and optimized ancilla readout \cite{heinsoo_rapid_2018}, or by using a single symmetrically coupled readout resonator \cite{kubica_erasure_2023}.
Additionally, ancilla reset and re-initialization of the dual-rail after an erasure event are still needed, and we expect implementations
can be closely adapted from standard transmon reset and leakage reduction protocols \cite{riste_feedback_2012, mcewen_removing_2021, marques_all-microwave_2023}.
}

\section{Robust operation at flexible operating points}

\begin{figure}
    \centering
    \includegraphics{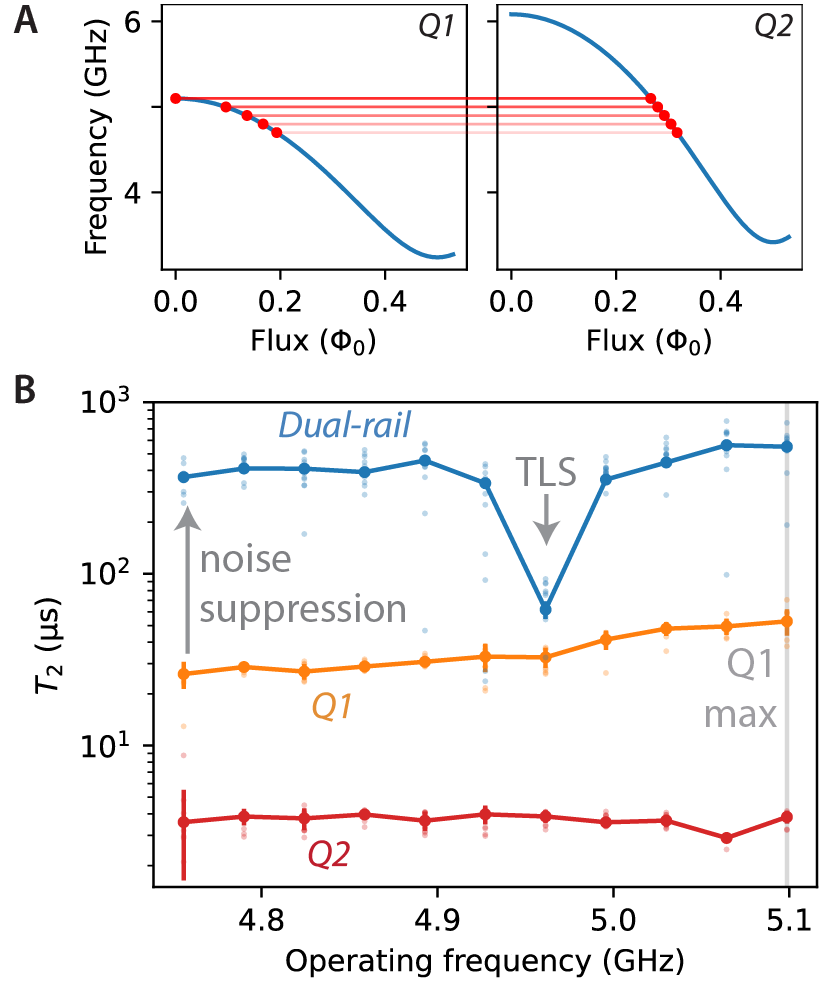}
    \caption{\textbf{Robust operation away from sweet spot.} (A) The dual-rail qubit can be operated anywhere the two transmons can be brought onto resonance, forming an engineered ``tunable" sweet spot. (B) At a range of operating points, we measure $T_2^{echo}$ of each underlying transmon individually as well as of the dual-rail qubit. With the exception of a narrow region around 4.96 GHz, the dual-rail qubit maintains a coherence of several hundred microseconds over most of this range ($366(20)~\mu$s to $550(30)~\mu$s), with the modest degradation likely due to increased effects of Q1's flux noise away from its sweet spot (Appendix~\ref{appendix:DualRailDecoherence}). These coherence times are over an order of magnitude larger than the higher-coherence transmon (Q1) and more than two orders of magnitude larger than Q2.
    The dip in dual-rail $T_2$ at 4.96 GHz is likely explained by a TLS coupled to Q2 which is near-resonant with the upper hybrid mode at this operating point; a similar reduction is present on Q2 directly at an offset operating point, as visible in supplementary datasets presented in Appendix~\ref{appendix:SupplementaryTunable}.
    Error bars are standard deviations of many fit results (individual dots) for Q1 and Q2, and are fit uncertainties for the  dual-rail qubit.
    }
    \label{fig:Fig5}
\end{figure}

The dual-rail qubit enables a large erasure noise bias as well as high-fidelity erasure detection. Importantly, these features are not fine tuned based on special operating frequencies, but instead can be preserved at a broad range of operating points. This robustness offers a significant advantage over standard  architectures in which tunable transmons are expected to operate close to their flux-insensitive  sweet spots, but may be unable to due to frequency collisions or parasitic modes / TLSs. While operating tunable transmons away from the sweet spot is possible, extensive dynamical decoupling is needed to recover $T_2$ performance, and this decoupling may complicate single-qubit or multi-qubit gates \cite{khodjasteh_dynamically_2009}. In the dual-rail qubit, however, this noise suppression is achieved passively, and carries over into the operation of single-qubit gates, erasure detection, and proposed implementations of multi-qubit gates \cite{kubica_erasure_2023}.

We test this robustness by parking the two dual-rail transmons at several operating points over a 350 MHz band from 4.75 to 5.1 GHz which is mutually accessible by each. We park each transmon individually at each of these operating points to first characterize their individual properties, and find that Q2 has fairly uniform coherence over this range since it is already far from its sweet-spot, while Q1's coherence degrades as it is tuned away from the sweet-spot (Fig.~\ref{fig:Fig5}). Across this full range, except for a narrow band at 4.96 GHz impacted by a TLS (discussed in Appendix~\ref{appendix:SupplementaryTunable}), the dual-rail qubit preserves $T_2^{echo}$ coherence of hundreds of microseconds, from $366(20)~\mu$s to $550(30)~\mu$s. 
While more investigations are needed to fully understand  coherence limitations, these measurements highlight the resilience of the dual-rail qubit to flux noise which normally precludes operating at highly flux sensitive points. Additional adjustments of the qubit parameters, such as tuning ranges and coupling strengths, may improve performance and robustness further.

\section{Discussion and outlook}
These results show that the dual-rail qubit is a promising, highly coherent building block for erasure-based error correction architectures. Coherence within the dual-rail subspace, even for transmons which are individually  noisy, is approaching the millisecond regime, while the erasure lifetime is set by the transmon $T_1$. 
Subsequent studies may complete the toolbox for operating an erasure surface code, including developing two-qubit gates between dual-rail pairs \cite{campbell_universal_2020},  re-initializing dual-rail qubits after erasure errors, and exploring novel erasure detection mechanisms \cite{kubica_erasure_2023}.

While the dual-rail transmon architecture is more complex than standard transmon architectures, an optimized approach can be taken which uses a single resonator for both readout and erasure detection \cite{kubica_erasure_2023}. In this setup, a dual-rail qubit involves only slightly more space on the processor and only one additional control line compared to standard transmons (Appendix~\ref{appendix:ResourceEstimation}).
For this modest increase in complexity, the conversion of $T_1$ decay to erasure errors enables leveraging of higher thresholds and effectively larger code distances as compared with Pauli errors, offering a positive prospect for accelerating the path to near-term below-threshold operation of error-corrected processors and for long-term logical qubit performance.
{Additionally, the broad effort towards scaling transmon systems and improving  transmon coherence will translate naturally into the dual-rail architecture; for example, improvement in transmon $T_1$ will result in lower erasure error rates.}

The passive noise suppression also opens  opportunities to build dual-rail qubits out of other components which have long $T_1$ but shorter $T_2$ coherence times. Candidate components would include transmon-like circuits which operate at lower values of $E_J/E_C$, phase qubits, or potentially fluxonium. Such physical qubits, which in some regimes are undesirable when used on their own due to short coherence, may form highly robust logical qubits in a dual-rail architecture with improved erasure lifetime while still maintaining strong phase coherence within the dual-rail subspace.

Finally, in addition to prospects for enhancing logical qubit performance, we note that the ability of erasure qubits to postselect against errors may already offer significant advantages for some tasks including quantum simulation \cite{mcardle2019error, scholl2023erasure}, sampling \cite{yang_post-selection_2023}, and non-verifiable algorithms~ \cite{simon1997power,mcardle2019error,botelho2022error}. For these applications, postselecting on the lack of erasure errors enables more accurate reconstruction of the probability distributions than would be achieved by conventional quantum processors even with lower overall error rates \cite{scholl2023erasure}. 

\emph{Note:} During completion of this manuscript, we became aware of related work on the dual-rail encoding in microwave cavities \cite{chou2023demonstrating}.

\section{Acknowledgements}
We thank the technical support from across the AWS Center for Quantum Computing, including the teams involved with theory, design, fabrication, device packaging, cryogenics, signals, software, procurement, and lab infrastructure. We also thank Simone Severini, Bill Vass, and AWS for supporting the quantum computing program.
Finally, we thank Jeff Thompson, Manuel Endres, and David Schuster for helpful discussions and feedback on the manuscript.

%

\appendix
\clearpage

\section{Superconducting device}
\label{Appendix:Processor}

\begin{figure*}
    \centering
    \includegraphics[width=5in]{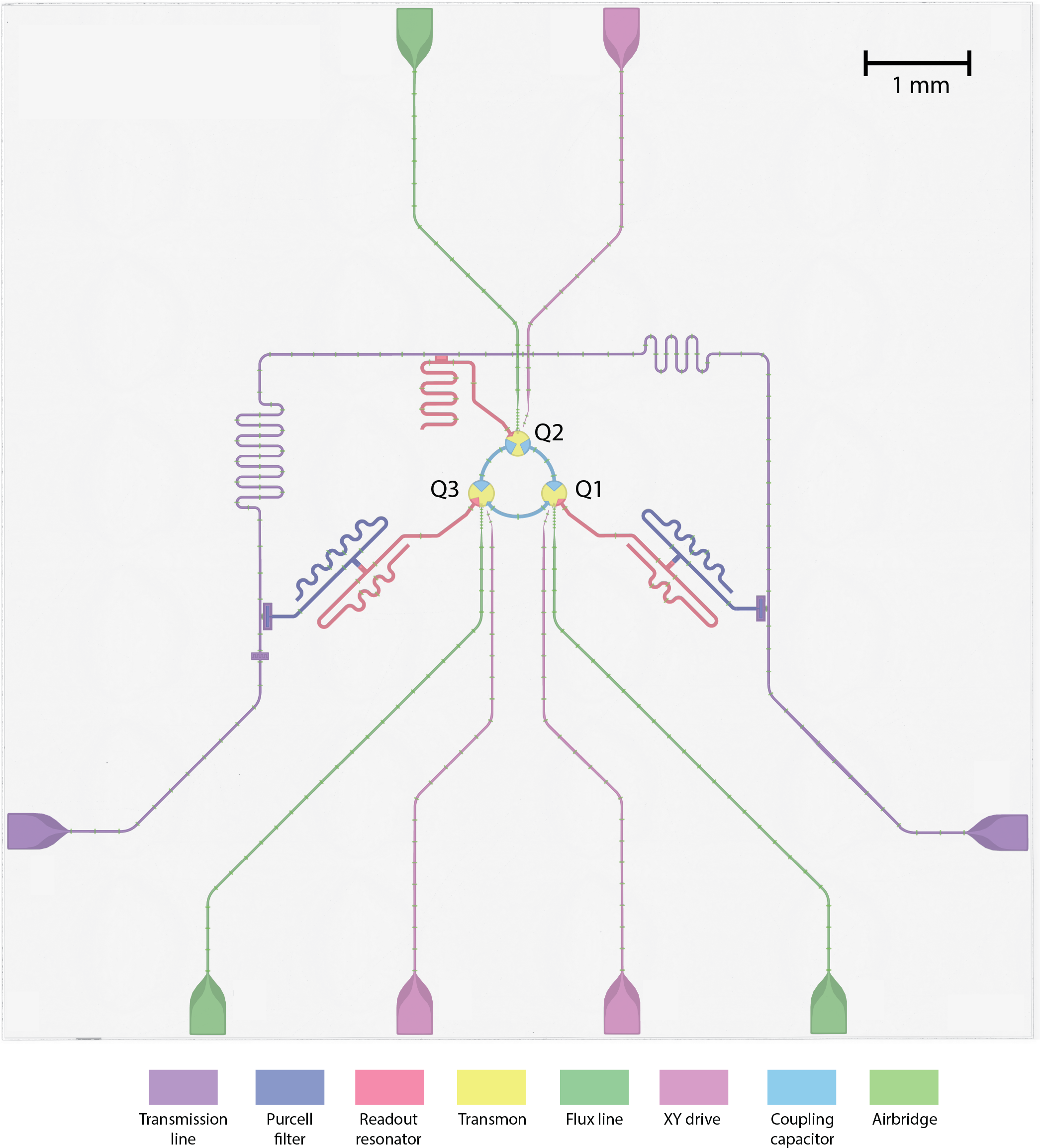}
    \caption{\textbf{Optical image of superconducting device.} False-color overlays indicate the functional purpose of each component needed to define the transmon qubits, control them with microwave and flux lines, and perform readout.}
    \label{fig:processor}
\end{figure*}
The superconducting quantum circuit is shown in Fig.~\ref{fig:processor}, and device parameters are summarized in Table~\ref{tab:DeviceProperties}. Each of the three transmons has a dedicated readout resonator, all of which are coupled to a single transmission line. Q1 and Q3 have additional Purcell filters on their resonators. { Transmon capacitive couplings are realized via wedge couplers \cite{krinner_realizing_2022}.}
\begin{table}
    \centering
    \begin{tabular}{c|c|c|c}
         Property & Q1 & Q2 & Q3 (ancilla) \\
         \hline
         $\omega_{min}/2\pi$ (GHz) & 3.1 & 3.3 & 2.5 \\
         $\omega_{max}/2\pi$ (GHz) & 5.1 & 6.1 & 3.95 \\
         $\omega_{idle}/2\pi$ (GHz) & 5.1 & 5.1 & 3.74 \\
          $\eta/2\pi$ (MHz) & 193 & 204 & 196 \\
          $T_1$ ($\mu$s) & 36(12) & 14(4) & 38(5) \\
         $T_2^*$ ($\mu$s) & 31(14) & 1.29(6) & 4.4(2)$^*$ \\
         $g_{12}/2\pi$ (MHz) & 90.1 & 90.1 & - \\
         $g_{13}/2\pi$ (MHz) & 8.4 & - & 8.4 \\
         $g_{23}/2\pi$ (MHz) & - & 81.7 & 81.7 \\
         $\omega_{RO}/2\pi$ (GHz) & 7.749 & 7.511 & 7.341 \\
         $\chi_{RO}/2\pi$ (MHz) & 3.73(7) & 0.32(1) & 2.53(3) \\
         $\kappa_{RO}/2\pi$ (MHz) & 9.3(4) & 0.87(2) & 6.7(2) \\
    \end{tabular}
    \caption{\textbf{Summary of device parameters.} Transmon frequencies and coherence times, coupling strengths $g_{ij}$, and readout resonator (RO)  frequencies $\omega_{RO}$ and $\kappa, \chi$ values, reported at the qubit idling points.  Coherence values are the median of $\sim 25$ measurements,
    with the listed uncertainty being the standard deviation of those measurements ($^*$except $T_2^*$ for Q3, which is from a single measurement). All coherence fits are to exponential decays which properly capture the qualitative behavior but neglect some small systematic deviations from exponential behavior such as beating in Ramsey $T_2^*$ fringes. The coupling $g_{12}$ is directly measured, while the couplings $g_{13}$ and $g_{23}$ are calculated  based on the measured ancilla dispersive shifts in Appendix~\ref{appendix:DispersiveShifts}.}
    \label{tab:DeviceProperties}
\end{table}

Our device is fabricated using electron-beam lithography (EBL) to pattern components onto a 100~nm aluminum ground plane. The ground plane is connected across coplanar waveguides through cross-overs, realized as aluminum air-bridges. The Josephson junctions are aluminum-based, fabricated using EBL followed by double-angle evaporation.
The junction electrodes are shorted to the base layer metallization using Al-based bandages.

\begin{figure}
    \centering
    \includegraphics[width=3.3in]{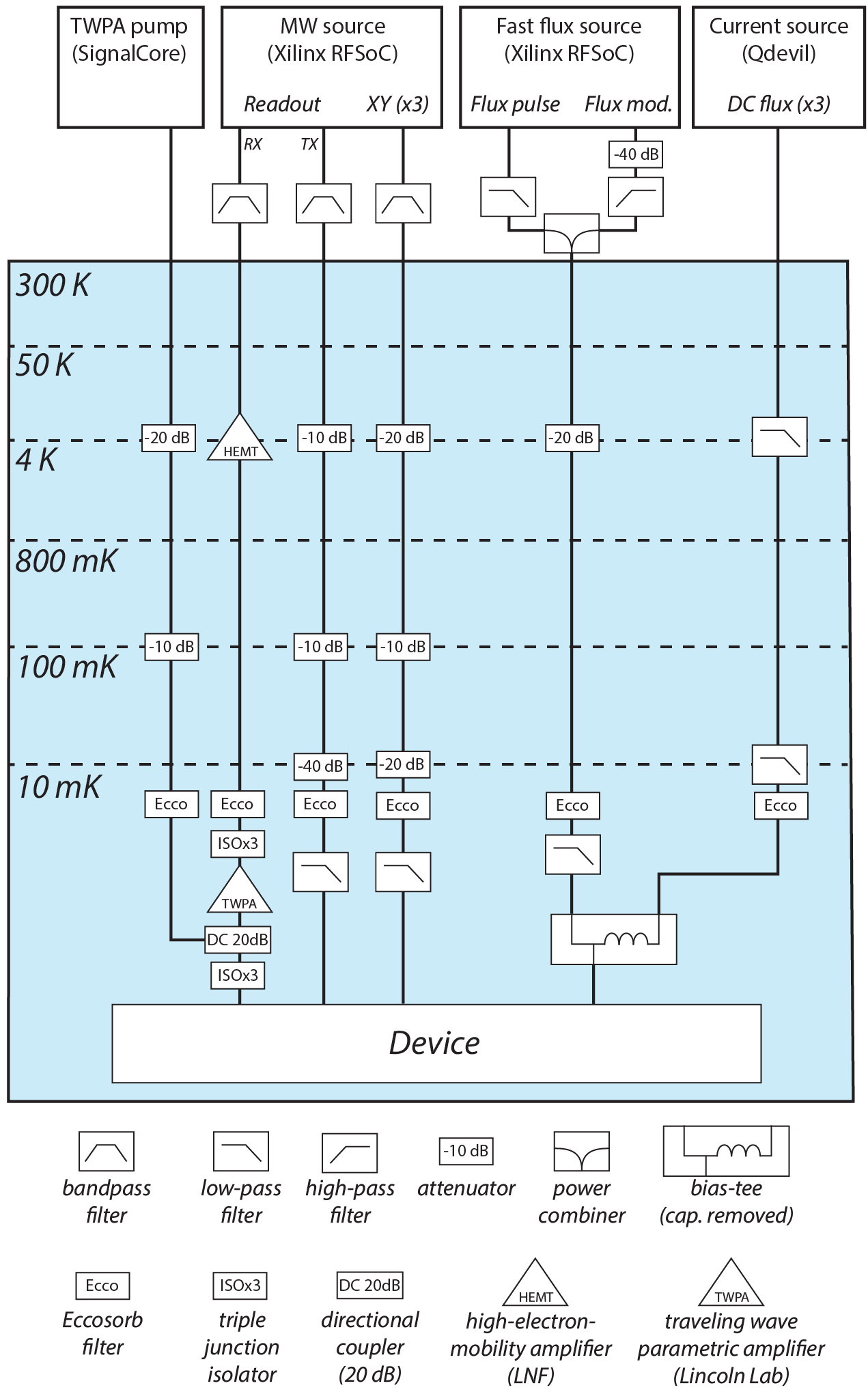}
    \caption{\textbf{Controls and cryogenic wiring.} We show the signal chain including components at various temperature stages within the dilution refrigerator. A single readout chain is used for the processor. Each qubit has an XY control line, as well as a slow-flux line which is combined with a fast-flux line on a bias-tee with the capacitor removed from the fast-flux side to enable low-frequency flux pulses. The fast-flux line for Q2 is generated by combining a signal which is low-pass filtered (cutoff $<22~\text{MHz}$), used for flux-assisted readout, along with another source which is attenuated and high-pass filtered (cutoff $>41~\text{MHz}$), used for flux modulation at $2g=2\pi\times180~\text{MHz}$ to drive single-qubit gates; this setup mitigates high-frequency control noise which could limit the dual-rail $T_1$. The fast-flux lines for the other qubits (Q1 and Q3) are not needed for these experiments and are terminated with $50~\Omega$ at room temperature (not shown). }
    \label{fig:cryostat}
\end{figure}

The device is wirebonded to a printed circuit board and cooled to $\sim 10~$mK at the base of a Bluefors dilution refrigerator. Microwave and baseband signals are delivered from custom control hardware based on a Xilinx RFSoC, with the signal chains shown in Fig.~\ref{fig:cryostat}.

\section{Transmon readout}
\label{appendix:Readout}
Fast transmon readout, particularly for the ancilla qubit, is critical for erasure qubit experiments. We use a traveling wave parametric amplifier (TWPA) \cite{macklin_nearquantum-limited_2015} from MIT Lincoln Laboratory to amplify readout signals and Purcell filters to facilitate fast readout without degrading qubit lifetimes. The ancilla readout consists of a 200~ns microwave tone on the transmission line, while we integrate the readout signal against a linear matched filter which extends for an additional 140~ns after the drive tone ends \cite{ryan_tomography_2015}.
We assign thresholds in the IQ plane to preferentially minimize false-positive rates (typically reaching $\lesssim 0.1\%$) at the cost of slightly increasing false-negative rates, which are mainly limited by the few-percent probability of $T_1$ decay during readout. This choice of threshold is helpful to minimize false-positive rates during erasure detection.

The dual-rail transmons Q1 and Q2 are read-out at the end of the circuit over $1~\mu$s. During this period, Q2 is flux-pulsed up to its maximum frequency at 6.1 GHz, while Q1 remains at 5.1 GHz. For the data in Fig.~\ref{fig:Fig5} in which we tune the operating point of the dual-rail qubit below 5.1 GHz, the final readout always occurs while Q2 is flux-pulsed back to its maximum frequency.

We observe that for some operating points, reading out the ancilla can stimulate transitions on the dual-rail qubit. We attribute these to measurement-induced state transitions (MIST) \cite{sank_measurement-induced_2016, khezri_measurement-induced_2023}. In particular, with the ancilla idling at 3.74 GHz, we find that if the ancilla is excited, there is a small $0.0293(3)\%$ chance that its readout will  excite the dual-rail pair from $\ket{00}$. This only occurs if the ancilla was actually excited, which already flags that an erasure error occurred on the dual-rail pair, and thus postselection against erasure handles these errors. We further note that other types of measurement-induced transitions, including fluctuating behavior and transitions when the ancilla is in $\ket{0}$, were observed at other operating points for the ancilla, including at its maximum frequency of 3.95 GHz, motivating the operation of the ancilla below its maximum in this work. In the future, such excitation events may be avoided by alternate allocation of frequencies, or can be addressed by resetting both the ancilla and data-qubits after detection of an erasure event.

\section{Dual-rail Hamiltonian and controls}
The Hamiltonian for the three transmons comprising the dual-rail and ancilla system is given as follows, treating the transmons as nonlinear oscillators:
\begin{equation}
H = \sum_{i=1}^3 (\omega_i a_i^\dag a_i - \frac{\eta_i}{2} a_i^\dag a_i^\dag a_i a_i) + \sum_{i < j} g_{ij} (a_i^\dag a_j + a_j^\dag a_i)
\end{equation}
with the values of these parameters provided in Table~\ref{tab:DeviceProperties}.
We work in a regime where the detuning between the ancilla and the transmons, $\Delta=\omega_3-\omega_2\simeq\omega_3-\omega_1$, is large compared with the anharmonicity $\eta =  \eta_1\approx \eta_2 \approx \eta_3$ and the inter-transmon couplings $g_{ij}$.
As such, the system is well described by an effective dispersive Hamiltonian coupling the ancilla and the dual rail qubit.

This can be derived using the standard black-box quantization approach:  one first diagonalizes the quadratic parts of the above Hamiltonian, and writes the nonlinearity in terms of the resulting eigenmodes (see for example Ref.~\cite{kubica_erasure_2023}).  Keeping only resonant terms, and projecting to the relevant states of the dual-rail qubit ($|00\rangle$, $|0_L\rangle$, $|1_L\rangle$), we have the following effective (rotating-frame) Hamiltonian:
\begin{equation}
    H  = \frac{1}{2}(\chi_0 |0_L\rangle \langle 0_L| + \chi_1 |1_L \rangle \langle 1_L|)\sigma_z^{anc},
\end{equation}
 where $\chi_0\approx\chi_1\approx 2\eta\frac{g_{23}^2}{\Delta^2}\approx 2\pi \times 1.5~$MHz (see Appendix~\ref{appendix:DispersiveShifts}). This amounts to half the usual dispersive coupling between a pair of transmons (Q2 and Q3), which can be understood from the fact that each dual-rail mode has half its weight in Q2.

The difference $\delta\chi=\chi_1-\chi_0$ arises due to subleading terms in $g_{12}/\Delta$ (which lead to a small difference in detuning between the ancilla and the two dual-rail modes), as well as the presence of the small additional coupling $g_{13}$ and potentially a detuning $\delta=\omega_1-\omega_2$.
One obtains
\begin{equation}
    \frac{\delta\chi}{\bar\chi} = 2\frac{g_{12}}{\Delta} +2\frac{g_{13}}{g_{23}} -\frac{\delta}{g_{12}},
\end{equation}
to first order in $\eta/\Delta$, $g_{ij}/\Delta$, $g_{13}/g_{23}$, and $\delta/g_{12}$, where $\bar\chi$ is the average of $\chi_0$ and $\chi_1$. 
The non-resonant terms in the expansion of the nonlinear Hamiltonian add an additional perturbative contribution to $\delta\chi/\bar{\chi}$ which is of second order in the above small parameters. In our parameter regime, $|\delta\chi/\bar{\chi}| \ll 1$, enabling the erasure-check approach illustrated in Fig.~\ref{fig:Fig1}e.

\textbf{Control fields:} The two main control fields for the dual-rail qubit are (1) microwave initialization ($|00\rangle \to |1_L\rangle$) and (2) flux modulation for driving single-qubit gates. Since the dual-rail modes are hybridized across Q1 and Q2, these fields can in principle be applied to either transmon. In particular, a charge drive applied to either transmon can couple from $\ket{00}$ to both logical states (as is visible in the spectroscopy signal in Fig.~\ref{fig:DualRailCalibration}a).

For single-qubit gates, we consider a simplified Hamiltonian for Q1 and Q2, given by $H=g(\sigma_1^+ \sigma_2^- + h.c)+\delta(t) \sigma_2^z$ describing a pair of coupled qubits with a time-dependent differential frequency \cite{campbell_universal_2020}. Modulating $\delta(t)$ at frequency $2g$ drives resonant rotations within the logical subspace. This frequency modulation can be realized by flux modulation on either transmon, but it is preferable to modulate the transmon whose frequency vs. flux sensitivity is most linear to avoid shifts in the average frequency during modulation.

\section{Dual-rail qubit calibration}
\label{Appendix:Calibration}
\begin{figure*}
    \centering
    \includegraphics{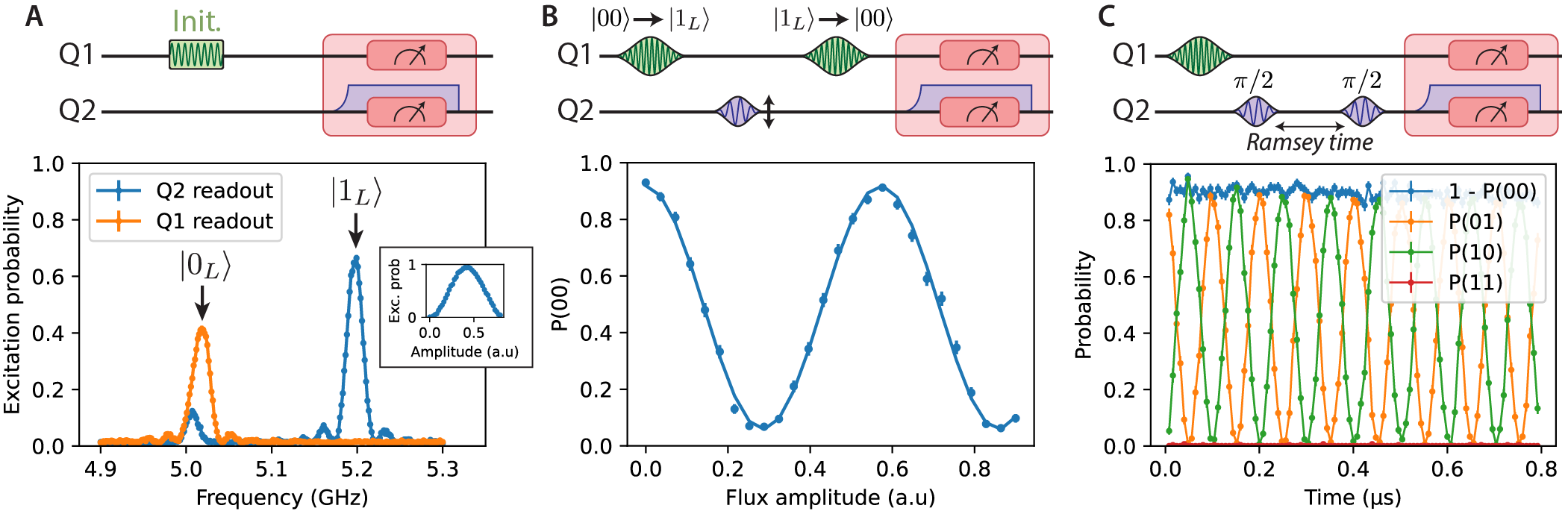}
    \caption{\textbf{Dual-rail calibration procedure}.
    (A) After parking the transmons on resonance, we perform microwave spectroscopy to identify the two hybrid modes $\ket{0_L}$ and $\ket{1_L}$ at $\omega_0 \pm g$. A small additional feature at $\omega_0 - \eta/2$ is observed, which is a two-photon resonance going to higher excited states; its near overlap with the transition to $\ket{0_L}$ is particular to these parameters where $g \sim \eta/2$, so for this device we choose to initialize $\ket{1_L}$. Inset: excitation probability as we scan the pulse amplitude on the $\ket{1_L}$ resonance.
    (B) We calibrate flux-modulation gates by applying a flux modulation pulse of variable amplitude in between two microwave pulses; this measurement detects population transfer from $\ket{1_L}$ to $\ket{0_L}$ induced by the flux pulse, and is used to approximately calibrate a $\pi/2$ or $\pi$ pulse within the dual-rail subspace.
    (C) A Ramsey experiment used to measure the dual-rail frequency relative to a reference frequency. We show here the full bitstring probabilities of the Q1 and Q2 readout; the $01$ and $10$ results are interpreted as representing population in $\ket{0_L}$ and $\ket{1_L}$, respectively, prior to the adiabatic separation of qubits during readout.
    }
    \label{fig:DualRailCalibration}
\end{figure*}
The dual-rail qubit calibration procedure is as follows:
\begin{enumerate}
    \item \textbf{Identify approximate operating points.} Spectroscopically measure the avoided crossing between the two transmons to identify which flux offsets bring the transmons on resonance (as shown in Fig.~\ref{fig:Fig1}b).
    \item \textbf{Initialize the dual-rail qubit.} With both transmons parked at the operating point, apply a microwave drive at one of the two hybrid mode frequencies and scan its amplitude to calibrate a $\pi$-pulse from $\ket{00}$ to $\ket{0_L}$ or $\ket{1_L}$ (Fig.~\ref{fig:DualRailCalibration}a).
    \item \textbf{Calibrate gates within the subspace.} To calibrate an approximate $\pi/2$ and $\pi$ pulse induced by flux modulation of one qubit, we perform two microwave $\pi-$pulses on the $\ket{00}$ to $\ket{1_L}$ transition with the flux modulation pulse applied in between to detect population transfer from $\ket{1_L}$ to $\ket{0_L}$. We scan the amplitude of the flux modulation pulse to obtain an approximate $\pi/2$ and $\pi$ pulse (Fig.~\ref{fig:DualRailCalibration}b).
    \item \textbf{Refine flux bias on transmons.} We subsequently refine the calibration by performing a Ramsey sequence on the dual-rail qubit to more precisely measure the dual-rail frequency (Fig.~\ref{fig:DualRailCalibration}c). We repeat this as a function of flux offset on one qubit to find the operating point which minimizes the dual-rail frequency, which optimizes the robustness of the dual-rail qubit to flux noise (example shown in Fig.~\ref{fig:FluxTuning}).
\end{enumerate}

\section{Refining single-qubit gate calibration}
\label{Appendix:GateCalibration}
The single-qubit gate is a flux-modulation pulse on Q2 at the dual-rail frequency $2g=2\pi\times180$~MHz. The pulse is 48~ns with a Gaussian envelope ($\sigma = 12~$ns). We calibrate the amplitude of the pulse by repeated application to optimize the rotation angle \cite{rudinger_experimental_2017}. We subsequently calibrate a $Z(\phi)$ phase correction with $\phi = 0.067~$rad to be applied after every $X90$ due to small shifts in the dual-rail frequency during the gate \cite{lucero_reduced_2010, mckay_efficient_2017}. $Z$ rotations are applied virtually by shifting the phase of the flux-modulation RF drive \cite{knill_algorithmic_2000, mckay_efficient_2017}. The speed of the gate can be optimized to approach the Larmor period of $\sim 6~\text{ns}$ set by the inverse of the qubit frequency $2g = 2\pi \times 180~$MHz, enabling similar gate speeds to that of transmons with typical $2\pi \times 200~\text{MHz}$ anharmonicity.

Other single-qubit gate schemes are also possible with RF qubits such as the dual-rail qubit, closely analogous to single-qubit gate schemes with heavy fluxonium \cite{weiss_fast_2022}. In particular, baseband flux pulses have been demonstrated to realize exact gates at the maximum speed given by the Larmor period of the qubit \cite{campbell_universal_2020}.
Such fast gates may be optimal for improving single-qubit gate fidelity, but require more complex calibration and preclude the use of phase-shifts on the RF field to realize virtual Z gates.

\section{Randomized benchmarking details and supplementary data}
\label{appendix:RBSequence}
\begin{figure*}
    \centering
    \includegraphics{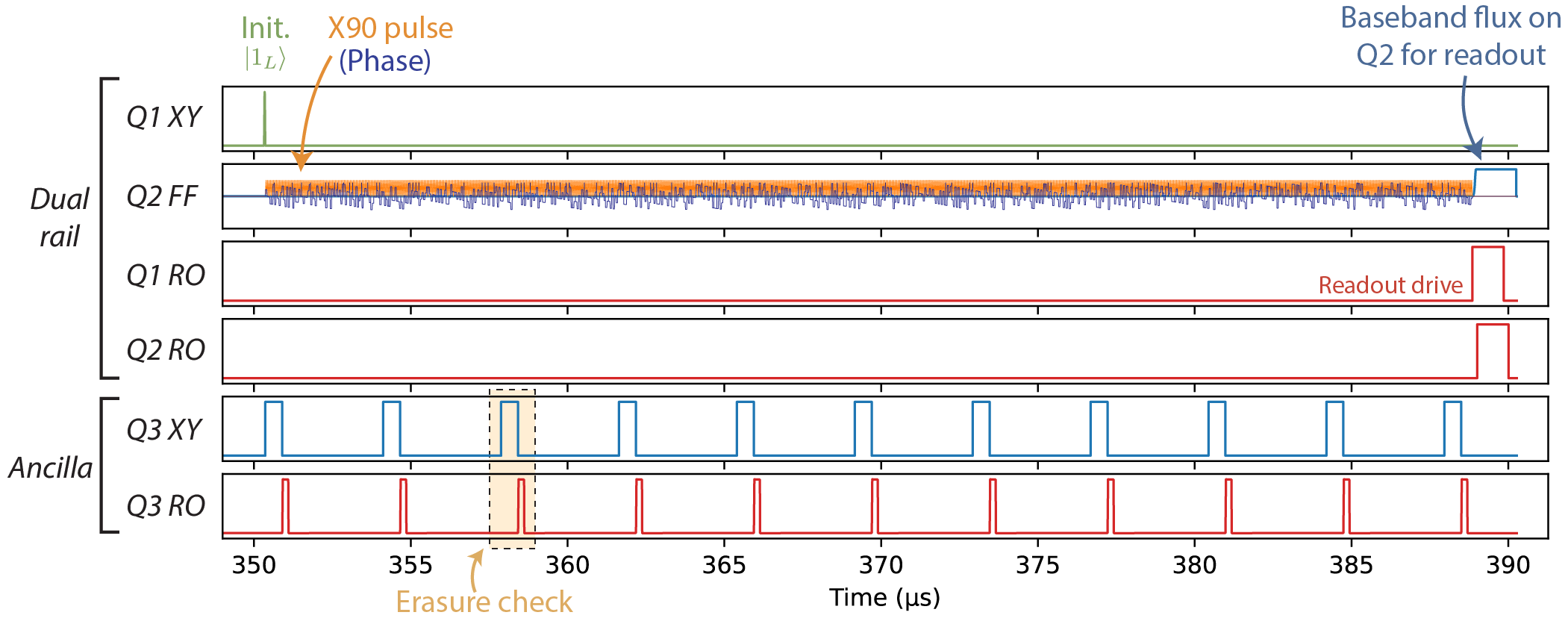}
    \caption{\textbf{Experimental sequence for randomized benchmarking}. The sequence begins with $\sim 350~\mu$s idle time for the transmons to relax into $\ket{0}$. A microwave pulse on Q1 initializes the dual-rail qubit in $\ket{1_L}$. A sequence of Cliffords is then applied, each composed of two X90 pulses with a particular phase shift of the drive field. Shown here is a sequence with 400 Cliffords (800 X90s). In parallel with this sequence, 11 erasure checks are performed, evenly spaced across the full circuit. Finally, a baseband flux pulse is applied to Q2 to adiabatically separate Q1 and Q2, during which these qubits are read-out.}
    \label{fig:PulseSequence}
\end{figure*}

We show in Fig.~\ref{fig:PulseSequence} an example timing diagram for a single-qubit randomized benchmarking sequence with 400 Cliffords. This sequence is representative of nearly all experiments presented in this work, with the main variability being the sequence of flux-modulation pulses being applied on the fast-flux line of Q2 to induce a particular choice and timing of single-qubit gates.

\begin{figure}
    \centering
    \includegraphics{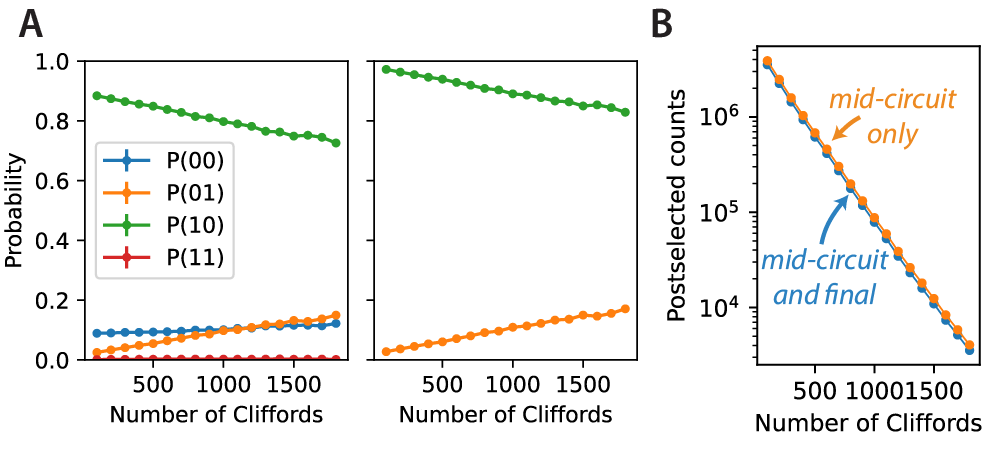}
    \caption{\textbf{Randomized benchmarking postselection protocols}.
    (A) The final dual-rail readout, postselected only on mid-circuit erasure checks (left panel), shows a gradual decline in the ideal outcome $P(10)$, as well as a background of $P(00) \approx 10\%$ due to readout errors. The  $P(10)$ trace does not decay to a known  offset, but a fit to $Ae^{-pN}$ is valid for short times and gives an error of $p = 1.10(2) \times 10^{-4}$ per Clifford. This result is close to the error rate evaluated with additional postselection against $\ket{00}$ and $\ket{11}$ in the final readout (right panel), which as reported in the main text has error $1.01(1)\times 10^{-4}$.
    (B) The erasure error per Clifford is computed based on the decay of postselection probability with increasing circuit depth, and gives nearly the same fitted erasure probability per Clifford when using just mid-circuit erasure checks ($0.437(3)\%$) as when using both mid-circuit checks and final readout as reported in the main text ($0.438(3)\%$).
    }
    \label{fig:RBPostselection}
\end{figure}
In the main text, the erasure error and residual error per gate are computed using postselection based on both the mid-circuit erasure checks and the final dual-rail readout. This evaluates most directly the gate fidelity within the dual-rail subspace, separating out the performance of mid-circuit erasure detection and the final readout. If we instead postselect only on the mid-circuit erasure checks, we find nearly the same gate error rates as shown in Fig.~\ref{fig:RBPostselection}.

\section{Dispersive shifts of the dual-rail qubit on the ancilla}
\label{appendix:DispersiveShifts}
The ancilla qubit's frequency is shifted depending on whether the dual-rail pair is in $\ket{00}$, $\ket{0_L}$, or $\ket{1_L}$. While we illustrate this shift with microwave spectroscopy in Fig.~\ref{fig:Fig1}e, we also use a Ramsey sequence to precisely characterize these shifts. Specifically, we perform a Ramsey experiment on the ancilla to measure its frequency after initializing the dual-rail qubit in its three possible states. We find that the dispersive shift on the ancilla when the dual-rail goes from $\ket{00}$ to $\ket{0_L}$ and $\ket{1_L}$ is $2\pi \times (1.514(4), 1.610(4))~$MHz, respectively. The slight differential shift between the two logical states is $2\pi \times 96(4)~$kHz.

\section{Deterministic phase shift induced by erasure check}
\label{appendix:ErasureCheckShift}
\begin{figure}
    \centering
    \includegraphics{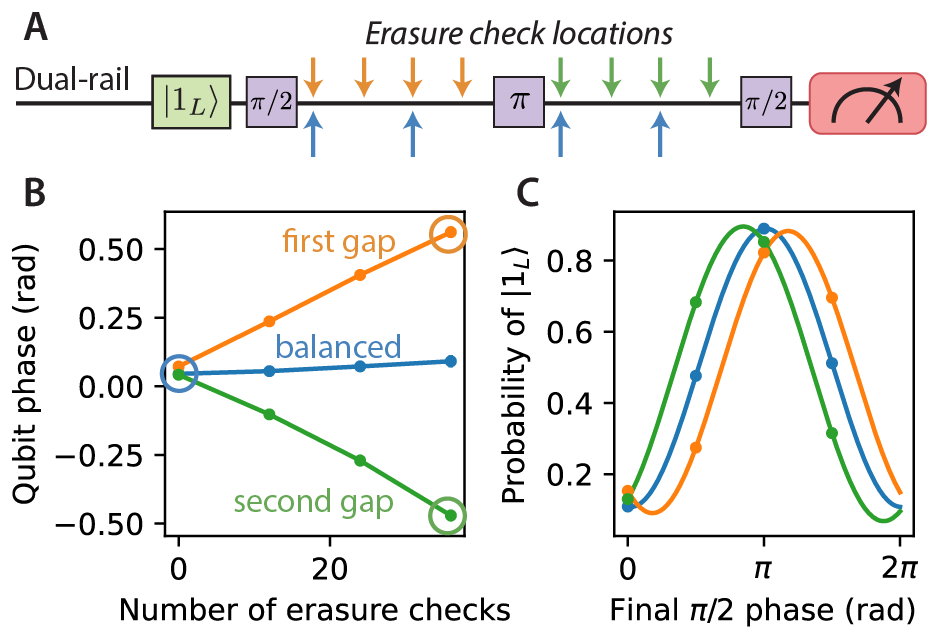}
    \caption{\textbf{Deterministic phase shift induced by erasure check.} (A) In a similar experiment to that presented in Fig.~\ref{fig:Fig4}, we perform a spin-echo and insert a variable number of erasure checks into either the first half, the second half, or both halves of the sequence. (B) We see that the phase of the dual-rail qubit is essentially independent of the number of erasure checks when balanced across both arms, but picks up a phase shift proportional to the number of checks when all checks are applied within one arm. The phase accumulation per check fits to 0.0142(3) rad. (C) We illustrate example phase-measurements of the points circled in (B), in which we scan the phase of the final $\pi/2$ pulse in the spin-echo sequence and fit to a sinusoidal model.}
    \label{fig:ErasureCheckPhaseShift}
\end{figure}
The ancilla-based erasure check induces a small deterministic phase shift on the dual-rail qubit, in addition to a potential dephasing error discussed in Appendix~\ref{appendix:ErasureCheckDephasing}. To characterize the deterministic phase shift, we repeat the experiment performed in Fig.~\ref{fig:Fig4}c in which we insert a variable number of erasure checks into a fixed-time spin echo experiment; rather than inserting them in a balanced way (equal number of checks in both halves of the spin-echo sequence), we  insert all checks in the first half or all checks in the second half (Fig.~\ref{fig:ErasureCheckPhaseShift}). By comparing the final dual-rail phase (measured by scanning the phase of the final $\pi/2$ pulse), we assess that the induced phase shift is $0.0142(3)~$rad per check. This deterministic shift is correctable in circuits with virtual Z gates.

\section{Bounding dephasing error induced by erasure checks}
\label{appendix:ErasureCheckDephasing}
In Fig.~\ref{fig:Fig4}c of the main text, we present a spin-echo experiment in which we measure the remaining phase coherence of the dual-rail qubit after $114~\mu$s as a function of how many erasure checks are inserted in the gap. As we add more erasure checks, we eliminate shots for which the system decayed to $\ket{00}$ and heated back into the subspace. This effect should plateau once there is a certain density of erasure checks, which we expect to be the plateau observed starting at $N \ge 16$ checks. Assuming this model, the residual decay from 16 to 128 checks  gives a dephasing error per check of $p_{err}=0.035\%$.

This analysis, however, is not rigorous as we cannot rule out that as we apply more postselection, the coherence should appear to continue to grow, and that this effect may be balanced by a larger amount of dephasing from more checks. We can bound such a model in the following way: first, we evaluate the worst-case scenario, in which the \emph{true} coherence is perfect after the $114~\mu$s evolution time, and that the measured phase coherence of $77.4(1.1)\%$ after 128 checks is purely due to dephasing from the checks: $0.774 \ge e^{-N p_{err}}$, from which  we conclude a rigorous upper bound on the error $p_{err} \le 0.2\%$.

This worst-case scenario is unrealistic, as it requires (1) perfect idling coherence of the dual-rail qubit in the absence of erasures and (2) the low coherence after $N=16$ erasure checks is fully explained by the imperfect postselection. We improve these assumptions as follows: first, we bound how much imperfect postselection can contribute at $N \ge 16$ based on independent $T_2$ coherence measurements where we use $11$ checks and more dynamical decoupling. In such experiments, we have  $\ge 85\%$ coherence remaining after $114~\mu$s, including readout and pulse errors; this means that the imperfect postselection can only contribute $\le 15\%$ of the loss of coherence for $N=16$; such a bound implies a slightly stronger upper bound for the erasure-check dephasing error of $p_{err} \le 0.16\%$. Even so, this bound assumes that in the absence of erasures, the dual-rail retains $95\%$ coherence after $114~\mu$s, consistent with an exponential decay time of  $T_2^{echo} = 2.2~$ms.
For a general $T_2^{echo}$, we would have that $ 0.774=e^{-\tau/T_2^{echo}}e^{-N p_{err}}$, or equivalently, { $p_{err} = \left[-\ln(0.774) - \tau / T_2^{echo}\right]/128$}. Plugging in our independently measured $T_2^{echo}=540~\mu$s presented in Fig.~\ref{fig:Fig2} which predicts that the true coherence remaining would only be $\sim 80\%$, consistent with the observed plateau,
this would give us a smaller error estimate of $p_{err}\approx0.035\%$. We hypothesize that this error may be related to measurement-induced state transitions while reading out the ancilla. In this work, we report a more conservative upper bound of $p_{err} < 0.1\%$, which could only be violated if $T_2^{echo} \gtrsim 900~\mu$s, and leave a more focused investigation for future work.

\section{Limits imposed by transmon heating}
\label{appendix:TransmonHeating}
\begin{figure}
    \centering
    \includegraphics{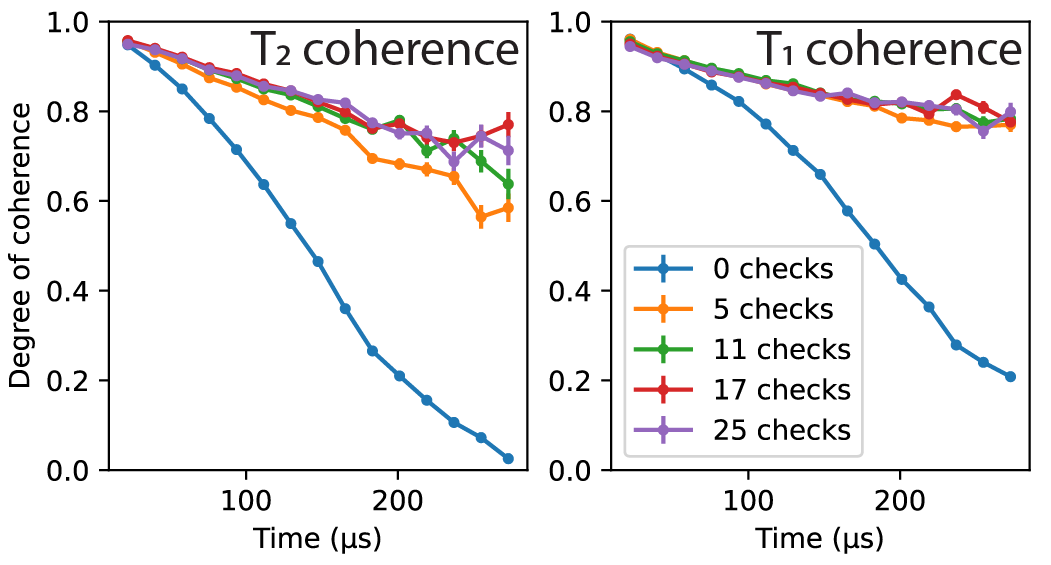}
    \caption{\textbf{Varying the number of erasure checks during coherence measurements.} When measuring dual-rail $T_2$ or $T_1$ with no erasure checks, the coherence appears to rapidly decay over $200-300~\mu$s due to leakage to $\ket{00}$ which then heats back into the dual-rail subspace. As we add more erasure checks (evenly spaced throughout the sequence for each variable length), we find that 11 erasure checks is enough to eliminate the contribution of such instances so that we can more accurately probe coherence within the subspace. }
    \label{fig:VaryingNumErasureChecks}
\end{figure}
The primary leakage mechanism of the dual-rail qubit is $T_1$ decay of underlying transmons that leaves the system in $\ket{00}$. However, the transmons individually have a slow heating timescale which is given by $T_{heat} \approx T_1 / p_{equil}$, where $p_{equil}$ is the thermal excitation probability of the individual transmons. We measure $p_{equil}=0.2(1)\%$ (using single-shot readout methods), so while the $T_1$ decay which leads to the erasure lifetime of the dual-rail is $\sim 30~\mu$s, the heating timescale is $\sim 15~$ms.

This heating mechanism gives rise to two important effects: firstly, population that has accumulated in $\ket{00}$ can return into the dual-rail subspace through heating. If we only postselect against leakage based on the final readout of the dual-rail pair, then such decay/reheating events, in which all phase coherence was lost, are not detectable and appear to limit the dual-rail coherence.
Surprisingly, despite the long $> 10~$ms timescale for heating, the timescale for dephasing due to decay/reheating is actually the much shorter time $T_1$.

To see why, we compare the coherent population which remains in the dual-rail subspace (without decaying) to the total population ending in the dual-rail subspace including decay/heating. The probability of never decaying is $e^{-t/T_1}$, while the probability of ending in the dual-rail subspace is given by thermal equilibration dynamics: $(1 - 2p_{equil})e^{-t / T_1} + 2p_{equil}$, where $2p_{equil}$ is double the single-transmon thermal population as there are two transmons which may be excited. The ratio between these defines the coherence function:
\begin{equation}
    C(t) = \frac{e^{-t / T_1}}{(1 - 2 p_{equil})e^{-t / T_1} + 2 p_{equil}}
\end{equation}
The dephasing time where $C(t) = 1/e$ is
\begin{equation}
    T_{deph} \approx T_1 \times \ln \frac{e-1}{2 p_{equil}}
\end{equation}
which is several times $T_1$ for typical values of $p_{equil}$ (ie., $T_{deph}\approx6T_1$ for $p_{equil}=0.2\%$), with only a logarithmic dependence on $p_{equil}$ or equivalently the heating timescale $T_{heat}$.

Mid-circuit erasure detection, however, effectively mitigates the contribution of this effect as long as the erasure checks are performed sufficiently frequently. 
Analytically, for $N$ perfect erasure checks evenly spaced in a total time $t$, the remaining coherence would be $C(t) \to C(t/N)^N$. Experimentally, we compare dual-rail coherence measurements with varying numbers of  erasure checks in Fig.~\ref{fig:VaryingNumErasureChecks} to ensure that we mitigate this contribution and are not limited by this mechanism after $\sim 300~\mu$s of evolution. In the main text, we use 11 erasure checks, evenly spaced during the free evolution time, for the data in all figures.

The other leakage mechanism is direct heating from the dual-rail subspace into the two-photon subspace spanned by $\ket{11}$, $\ket{02}$, and $\ket{20}$, which, due to the additional channels for heating, has a 3x combined rate compared to the heating rate $\Gamma_{0 \to 1}$ of a single transmon (ie., from $\ket{01}$ the heating into $\ket{11}$ occurs at rate $\Gamma_{0 \to 1}$ while the heating into $\ket{02}$ occurs at $2\Gamma_{0 \to 1}$ due to the larger matrix element). Such heating events will be followed by decay back into the dual-rail subspace and will similarly appear as decoherence with timescale $T_2 \sim T_{heat} / 3 \sim 5~$ms. This will typically set an upper limit on how much coherence can be observed within the dual-rail subspace, as such events are not caught by mid-circuit erasure detection as currently implemented. To circumvent this limit, one could expand the erasure detection protocol to detect not only leakage to $\ket{00}$ but also leakage to the two-photon subspace. Such an approach is a feasible extension of the ancilla qubit method described in this work, and may reduce the requirements on transmon temperature to enable working at higher cryostat temperatures.

\section{Telegraph noise in the dual-rail qubit}
\label{appendix:TelegraphNoise}
\begin{figure*}
    \centering
    \includegraphics[width=\textwidth]{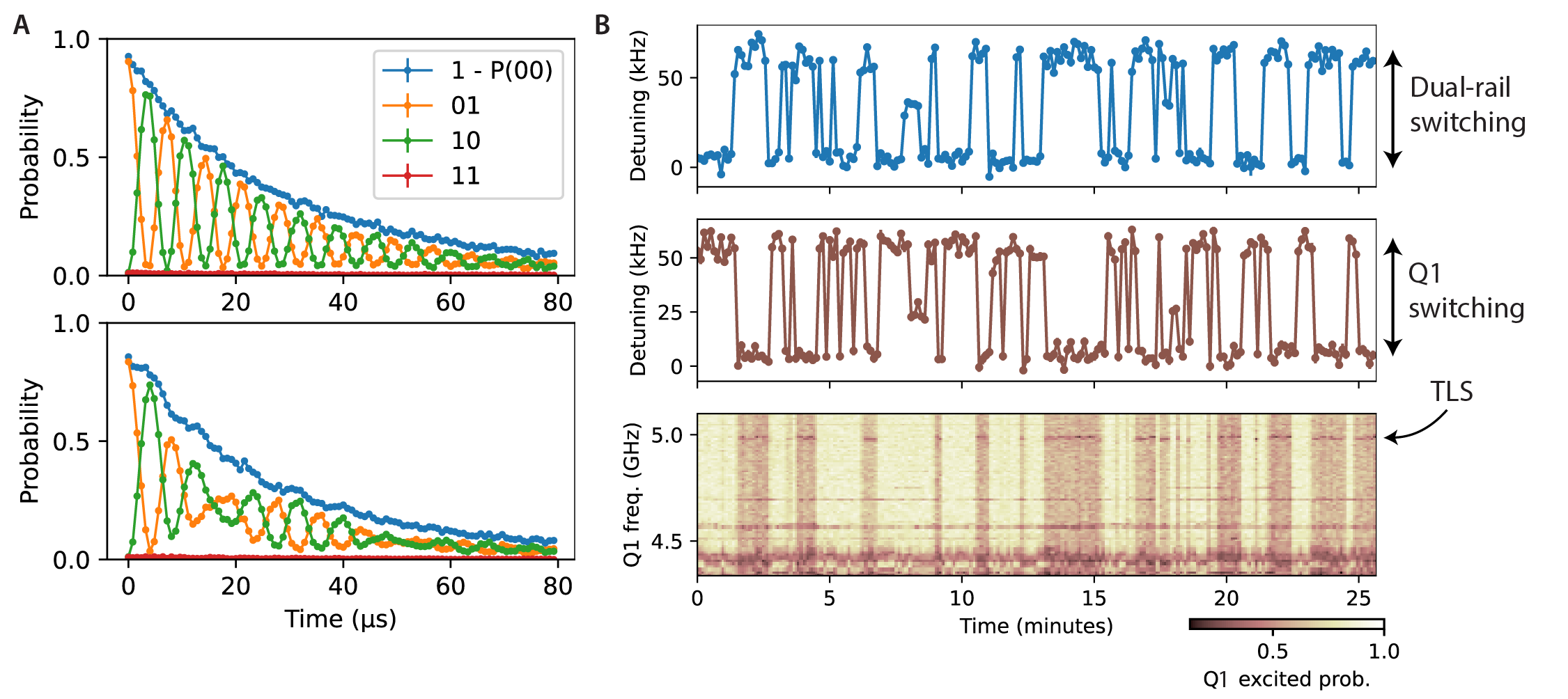}
    \caption{\textbf{Telegraph noise on the dual-rail qubit.} (A) We show two typical Ramsey $T_2^*$ measurements, integrated for $\sim 1~$minute. Within the envelope of not decaying to $\ket{00}$, we see fringes which track the phase coherence of the dual-rail qubit. We find in some cases stable fringes (upper plot) and in other cases decay-revival behavior (lower plot). (B) By repeatedly measuring the dual-rail frequency over shorter time intervals, we identify telegraph-like switching in the frequency. Integrating a Ramsey experiment for longer than this switching time gives rise to the beating effect seen in (A). By interleaving this time-resolved measurement with similar time-resolved measurements of Q1's frequency and flux-pulse spectroscopy measurements on Q1, we note that this switching is present in similar scales on both the dual-rail and Q1 and is tied to the toggling of a TLS-like mode at 4.98 GHz. When the TLS is present, it reduces the readout fidelity of Q1 leading to a reduced contrast in the spectroscopy slice.}
    \label{fig:RamseyBeating}
\end{figure*}
The main text presents dual-rail coherence with dynamical decoupling. Without dynamical decoupling, the dual-rail Ramsey coherence exhibits fluctuating behavior, sometimes showing ``beat-note" like decay and revival (see Fig.~\ref{fig:RamseyBeating}a for two representative Ramsey measurements averaged over $\sim 1$~minute).

By instead using short Ramsey experiments to measure the dual-rail frequency in second-scale intervals, we directly observe that this behavior is explained by telegraph noise, with a switching amplitude of $\sim 60~$kHz and switching timescale of 10s of seconds (Fig.~\ref{fig:RamseyBeating}b, upper panel). Interleaving such measurements with Ramsey measurements on Q1 when parked alone at the same operating point, we find that both exhibit similar scale and time-correlated switching (Fig.~\ref{fig:RamseyBeating}b, middle panel).

We attribute this to a flickering TLS mode at  4.98 GHz by interleaving the above frequency measurements with a spectroscopy measurement on Q1 in which we initialize Q1 in $\ket{1}$ and flux-pulse it to a variable frequency for $1~\mu$s and observe if the excitation is lost \cite{krinner_realizing_2022}. We find a clear correlation between the presence of a dark mode at 4.98 GHz with the frequency shift observed both on Q1 alone and on the dual-rail qubit.

While this telegraph noise is sufficiently slow that dynamical decoupling easily addresses it, interactions between the dual-rail qubit and proximal TLS modes warrant further study to understand what types of impacts TLSs may have on dual-rail qubits beyond just dispersive shifts and also potentially to probe TLS physics with a complementary toolbox to transmon probes (see Appendix~\ref{appendix:TLS} for further discussion).

\section{Dual-rail qubit stability over time}
\label{appendix:DualRailStability}

\begin{figure*}
    \centering
    \includegraphics[width=\textwidth]{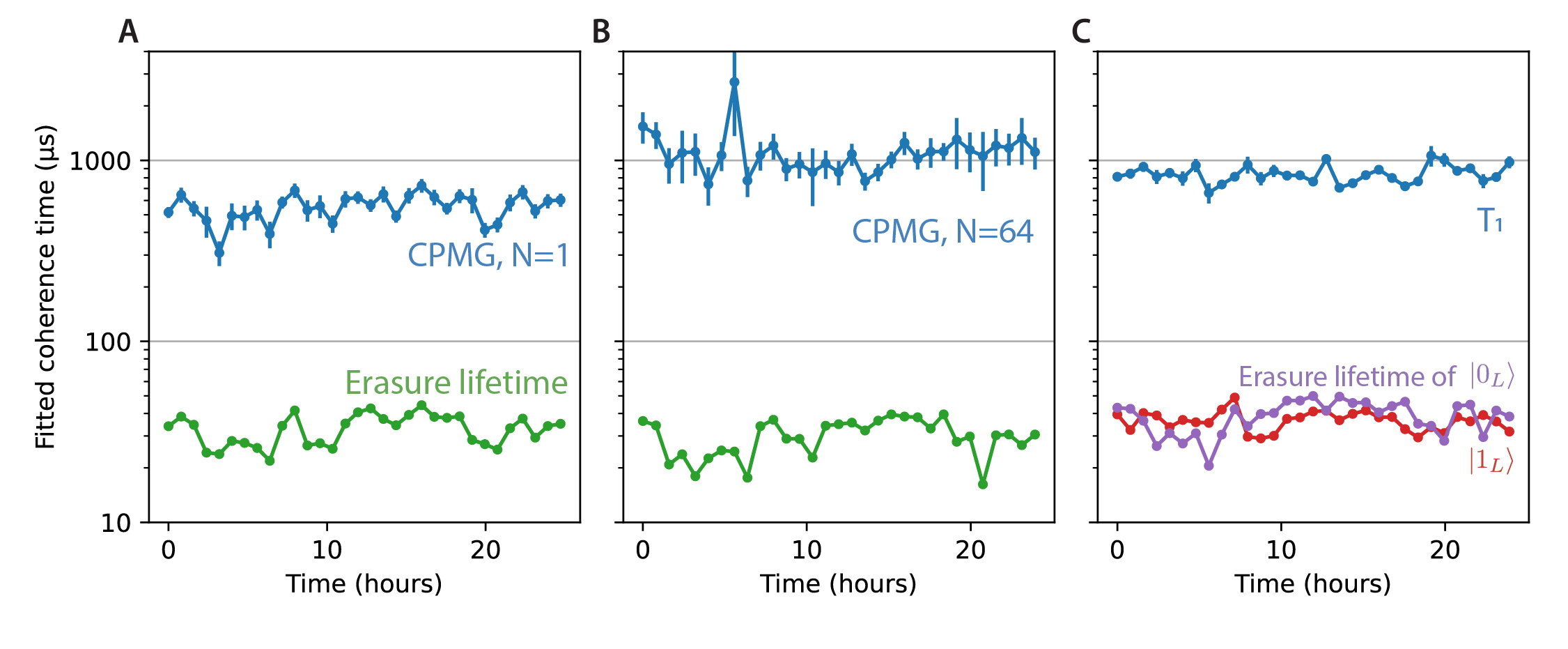}
    \caption{\textbf{Stability of dual-rail coherence metrics.} The coherence measurements presented in Fig.~\ref{fig:Fig2} were averaged over data acquired over $\sim 24$~hours. Here, we analyze subsets of the same dataset in a time-resolved way to track the fitted coherence metrics during that period. We show three representative metrics: (A) CPMG with one $\pi$ pulse, (B) CPMG with 64 $\pi$ pulses, and (C) $T_1$ within the dual-rail subspace. In each plot, we additionally show the erasure lifetime extracted during each measurement. For the $T_1$ measurement, this erasure lifetime is extracted separately for the two logical states and found to be fluctuating differently for the two states, despite the measurements being taken in an interleaved fashion. Large discrepancies between the erasure lifetimes of the two logical states can lead to dephasing of the dual-rail qubit, but this effect is mitigated by dynamical decoupling.}
    \label{fig:CoherenceStability}
\end{figure*}

While the dual-rail frequency flickers on slow timescales due to telegraph noise, the behavior with even minimal dynamical decoupling (ie., a single echo pulse) shows much more stable performance over long time intervals. The coherence data presented in the main text Fig.~\ref{fig:Fig2} was averaged over $\sim 24$~hours, and in Fig.~\ref{fig:CoherenceStability} we show the  data in discrete chunks during this window.

From the same datasets, we similarly extract the erasure lifetime over the measurement period, and find moderate fluctuations between 20 and 40$~\mu$s, consistent with typical $T_1$ fluctuations for individual transmons. Somewhat surprisingly, we note that the two dual-rail logical states exhibit different erasure lifetimes, each of which fluctuate seemingly independently. While naively we would assume that each mode would have the same erasure lifetime given by the average $T_1$ decay rate of the two constituent transmons, this differential may be due to frequency-dependent effects which are differently resonant with the two hybrid modes.

\section{Dual-rail coherence and noise suppression}
\label{appendix:CoherenceAnalysis}
The energy gap of the dual-rail qubit is determined primarily by the capacitive coupling strength $g$ between transmons, and is only weakly sensitive to the detuning between the transmons. For transmon frequencies $\omega_1, \omega_2$, the dual-rail energy is given by
\begin{equation}
    E_{DR}=\sqrt{(2g)^2+(\omega_1 - \omega_2)^2} \approx 2g + \frac{(\omega_1 - \omega_2)^2}{4g}
\end{equation}
thus offering quadratic suppression of the detuning between transmons. However, the Hamiltonian coupling strength $g$, produced by a fixed capacitance between the qubits, itself scales with qubit frequency as $g \propto \sqrt{\omega_1 \omega_2}$. Defining a reference operating frequency for the qubits $\omega_0$ with corresponding coupling strength $g_0$, and with relative detunings $\delta_{1,2} = \omega_{1,2} - \omega_0$, the energy gap takes the following form assuming $\delta_{1,2} \ll g_0 \ll \omega_0$:
\begin{equation}
\label{eq:dual_rail_full_energy_gap}
    E_{DR}=2g_0 + \frac{g_0}{\omega_0}(\delta_1+\delta_2) + \frac{(\delta_1 - \delta_2)^2}{4 g_0}
\end{equation}

When operating at $\delta_1 = \delta_2 = 0$, the final term offers quadratic suppression, but the overall energy gap retains a linear sensitivity to qubit frequency noise scaled down by $\partial E_{DR} / \partial \delta_{1,2}=g_0/\omega_0 \approx 0.018 \approx 1/57$.
If the qubits are limited by $1/f$ noise, this would predict that the dual-rail coherence is scaled linearly with this factor.

By introducing a small detuning, however, the dual-rail energy gap can become linearly insensitive to frequency noise on one of the two qubits ($\partial E_{DR} / \partial \delta_1 = 0$), at the cost of a larger sensitivity to the other $(\partial E_{DR} / \partial \delta_2 = 2g_0 / \omega_0$). This is done by parking at $\delta_1 = \delta_2 - 2g_0^2/\omega_0$; ie., one qubit is parked below the other qubit by an amount $2g_0^2 / \omega_0 \approx 2\pi \times 3.2~$MHz for the current device parameters. Letting $\epsilon_1, \epsilon_2$ now be deviations from these target operating points, we have:
\begin{equation}
\label{eq:dual_rail_gap_optimal}
    E_{DR}=\left[2g_0 - \frac{g_0^3}{\omega_0^2}\right] + \frac{2g_0 \epsilon_2}{\omega_0} + \frac{(\epsilon_1 - \epsilon_2)^2}{4g_0}
\end{equation}
This is a preferable operating point if one qubit is noisier than the other, which is in practice often the case.

\begin{figure}
    \centering
    \includegraphics{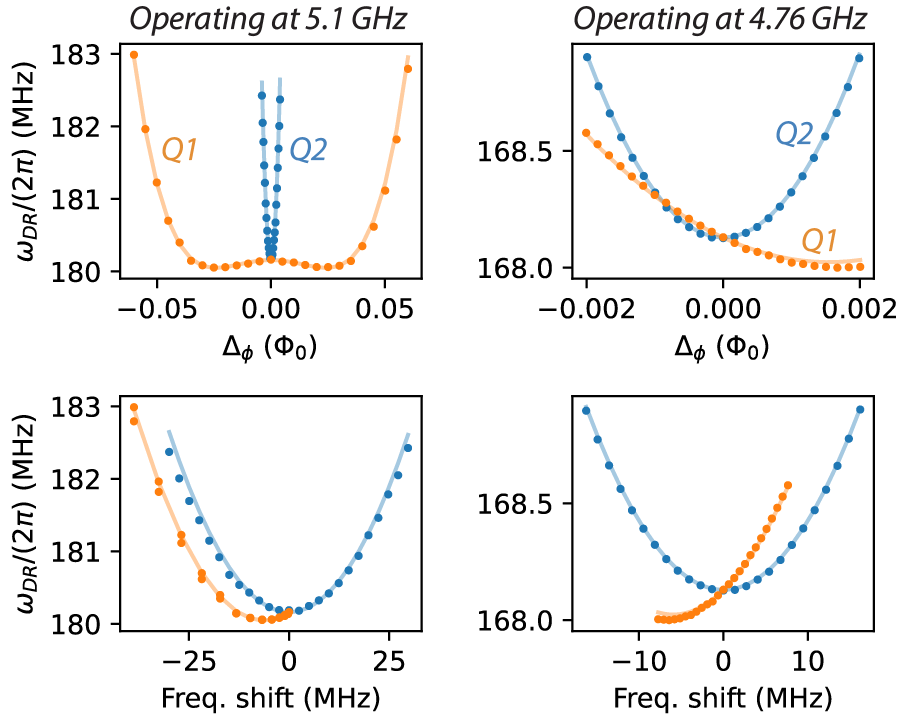}
    \caption{\textbf{Measuring dependence of dual-rail frequency on transmon flux}. At the two extremal operating points from Fig.~\ref{fig:Fig5}, we measure how the dual-rail frequency shifts as we shift the flux on one of the two transmons away from its calibrated operating point (upper row). Each point here is measured with a Ramsey experiment on the dual-rail qubit. The functional form is dependent both on how the dual-rail frequency changes with underlying transmon frequency and also how the transmon frequency shifts with flux. By independently calibrating how the transmon frequencies change in this range, we plot the same data with an altered x-axis that reflects instead how much the transmon frequency shifted (bottom row). In both rows, the transparent trace is the analytic model from  eq.~\eqref{eq:dual_rail_gap_optimal}, with the only adjusted parameters being the calibrated dual-rail energy gap at each operating point.
    }
    \label{fig:FluxTuning}
\end{figure}

This type of local optimum is found automatically through a calibration procedure in which one transmon is parked at a target operating point (ideally, its flux sweet-spot), and then the other transmon is tuned near resonance, where the dual-rail energy gap is measured as a function of flux on the scanning transmon. The operating flux is selected as the one which minimizes the dual-rail energy gap, thus minimizing sensitivity to flux noise. After calibrating in this way, we check the validity of this model by measuring the dual-rail frequency as we scan flux on each qubit and find excellent agreement (Fig.~\ref{fig:FluxTuning}).

We use this strategy at each operating point studied in Fig.~\ref{fig:Fig5}. As a result, we expect the dominant contribution from transmon frequency noise to dual-rail dephasing to come from the linear suppression on Q1. Quantitative estimates of these effects are provided in Appendix~\ref{appendix:DualRailDecoherence}.

\section{Mechanisms for dual-rail decoherence}
\label{appendix:DualRailDecoherence}
Here, we enumerate several mechanisms which can contribute to dual-rail dephasing and where possible estimate the associated timescales.

\subsection{Flux noise} Flux noise induces frequency noise on the individual transmons. The transmon frequency noise can affect the dual-rail qubit in two ways: (1) the dual-rail energy gap is sensitive to frequency noise on the underlying qubits according to eq.~\eqref{eq:dual_rail_gap_optimal}, which can cause dephasing of the dual-rail qubit, and (2) high-frequency noise at the dual-rail energy gap $2g$ can cause $T_1$-limiting bit-flip transitions within the dual-rail subspace. The rate of such decoherence mechanisms depends heavily on the noise spectrum; while we do not have a full model for the qubit noise spectrum, we assess the influence of two paradigmatic noise processes below: $1/f$ noise and Johnson-Nyquist noise.

\subsubsection{$1/f$ noise}
To assess the possible influence of $1/f$ noise, we consider a noise spectrum which is scaled in amplitude to give a particular coherence time when applied to a single transmon alone. We focus on the $T_\phi^{echo}$ dephasing time, that is the pure dephasing time in a spin-echo experiment on a single transmon, which is preferable to analyze over the Ramsey dephasing time as it is insensitive to the divergence of $1/f$ at low frequencies.

Firstly, we consider bit-flips induced by $1/f$ noise extending up to $2g = 2\pi \times 180$~MHz. We consider a two-sided frequency noise spectrum  $S(\omega) = 2\pi A^2 / |\omega|$ which gives rise to a spin-echo dephasing time for a single transmon of $T_\phi^{echo} = [A\sqrt{\ln 2}]^{-1}$ \cite{bylander_noise_2011}. With the same noise applied to one transmon within a dual-rail pair, the dual-rail $T_1$ would be limited as $T_1^{DR} = 2 / S(2g)$ \cite{noise_spectrum_footnote}; using $A = [T_\phi^{echo} \sqrt{\ln 2}]^{-1}$, we see that the dual-rail $T_1^{DR}$ is  related to the single-transmon coherence as \begin{equation}
    T_1^{DR}=\frac{2 \ln 2}{\pi}g (T_\phi^{echo})^2
\end{equation}
For a noise amplitude which gives reasonable qubit coherences of $T_\phi^{echo} > 3~\mu$s (a lower bound for each measured transmon in the dual-rail pair), the limit on the dual-rail coherence would be $T_1^{DR} > 2.2~$ms. This is likely a conservative upper bound on the true $1/f$ noise contributions, as the noisier qubit  Q2's $T_2^{echo}$ is likely limited by extra low-frequency noise from the control hardware which is low-pass filtered to $< 2\pi \times 20~$MHz and would not extend high enough in frequency to reach $2g$; as such, we expect this noise contribution not to limit our measured dual-rail $T_1$.

Secondly, we consider the dephasing induced by frequency noise on Q1 which is linearly suppressed by a factor of $2g_0 / \omega_0 \approx 1/28$. For $1/f$ noise, the dual-rail coherence would be scaled by this linear factor, with $T_2^{DR} \approx 28 \times T_\phi^{echo}$.
Q1 exhibits a low pure-dephasing rate at its sweet spot, with $T_\phi^{echo} > 100~\mu$s; if this were due to $1/f$ noise, it would correspond to a dual-rail dephasing time of $> 2.8~$ms, again not limiting for current experiments. In the furthest operating point where  Q1 is on its slope, it exhibits a dephasing time of $T_\phi^{echo} \approx 37(9)~\mu$s, which would correspond to a dual-rail dephasing time of $T_2^{DR} \approx 1~$ms. While this is not limiting, this additional contribution is consistent with the decrease in $T_2^{DR}$ measured from the best operating point ($T_2^{DR} \approx 550(30)~\mu$s) to the worst operating point ($T_2^{DR} \approx 366(20)~\mu$s) as shown in Fig.~\ref{fig:SupplementaryTunableData}.

Finally, we consider the dephasing induced by frequency noise on Q2 which is quadratically suppressed. We numerically simulate $1/f$ noise, with an amplitude scaled according to the measured $T_2^{Q2}=3~\mu$s, and find that the contributions to dual-rail dephasing are at the level of $\sim 35~$ms. While this $1/f$ model likely is not an accurate description of Q2's noise spectrum, particularly due the presence of  noise from the control hardware, this illustrates the efficacy of noise suppression in the dual-rail qubit, and we leave a further detailed study of noise spectra for future work.

\subsubsection{Thermal Johnson-Nyquist noise in flux lines}
\label{appendix:JohnsonNoise}
Our flux lines are thermalized at 4~K without significant further attenuation at lower temperature stages. As such, we expect thermal Johnson-Nyquist noise at this characteristic temperature to be carried down the flux lines and affect the qubits. The induced two-sided spectrum describing transmon frequency noise due to equilibrium current fluctuations is, for $\omega \ll k_BT$,
\begin{equation}
    S(\omega)\approx2 \left( \frac{ \partial \omega_T}{\partial \Phi_e} \right)^2 M^2 \frac{ k_B T}{Z}
\end{equation}
where $\partial \omega_T / \partial \Phi_e$ is the sensitivity of the transmon frequency with respect to external flux, $M$ is the mutual inductance to the flux line, $k_B$ is the Boltzmann constant, $T$ is the effective noise temperature, and $Z$ is the transmission line impedance. The noise spectrum at $\pm 2g$ drives bit-flips within the dual-rail subspace with a combined $T_1$ limit  of $T_1^{DR} = 2 / S(2g)$ \cite{noise_spectrum_footnote}.

This noise affects both qubits according to their slopes $\partial \omega_T / \partial \Phi_e$ and their mutual inductances which are both $M \approx 1.28~\text{pH}$. At the main operating point, Q1's slope vanishes, leading to a contribution just from Q2 which has slope $2\pi\times7.8~\text{GHz}/\Phi_0$. Using $T = 4~$K and $Z = 50~\Omega$, this predicts a limit of $T_1^{DR} \approx 1~$ms, which is roughly consistent with the measured $T_1$. Further study is warranted to evaluate more directly whether this is the dominant error mechanism by altering the flux line configuration.

\subsection{Photon fluctuations in readout resonators}

Thermal fluctuations of photons in readout resonators have been known to limit qubit coherence in superconducting architectures. While we do not have a precise estimate for the thermal equilibrium population in the resonators, we can bound such population based on how it would limit the coherence of individual transmons. In particular, the transmon dephasing induced by thermal fluctuations in a coupled resonator is given by $\Gamma_\phi = 4\chi^2 \eta \bar{n} / \kappa$, where $\chi$ is the dispersive coupling between the transmon and the resonator, $\kappa$ is the resonator linewidth, $\bar{n}$ is the thermal population, and $\eta = \kappa^2 / (\kappa^2 + 4\chi^2)$ \cite{krantz_quantum_2019}. Using the readout parameters of Table~\ref{tab:DeviceProperties}, we bound thermal population at $\bar{n} \lesssim 0.001$, which would cause a dephasing time of $T_\phi = 44~\mu$s on Q1.

At this level of thermal population, we calculate how these fluctuations would affect dual-rail $T_1^{DR}$ and $T_2^{DR}$. These fluctuations induce a Lorentzian noise spectrum on the qubits, $S(\omega) = \frac{8\chi^2 \kappa \bar{n}}{\omega^2 + \kappa^2}$ \cite{krantz_quantum_2019}, whose contribution at the energy gap $S(2g)$ causes bit-flip errors in the dual-rail subspace. We calculate that this contribution for $\bar{n}=0.001$ would impose $T_1^{DR}=2/S(2g) \approx10~$ms \cite{noise_spectrum_footnote}.

To estimate the induced dephasing on the dual-rail qubit, we numerically simulate the master equation for two coupled qubits, each dispersively coupled to a readout resonator with Hamiltonian
\begin{equation}
    H = g(\sigma^-_{1} \sigma^+_{2} + h.c) + \sum_{i=1,2} \chi_i \sigma_i^{z} a_{i}^\dag a_{i}
\end{equation}
along with collapse operators $C_{i}^\downarrow = a_i \sqrt{\kappa_i(1+\bar{n})}$ and $C_{i}^\uparrow = a_i^\dag \sqrt{\kappa \bar{n}}$. In this model, we find that the induced dual-rail dephasing nearly saturates at the $2\times T_1^{DR}$ limit imposed by the high-frequency part of the noise spectrum, $S(2g)$. As such, we expect similarly that this dephasing is at a scale of $> 10~$ms and thus not limiting.

\subsection{Interaction with TLSs}
\label{appendix:TLS}
Parasitic two-level systems (TLSs) which are coupled to either of the two transmons comprising the dual-rail qubit can affect its coherence. We consider a simple model with two transmons (described as bosonic modes $a_1,a_2$) and a TLS described by Pauli operators $\sigma_\text{TLS}^{x,y,z}$. While a TLS may dispersively shift its coupled transmon by a characteristic $\chi \sigma_z a^\dag a$, it is insufficient to simply analyze this frequency shift in the context of dual-rail frequency sensitivity, as described by eq.~\eqref{eq:dual_rail_gap_optimal}. Instead, it is important to evaluate how the TLS couples to the dual-rail logical states.

We start with the following Hamiltonian in the rotating frame at the transmon frequency (excluding transmon anharmonicity for simplicity):
\begin{equation}
    H_{\text{TLS}} = g(a_1^\dag a_2 + h.c) + \frac{\Delta}{2} \sigma_z + \lambda (a_1^\dag \sigma_- + h.c)
\end{equation}

Rewriting in terms of the dual-rail eigenmodes $d_\pm = \frac{1}{\sqrt{2}}(a_1 \pm a_2)$:
\begin{equation}
    H = g(d_+^\dag d_+ - d_-^\dag d_-) + \frac{\Delta}{2} \sigma_z +\frac{\lambda}{\sqrt{2}} \left( (d_+^\dag + d_-^\dag) \sigma_- + h.c \right)
\end{equation}

The TLS is now coupled equally to the two dual-rail eigenmodes at frequency $\pm g$, but possibly with a different detuning to the two modes.

Several effects may emerge from this model: firstly, if the TLS has a short lifetime and is nearly resonant with one mode, it can cause one logical state to have a faster decay rate out of the dual-rail subspace. This is visible in Fig.~\ref{fig:SupplementaryTunableData} when the dual-rail qubit is parked at 4.96~GHz, while its upper mode is nearly resonant with a TLS which is observed directly on Q2 when parked at this upper hybrid mode frequency; specifically, we see that the upper mode has a short erasure lifetime.

Secondly, if not right on resonance with a dual-rail mode, the TLS becomes weakly hybridized with the modes, inducing dispersive shifts. 
Letting $\Delta_\pm = \Delta \pm g$ be the detuning to each mode; then the dispersive shifts are $\chi_\pm = \lambda^2 / 2\Delta_\pm$. The difference between these shifts, $\chi_{DR} = \chi_+ - \chi_-$, is what causes dual-rail dephasing, as this describes the effective dispersive coupling to the dual-rail qubit. Importantly, $|\chi_{DR}| = \frac{2\lambda^2 g}{\Delta^2 - g^2}$ can be of the same order of magnitude as the dispersive coupling directly from the TLS onto a single isolated transmon, $\lambda^2 / \Delta$.

Finally, the hybridization between the dual-rail modes and the TLS can also cause the dual-rail qubit to inherit noise processes associated with the TLS. Slow switching can induce telegraph noise on the dual-rail qubit associated with toggling dispersive shifts, as observed experimentally and described in Fig.~\ref{fig:RamseyBeating}. (Notably, in this figure it is shown that the telegraph noise on the dual-rail qubit is of similar scale to that on the individual transmon, as compatible with the above analysis). Faster noise processes, including Markovian dephasing of the TLS, would include high-frequency noise components that can drive transitions between logical states, limiting the dual-rail $T_1^{DR}$, as well as noise that contributes to direct dual-rail dephasing.

Further studies of how dual-rail qubits interact with TLSs are warranted, as this approach may offer complementary tools to characterize the properties of TLSs beyond what is available using single transmon probes. For quantum computing applications, however, the dual-rail qubit primarily benefits by being able to tune its operating point to dodge TLSs.

\section{Supplementary data across tunable operating points}
\label{appendix:SupplementaryTunable}
\begin{figure*}
    \centering
    \includegraphics[width=\textwidth]{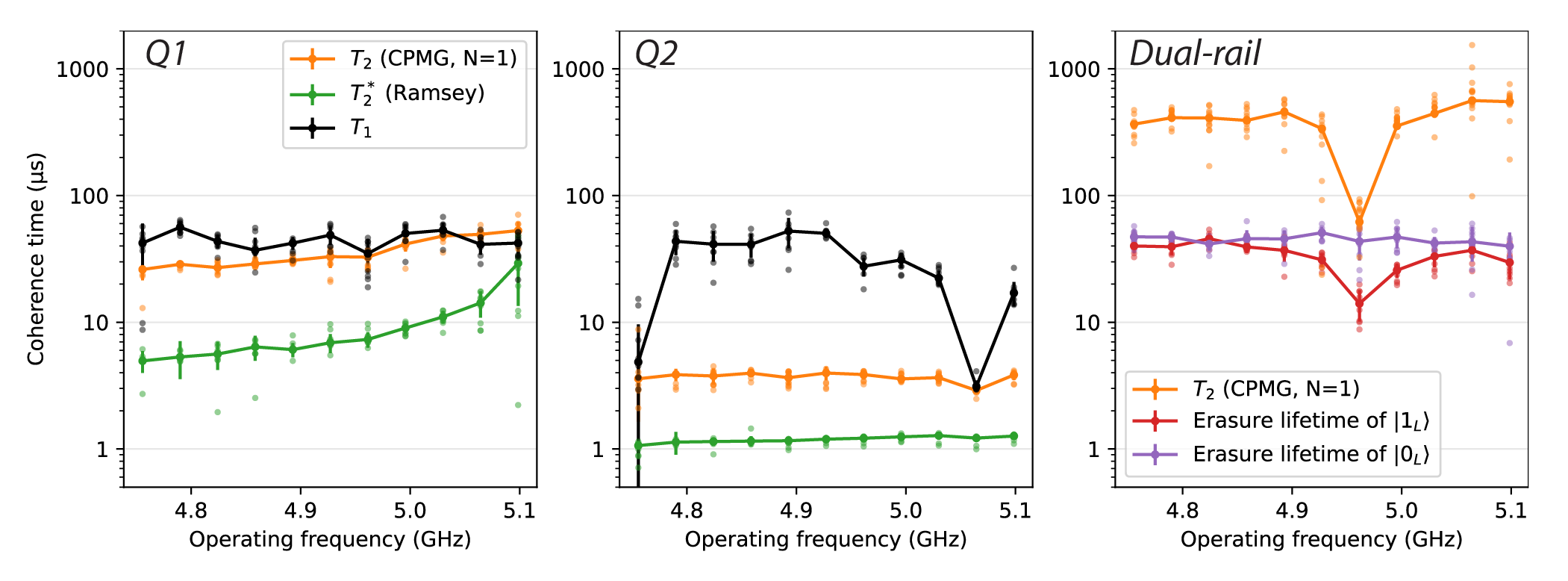}
    \caption{\textbf{Supplementary coherence data across tunable operating band.} We show $T_2^*$, $T_2^{echo}$, and $T_1$ for both transmons Q1 and Q2 at their range of operating points. For the dual-rail qubit, we show the $T_2^{echo}$ as well as the erasure lifetime for each logical state. We find that the dip in dual-rail coherence at 4.96 GHz corresponds to a drop in the $\ket{1_L}$ erasure lifetime; this upper hybrid mode is at $4.96~\text{GHz} + 90~\text{MHz} = 5.05~\text{GHz}$. We attribute this feature to a TLS coupled to Q2, which is independently seen as the sharp drop in Q2's $T_1$ around 5.05 GHz. All individual points are fit results for single measurements; for the transmon data for Q1 and Q2, the average result (errorbar) is the median of the individual datasets (standard deviation). For the dual-rail $T_2$ data, the average (errorbar) is the fit result of all datasets averaged together (fit uncertainty).}
    \label{fig:SupplementaryTunableData}
\end{figure*}
To supplement the core coherence metrics across the tunable operating range presented in Fig.~\ref{fig:Fig5}, we show in Fig.~\ref{fig:SupplementaryTunableData} a more complete set of metrics across this range, including $T_1, T_2^*$, and $T_2^{echo}$ for the individual transmons and the erasure lifetimes for the dual-rail qubit.

We note that  the dip in dual-rail coherence which is observed at $\sim 4.96$~GHz is not matched with a dip in coherence in either Q1 or Q2 at those operating points. Instead, we attribute the dual-rail dip to a marked drop in Q2's $T_1$ at around 5.07 GHz. We hypothesize that these two dips are related because while the dual-rail is operated at 4.96 GHz, its eigenmodes are present at $4.96~\text{GHZ} \pm 90~\text{MHz}$, putting the upper mode near resonance with the TLS in Q2's spectrum. This is consistent with the notable reduction in the $\ket{1_L}$-state erasure lifetime at this operating point.

\section{Control line and chip resource estimation}
\label{appendix:ResourceEstimation}
While the dual-rail qubit offers significant advantages over standard transmons for quantum error correction, it comes at higher cost in terms of chip complexity and control lines. In this section, we enumerate such resource costs for two different possible dual-rail architectures and compare to two transmon-based architectures, with the summary presented in Table~\ref{table:ResourceCosts}.
In all schemes, including dual-rail and transmon-based, we neglect resource costs for tunable couplers, as similar coupler styles are possible in all architectures.

\begin{enumerate}
    \item \textbf{Dual-rail with ancilla qubits:} This scheme is the approach adopted in this paper, in which each dual-rail qubit is composed of two tunable transmons. An ancilla transmon (which could be fixed or tunable, but we assume tunable to maximize flexibility) which is coupled to this pair is used to detect erasure errors. All three transmons require a separate readout resonator. XY lines are required only for the ancilla and one of the dual-rail pair. Slow flux tuning would be needed for all, but fast-flux would only be needed on one of the dual-rail pair, both for modulation to drive single-qubit gates and also baseband to separate the pair for readout.
    \item \textbf{Dual-rail with symmetric readout resonator:} As described in Ref.~\cite{kubica_erasure_2023}, erasure detection can also be done using a readout resonator which is symmetrically coupled to the dual-rail pair. This same resonator can also be used to read-out the final state of the dual-rail pair, either by shelving population outside of the dual-rail subspace (ie., into $\ket{11}$) or by separating the transmons again using baseband flux. As a result, only one readout resonator is needed for the dual-rail qubit in this scheme, along with one XY line for either of the two transmons, slow flux for both, and fast-flux for just one.
    \item \textbf{Tunable transmons:} A tunable-transmon approach, as demonstrated in Ref.~\cite{google2023suppressing}, involves one readout resonator for each transmon, as well as an XY line and fast-flux line, where in some cases the XY and Z signals can be combined.
    \item \textbf{Fixed transmons:} Another standard fixed-transmon approach, as demonstrated in Ref.~\cite{jurcevic_demonstration_2021}, similarly involves one readout resonator and XY line for each transmon, with no flux control.
\end{enumerate}

\begin{table}[h]
    \centering
    \begin{tabular}{| r | p{1.3cm} | p{1.3cm} | p{1.3cm} | p{1.3cm } |}
         \hline
         Quantity per qubit & Dual-rail, \newline ancilla qubit & Dual-rail, symm. resonator & Tunable transmon & Fixed transmon \\
         \hline
         Readout resonators & 3 & 1 & 1 & 1 \\
         XY lines & 2 & 1 & 1 & 1 \\
         Z lines (total) & 3 & 2 & 1 & 0 \\
         Z lines (fast flux) & 1 & 1 & 1 & 0 \\
        
        Josephson junctions & 6 & 4 & 2 & 1 \\
        \hline
    \end{tabular}
    \caption{\textbf{Comparison of resource costs per qubit across transmon-based architectures.} These resource costs are based on the discussion of each architecture in Appendix~\ref{appendix:ResourceEstimation}.}
    \label{table:ResourceCosts}
\end{table}


\begin{thebibliography}{58}%
\makeatletter
\providecommand \@ifxundefined [1]{%
 \@ifx{#1\undefined}
}%
\providecommand \@ifnum [1]{%
 \ifnum #1\expandafter \@firstoftwo
 \else \expandafter \@secondoftwo
 \fi
}%
\providecommand \@ifx [1]{%
 \ifx #1\expandafter \@firstoftwo
 \else \expandafter \@secondoftwo
 \fi
}%
\providecommand \natexlab [1]{#1}%
\providecommand \enquote  [1]{``#1''}%
\providecommand \bibnamefont  [1]{#1}%
\providecommand \bibfnamefont [1]{#1}%
\providecommand \citenamefont [1]{#1}%
\providecommand \href@noop [0]{\@secondoftwo}%
\providecommand \href [0]{\begingroup \@sanitize@url \@href}%
\providecommand \@href[1]{\@@startlink{#1}\@@href}%
\providecommand \@@href[1]{\endgroup#1\@@endlink}%
\providecommand \@sanitize@url [0]{\catcode `\\12\catcode `\$12\catcode
  `\&12\catcode `\#12\catcode `\^12\catcode `\_12\catcode `\%12\relax}%
\providecommand \@@startlink[1]{}%
\providecommand \@@endlink[0]{}%
\providecommand \url  [0]{\begingroup\@sanitize@url \@url }%
\providecommand \@url [1]{\endgroup\@href {#1}{\urlprefix }}%
\providecommand \urlprefix  [0]{URL }%
\providecommand \Eprint [0]{\href }%
\providecommand \doibase [0]{https://doi.org/}%
\providecommand \selectlanguage [0]{\@gobble}%
\providecommand \bibinfo  [0]{\@secondoftwo}%
\providecommand \bibfield  [0]{\@secondoftwo}%
\providecommand \translation [1]{[#1]}%
\providecommand \BibitemOpen [0]{}%
\providecommand \bibitemStop [0]{}%
\providecommand \bibitemNoStop [0]{.\EOS\space}%
\providecommand \EOS [0]{\spacefactor3000\relax}%
\providecommand \BibitemShut  [1]{\csname bibitem#1\endcsname}%
\let\auto@bib@innerbib\@empty
\bibitem [{\citenamefont {{Google Quantum AI}}(2023)}]{google2023suppressing}%
  \BibitemOpen
  \bibfield  {author} {\bibinfo {author} {\bibnamefont {{Google Quantum AI}}},\
  }\bibfield  {title} {\bibinfo {title} {Suppressing quantum errors by scaling
  a surface code logical qubit},\ }\href
  {https://doi.org/10.1038/s41586-022-05434-1} {\bibfield  {journal} {\bibinfo
  {journal} {Nature}\ }\textbf {\bibinfo {volume} {614}},\ \bibinfo {pages}
  {676} (\bibinfo {year} {2023})}\BibitemShut {NoStop}%
\bibitem [{\citenamefont {Clark}\ \emph {et~al.}(2021)\citenamefont {Clark},
  \citenamefont {Tinkey}, \citenamefont {Sawyer}, \citenamefont {Meier},
  \citenamefont {Burkhardt}, \citenamefont {Seck}, \citenamefont {Shappert},
  \citenamefont {Guise}, \citenamefont {Volin}, \citenamefont {Fallek},
  \citenamefont {Hayden}, \citenamefont {Rellergert},\ and\ \citenamefont
  {Brown}}]{clark2021high}%
  \BibitemOpen
  \bibfield  {author} {\bibinfo {author} {\bibfnamefont {C.~R.}\ \bibnamefont
  {Clark}}, \bibinfo {author} {\bibfnamefont {H.~N.}\ \bibnamefont {Tinkey}},
  \bibinfo {author} {\bibfnamefont {B.~C.}\ \bibnamefont {Sawyer}}, \bibinfo
  {author} {\bibfnamefont {A.~M.}\ \bibnamefont {Meier}}, \bibinfo {author}
  {\bibfnamefont {K.~A.}\ \bibnamefont {Burkhardt}}, \bibinfo {author}
  {\bibfnamefont {C.~M.}\ \bibnamefont {Seck}}, \bibinfo {author}
  {\bibfnamefont {C.~M.}\ \bibnamefont {Shappert}}, \bibinfo {author}
  {\bibfnamefont {N.~D.}\ \bibnamefont {Guise}}, \bibinfo {author}
  {\bibfnamefont {C.~E.}\ \bibnamefont {Volin}}, \bibinfo {author}
  {\bibfnamefont {S.~D.}\ \bibnamefont {Fallek}}, \bibinfo {author}
  {\bibfnamefont {H.~T.}\ \bibnamefont {Hayden}}, \bibinfo {author}
  {\bibfnamefont {W.~G.}\ \bibnamefont {Rellergert}},\ and\ \bibinfo {author}
  {\bibfnamefont {K.~R.}\ \bibnamefont {Brown}},\ }\bibfield  {title} {\bibinfo
  {title} {High-{Fidelity} {Bell}-{State} {Preparation} with {Ca} + 40
  {Optical} {Qubits}},\ }\href {https://doi.org/10.1103/PhysRevLett.127.130505}
  {\bibfield  {journal} {\bibinfo  {journal} {Physical Review Letters}\
  }\textbf {\bibinfo {volume} {127}},\ \bibinfo {pages} {130505} (\bibinfo
  {year} {2021})}\BibitemShut {NoStop}%
\bibitem [{\citenamefont {Ryan-Anderson}\ \emph {et~al.}(2021)\citenamefont
  {Ryan-Anderson}, \citenamefont {Bohnet}, \citenamefont {Lee}, \citenamefont
  {Gresh}, \citenamefont {Hankin}, \citenamefont {Gaebler}, \citenamefont
  {Francois}, \citenamefont {Chernoguzov}, \citenamefont {Lucchetti},
  \citenamefont {Brown}, \citenamefont {Gatterman}, \citenamefont {Halit},
  \citenamefont {Gilmore}, \citenamefont {Gerber}, \citenamefont {Neyenhuis},
  \citenamefont {Hayes},\ and\ \citenamefont {Stutz}}]{ryan2021realization}%
  \BibitemOpen
  \bibfield  {author} {\bibinfo {author} {\bibfnamefont {C.}~\bibnamefont
  {Ryan-Anderson}}, \bibinfo {author} {\bibfnamefont {J.}~\bibnamefont
  {Bohnet}}, \bibinfo {author} {\bibfnamefont {K.}~\bibnamefont {Lee}},
  \bibinfo {author} {\bibfnamefont {D.}~\bibnamefont {Gresh}}, \bibinfo
  {author} {\bibfnamefont {A.}~\bibnamefont {Hankin}}, \bibinfo {author}
  {\bibfnamefont {J.}~\bibnamefont {Gaebler}}, \bibinfo {author} {\bibfnamefont
  {D.}~\bibnamefont {Francois}}, \bibinfo {author} {\bibfnamefont
  {A.}~\bibnamefont {Chernoguzov}}, \bibinfo {author} {\bibfnamefont
  {D.}~\bibnamefont {Lucchetti}}, \bibinfo {author} {\bibfnamefont
  {N.}~\bibnamefont {Brown}}, \bibinfo {author} {\bibfnamefont
  {T.}~\bibnamefont {Gatterman}}, \bibinfo {author} {\bibfnamefont
  {S.}~\bibnamefont {Halit}}, \bibinfo {author} {\bibfnamefont
  {K.}~\bibnamefont {Gilmore}}, \bibinfo {author} {\bibfnamefont
  {J.}~\bibnamefont {Gerber}}, \bibinfo {author} {\bibfnamefont
  {B.}~\bibnamefont {Neyenhuis}}, \bibinfo {author} {\bibfnamefont
  {D.}~\bibnamefont {Hayes}},\ and\ \bibinfo {author} {\bibfnamefont
  {R.}~\bibnamefont {Stutz}},\ }\bibfield  {title} {\bibinfo {title}
  {Realization of {Real}-{Time} {Fault}-{Tolerant} {Quantum} {Error}
  {Correction}},\ }\href {https://doi.org/10.1103/PhysRevX.11.041058}
  {\bibfield  {journal} {\bibinfo  {journal} {Physical Review X}\ }\textbf
  {\bibinfo {volume} {11}},\ \bibinfo {pages} {041058} (\bibinfo {year}
  {2021})}\BibitemShut {NoStop}%
\bibitem [{\citenamefont {Zajac}\ \emph {et~al.}(2021)\citenamefont {Zajac},
  \citenamefont {Stehlik}, \citenamefont {Underwood}, \citenamefont {Phung},
  \citenamefont {Blair}, \citenamefont {Carnevale}, \citenamefont {Klaus},
  \citenamefont {Keefe}, \citenamefont {Carniol}, \citenamefont {Kumph},
  \citenamefont {Steffen},\ and\ \citenamefont {Dial}}]{zajac2021spectator}%
  \BibitemOpen
  \bibfield  {author} {\bibinfo {author} {\bibfnamefont {D.~M.}\ \bibnamefont
  {Zajac}}, \bibinfo {author} {\bibfnamefont {J.}~\bibnamefont {Stehlik}},
  \bibinfo {author} {\bibfnamefont {D.~L.}\ \bibnamefont {Underwood}}, \bibinfo
  {author} {\bibfnamefont {T.}~\bibnamefont {Phung}}, \bibinfo {author}
  {\bibfnamefont {J.}~\bibnamefont {Blair}}, \bibinfo {author} {\bibfnamefont
  {S.}~\bibnamefont {Carnevale}}, \bibinfo {author} {\bibfnamefont
  {D.}~\bibnamefont {Klaus}}, \bibinfo {author} {\bibfnamefont {G.~A.}\
  \bibnamefont {Keefe}}, \bibinfo {author} {\bibfnamefont {A.}~\bibnamefont
  {Carniol}}, \bibinfo {author} {\bibfnamefont {M.}~\bibnamefont {Kumph}},
  \bibinfo {author} {\bibfnamefont {M.}~\bibnamefont {Steffen}},\ and\ \bibinfo
  {author} {\bibfnamefont {O.~E.}\ \bibnamefont {Dial}},\ }\href@noop {}
  {\bibinfo {title} {Spectator errors in tunable coupling architectures}},\
  \bibinfo {howpublished} {arXiv:2108.11221} (\bibinfo {year}
  {2021})\BibitemShut {NoStop}%
\bibitem [{\citenamefont {Evered}\ \emph {et~al.}(2023)\citenamefont {Evered},
  \citenamefont {Bluvstein}, \citenamefont {Kalinowski}, \citenamefont {Ebadi},
  \citenamefont {Manovitz}, \citenamefont {Zhou}, \citenamefont {Li},
  \citenamefont {Geim}, \citenamefont {Wang}, \citenamefont {Maskara},
  \citenamefont {Levine}, \citenamefont {Semeghini}, \citenamefont {Greiner},
  \citenamefont {Vuletić},\ and\ \citenamefont {Lukin}}]{evered2023high}%
  \BibitemOpen
  \bibfield  {author} {\bibinfo {author} {\bibfnamefont {S.~J.}\ \bibnamefont
  {Evered}}, \bibinfo {author} {\bibfnamefont {D.}~\bibnamefont {Bluvstein}},
  \bibinfo {author} {\bibfnamefont {M.}~\bibnamefont {Kalinowski}}, \bibinfo
  {author} {\bibfnamefont {S.}~\bibnamefont {Ebadi}}, \bibinfo {author}
  {\bibfnamefont {T.}~\bibnamefont {Manovitz}}, \bibinfo {author}
  {\bibfnamefont {H.}~\bibnamefont {Zhou}}, \bibinfo {author} {\bibfnamefont
  {S.~H.}\ \bibnamefont {Li}}, \bibinfo {author} {\bibfnamefont {A.~A.}\
  \bibnamefont {Geim}}, \bibinfo {author} {\bibfnamefont {T.~T.}\ \bibnamefont
  {Wang}}, \bibinfo {author} {\bibfnamefont {N.}~\bibnamefont {Maskara}},
  \bibinfo {author} {\bibfnamefont {H.}~\bibnamefont {Levine}}, \bibinfo
  {author} {\bibfnamefont {G.}~\bibnamefont {Semeghini}}, \bibinfo {author}
  {\bibfnamefont {M.}~\bibnamefont {Greiner}}, \bibinfo {author} {\bibfnamefont
  {V.}~\bibnamefont {Vuletić}},\ and\ \bibinfo {author} {\bibfnamefont
  {M.~D.}\ \bibnamefont {Lukin}},\ }\bibfield  {title} {\bibinfo {title}
  {High-fidelity parallel entangling gates on a neutral-atom quantum
  computer},\ }\href {https://doi.org/10.1038/s41586-023-06481-y} {\bibfield
  {journal} {\bibinfo  {journal} {Nature}\ }\textbf {\bibinfo {volume} {622}},\
  \bibinfo {pages} {268} (\bibinfo {year} {2023})}\BibitemShut {NoStop}%
\bibitem [{\citenamefont {Guillaud}\ and\ \citenamefont
  {Mirrahimi}(2019)}]{guillaud2019repetition}%
  \BibitemOpen
  \bibfield  {author} {\bibinfo {author} {\bibfnamefont {J.}~\bibnamefont
  {Guillaud}}\ and\ \bibinfo {author} {\bibfnamefont {M.}~\bibnamefont
  {Mirrahimi}},\ }\bibfield  {title} {\bibinfo {title} {Repetition {Cat}
  {Qubits} for {Fault}-{Tolerant} {Quantum} {Computation}},\ }\href
  {https://doi.org/10.1103/PhysRevX.9.041053} {\bibfield  {journal} {\bibinfo
  {journal} {Physical Review X}\ }\textbf {\bibinfo {volume} {9}},\ \bibinfo
  {pages} {041053} (\bibinfo {year} {2019})}\BibitemShut {NoStop}%
\bibitem [{\citenamefont {Tuckett}\ \emph {et~al.}(2019)\citenamefont
  {Tuckett}, \citenamefont {Darmawan}, \citenamefont {Chubb}, \citenamefont
  {Bravyi}, \citenamefont {Bartlett},\ and\ \citenamefont
  {Flammia}}]{PhysRevX.9.041031}%
  \BibitemOpen
  \bibfield  {author} {\bibinfo {author} {\bibfnamefont {D.~K.}\ \bibnamefont
  {Tuckett}}, \bibinfo {author} {\bibfnamefont {A.~S.}\ \bibnamefont
  {Darmawan}}, \bibinfo {author} {\bibfnamefont {C.~T.}\ \bibnamefont {Chubb}},
  \bibinfo {author} {\bibfnamefont {S.}~\bibnamefont {Bravyi}}, \bibinfo
  {author} {\bibfnamefont {S.~D.}\ \bibnamefont {Bartlett}},\ and\ \bibinfo
  {author} {\bibfnamefont {S.~T.}\ \bibnamefont {Flammia}},\ }\bibfield
  {title} {\bibinfo {title} {Tailoring surface codes for highly biased noise},\
  }\href {https://doi.org/10.1103/PhysRevX.9.041031} {\bibfield  {journal}
  {\bibinfo  {journal} {Phys. Rev. X}\ }\textbf {\bibinfo {volume} {9}},\
  \bibinfo {pages} {041031} (\bibinfo {year} {2019})}\BibitemShut {NoStop}%
\bibitem [{\citenamefont {Darmawan}\ \emph {et~al.}(2021)\citenamefont
  {Darmawan}, \citenamefont {Brown}, \citenamefont {Grimsmo}, \citenamefont
  {Tuckett},\ and\ \citenamefont {Puri}}]{darmawan_practical_2021}%
  \BibitemOpen
  \bibfield  {author} {\bibinfo {author} {\bibfnamefont {A.~S.}\ \bibnamefont
  {Darmawan}}, \bibinfo {author} {\bibfnamefont {B.~J.}\ \bibnamefont {Brown}},
  \bibinfo {author} {\bibfnamefont {A.~L.}\ \bibnamefont {Grimsmo}}, \bibinfo
  {author} {\bibfnamefont {D.~K.}\ \bibnamefont {Tuckett}},\ and\ \bibinfo
  {author} {\bibfnamefont {S.}~\bibnamefont {Puri}},\ }\bibfield  {title}
  {\bibinfo {title} {Practical {Quantum} {Error} {Correction} with the {XZZX}
  {Code} and {Kerr}-{Cat} {Qubits}},\ }\href
  {https://doi.org/10.1103/PRXQuantum.2.030345} {\bibfield  {journal} {\bibinfo
   {journal} {PRX Quantum}\ }\textbf {\bibinfo {volume} {2}},\ \bibinfo {pages}
  {030345} (\bibinfo {year} {2021})}\BibitemShut {NoStop}%
\bibitem [{\citenamefont {Cong}\ \emph {et~al.}(2022)\citenamefont {Cong},
  \citenamefont {Levine}, \citenamefont {Keesling}, \citenamefont {Bluvstein},
  \citenamefont {Wang},\ and\ \citenamefont
  {Lukin}}]{cong_hardware-efficient_2022}%
  \BibitemOpen
  \bibfield  {author} {\bibinfo {author} {\bibfnamefont {I.}~\bibnamefont
  {Cong}}, \bibinfo {author} {\bibfnamefont {H.}~\bibnamefont {Levine}},
  \bibinfo {author} {\bibfnamefont {A.}~\bibnamefont {Keesling}}, \bibinfo
  {author} {\bibfnamefont {D.}~\bibnamefont {Bluvstein}}, \bibinfo {author}
  {\bibfnamefont {S.-T.}\ \bibnamefont {Wang}},\ and\ \bibinfo {author}
  {\bibfnamefont {M.~D.}\ \bibnamefont {Lukin}},\ }\bibfield  {title} {\bibinfo
  {title} {Hardware-{Efficient}, {Fault}-{Tolerant} {Quantum} {Computation}
  with {Rydberg} {Atoms}},\ }\href {https://doi.org/10.1103/PhysRevX.12.021049}
  {\bibfield  {journal} {\bibinfo  {journal} {Physical Review X}\ }\textbf
  {\bibinfo {volume} {12}},\ \bibinfo {pages} {021049} (\bibinfo {year}
  {2022})}\BibitemShut {NoStop}%
\bibitem [{\citenamefont {Chamberland}\ \emph {et~al.}(2022)\citenamefont
  {Chamberland}, \citenamefont {Noh}, \citenamefont {Arrangoiz-Arriola},
  \citenamefont {Campbell}, \citenamefont {Hann}, \citenamefont {Iverson},
  \citenamefont {Putterman}, \citenamefont {Bohdanowicz}, \citenamefont
  {Flammia}, \citenamefont {Keller}, \citenamefont {Refael}, \citenamefont
  {Preskill}, \citenamefont {Jiang}, \citenamefont {Safavi-Naeini},
  \citenamefont {Painter},\ and\ \citenamefont
  {Brand\~ao}}]{chamberland_building_2022}%
  \BibitemOpen
  \bibfield  {author} {\bibinfo {author} {\bibfnamefont {C.}~\bibnamefont
  {Chamberland}}, \bibinfo {author} {\bibfnamefont {K.}~\bibnamefont {Noh}},
  \bibinfo {author} {\bibfnamefont {P.}~\bibnamefont {Arrangoiz-Arriola}},
  \bibinfo {author} {\bibfnamefont {E.~T.}\ \bibnamefont {Campbell}}, \bibinfo
  {author} {\bibfnamefont {C.~T.}\ \bibnamefont {Hann}}, \bibinfo {author}
  {\bibfnamefont {J.}~\bibnamefont {Iverson}}, \bibinfo {author} {\bibfnamefont
  {H.}~\bibnamefont {Putterman}}, \bibinfo {author} {\bibfnamefont {T.~C.}\
  \bibnamefont {Bohdanowicz}}, \bibinfo {author} {\bibfnamefont {S.~T.}\
  \bibnamefont {Flammia}}, \bibinfo {author} {\bibfnamefont {A.}~\bibnamefont
  {Keller}}, \bibinfo {author} {\bibfnamefont {G.}~\bibnamefont {Refael}},
  \bibinfo {author} {\bibfnamefont {J.}~\bibnamefont {Preskill}}, \bibinfo
  {author} {\bibfnamefont {L.}~\bibnamefont {Jiang}}, \bibinfo {author}
  {\bibfnamefont {A.~H.}\ \bibnamefont {Safavi-Naeini}}, \bibinfo {author}
  {\bibfnamefont {O.}~\bibnamefont {Painter}},\ and\ \bibinfo {author}
  {\bibfnamefont {F.~G.}\ \bibnamefont {Brand\~ao}},\ }\bibfield  {title}
  {\bibinfo {title} {Building a {Fault}-{Tolerant} {Quantum} {Computer} {Using}
  {Concatenated} {Cat} {Codes}},\ }\href
  {https://doi.org/10.1103/PRXQuantum.3.010329} {\bibfield  {journal} {\bibinfo
   {journal} {PRX Quantum}\ }\textbf {\bibinfo {volume} {3}},\ \bibinfo {pages}
  {010329} (\bibinfo {year} {2022})}\BibitemShut {NoStop}%
\bibitem [{\citenamefont {Lescanne}\ \emph {et~al.}(2020)\citenamefont
  {Lescanne}, \citenamefont {Villiers}, \citenamefont {Peronnin}, \citenamefont
  {Sarlette}, \citenamefont {Delbecq}, \citenamefont {Huard}, \citenamefont
  {Kontos}, \citenamefont {Mirrahimi},\ and\ \citenamefont
  {Leghtas}}]{lescanne_exponential_2020}%
  \BibitemOpen
  \bibfield  {author} {\bibinfo {author} {\bibfnamefont {R.}~\bibnamefont
  {Lescanne}}, \bibinfo {author} {\bibfnamefont {M.}~\bibnamefont {Villiers}},
  \bibinfo {author} {\bibfnamefont {T.}~\bibnamefont {Peronnin}}, \bibinfo
  {author} {\bibfnamefont {A.}~\bibnamefont {Sarlette}}, \bibinfo {author}
  {\bibfnamefont {M.}~\bibnamefont {Delbecq}}, \bibinfo {author} {\bibfnamefont
  {B.}~\bibnamefont {Huard}}, \bibinfo {author} {\bibfnamefont
  {T.}~\bibnamefont {Kontos}}, \bibinfo {author} {\bibfnamefont
  {M.}~\bibnamefont {Mirrahimi}},\ and\ \bibinfo {author} {\bibfnamefont
  {Z.}~\bibnamefont {Leghtas}},\ }\bibfield  {title} {\bibinfo {title}
  {Exponential suppression of bit-flips in a qubit encoded in an oscillator},\
  }\href {https://doi.org/10.1038/s41567-020-0824-x} {\bibfield  {journal}
  {\bibinfo  {journal} {Nature Physics}\ }\textbf {\bibinfo {volume} {16}},\
  \bibinfo {pages} {509} (\bibinfo {year} {2020})}\BibitemShut {NoStop}%
\bibitem [{\citenamefont {Sivak}\ \emph {et~al.}(2023)\citenamefont {Sivak},
  \citenamefont {Eickbusch}, \citenamefont {Royer}, \citenamefont {Singh},
  \citenamefont {Tsioutsios}, \citenamefont {Ganjam}, \citenamefont {Miano},
  \citenamefont {Brock}, \citenamefont {Ding}, \citenamefont {Frunzio},
  \citenamefont {Girvin}, \citenamefont {Schoelkopf},\ and\ \citenamefont
  {Devoret}}]{sivak_real-time_2023}%
  \BibitemOpen
  \bibfield  {author} {\bibinfo {author} {\bibfnamefont {V.~V.}\ \bibnamefont
  {Sivak}}, \bibinfo {author} {\bibfnamefont {A.}~\bibnamefont {Eickbusch}},
  \bibinfo {author} {\bibfnamefont {B.}~\bibnamefont {Royer}}, \bibinfo
  {author} {\bibfnamefont {S.}~\bibnamefont {Singh}}, \bibinfo {author}
  {\bibfnamefont {I.}~\bibnamefont {Tsioutsios}}, \bibinfo {author}
  {\bibfnamefont {S.}~\bibnamefont {Ganjam}}, \bibinfo {author} {\bibfnamefont
  {A.}~\bibnamefont {Miano}}, \bibinfo {author} {\bibfnamefont {B.~L.}\
  \bibnamefont {Brock}}, \bibinfo {author} {\bibfnamefont {A.~Z.}\ \bibnamefont
  {Ding}}, \bibinfo {author} {\bibfnamefont {L.}~\bibnamefont {Frunzio}},
  \bibinfo {author} {\bibfnamefont {S.~M.}\ \bibnamefont {Girvin}}, \bibinfo
  {author} {\bibfnamefont {R.~J.}\ \bibnamefont {Schoelkopf}},\ and\ \bibinfo
  {author} {\bibfnamefont {M.~H.}\ \bibnamefont {Devoret}},\ }\bibfield
  {title} {\bibinfo {title} {Real-time quantum error correction beyond
  break-even},\ }\href {https://doi.org/10.1038/s41586-023-05782-6} {\bibfield
  {journal} {\bibinfo  {journal} {Nature}\ }\textbf {\bibinfo {volume} {616}},\
  \bibinfo {pages} {50} (\bibinfo {year} {2023})}\BibitemShut {NoStop}%
\bibitem [{\citenamefont {Wu}\ \emph {et~al.}(2022)\citenamefont {Wu},
  \citenamefont {Kolkowitz}, \citenamefont {Puri},\ and\ \citenamefont
  {Thompson}}]{wu_erasure_2022}%
  \BibitemOpen
  \bibfield  {author} {\bibinfo {author} {\bibfnamefont {Y.}~\bibnamefont
  {Wu}}, \bibinfo {author} {\bibfnamefont {S.}~\bibnamefont {Kolkowitz}},
  \bibinfo {author} {\bibfnamefont {S.}~\bibnamefont {Puri}},\ and\ \bibinfo
  {author} {\bibfnamefont {J.~D.}\ \bibnamefont {Thompson}},\ }\bibfield
  {title} {\bibinfo {title} {Erasure conversion for fault-tolerant quantum
  computing in alkaline earth {Rydberg} atom arrays},\ }\href
  {https://doi.org/10.1038/s41467-022-32094-6} {\bibfield  {journal} {\bibinfo
  {journal} {Nature Communications}\ }\textbf {\bibinfo {volume} {13}},\
  \bibinfo {pages} {4657} (\bibinfo {year} {2022})}\BibitemShut {NoStop}%
\bibitem [{\citenamefont {Kubica}\ \emph {et~al.}(2023)\citenamefont {Kubica},
  \citenamefont {Haim}, \citenamefont {Vaknin}, \citenamefont {Levine},
  \citenamefont {Brandão},\ and\ \citenamefont
  {Retzker}}]{kubica_erasure_2023}%
  \BibitemOpen
  \bibfield  {author} {\bibinfo {author} {\bibfnamefont {A.}~\bibnamefont
  {Kubica}}, \bibinfo {author} {\bibfnamefont {A.}~\bibnamefont {Haim}},
  \bibinfo {author} {\bibfnamefont {Y.}~\bibnamefont {Vaknin}}, \bibinfo
  {author} {\bibfnamefont {H.}~\bibnamefont {Levine}}, \bibinfo {author}
  {\bibfnamefont {F.}~\bibnamefont {Brandão}},\ and\ \bibinfo {author}
  {\bibfnamefont {A.}~\bibnamefont {Retzker}},\ }\bibfield  {title} {\bibinfo
  {title} {Erasure {Qubits}: {Overcoming} the ${T}_1$ {Limit} in
  {Superconducting} {Circuits}},\ }\href
  {https://doi.org/10.1103/PhysRevX.13.041022} {\bibfield  {journal} {\bibinfo
  {journal} {Physical Review X}\ }\textbf {\bibinfo {volume} {13}},\ \bibinfo
  {pages} {041022} (\bibinfo {year} {2023})}\BibitemShut {NoStop}%
\bibitem [{\citenamefont {Kubica}\ and\ \citenamefont
  {Retzker}(2021)}]{DualRailPatent}%
  \BibitemOpen
  \bibfield  {author} {\bibinfo {author} {\bibfnamefont {A.}~\bibnamefont
  {Kubica}}\ and\ \bibinfo {author} {\bibfnamefont {A.}~\bibnamefont
  {Retzker}},\ }\href
  {https://image-ppubs.uspto.gov/dirsearch-public/print/downloadPdf/11748652}
  {\bibinfo {title} {Heralding of amplitude damping decay noise for quantum
  error correction}},\ \bibinfo {note} {{}U.S. Patent
  11,748,652 (2021)}\BibitemShut {NoStop}%
\bibitem [{\citenamefont {Teoh}\ \emph {et~al.}(2023)\citenamefont {Teoh},
  \citenamefont {Winkel}, \citenamefont {Babla}, \citenamefont {Chapman},
  \citenamefont {Claes}, \citenamefont {De~Graaf}, \citenamefont {Garmon},
  \citenamefont {Kalfus}, \citenamefont {Lu}, \citenamefont {Maiti},
  \citenamefont {Sahay}, \citenamefont {Thakur}, \citenamefont {Tsunoda},
  \citenamefont {Xue}, \citenamefont {Frunzio}, \citenamefont {Girvin},
  \citenamefont {Puri},\ and\ \citenamefont {Schoelkopf}}]{teoh2023dual}%
  \BibitemOpen
  \bibfield  {author} {\bibinfo {author} {\bibfnamefont {J.~D.}\ \bibnamefont
  {Teoh}}, \bibinfo {author} {\bibfnamefont {P.}~\bibnamefont {Winkel}},
  \bibinfo {author} {\bibfnamefont {H.~K.}\ \bibnamefont {Babla}}, \bibinfo
  {author} {\bibfnamefont {B.~J.}\ \bibnamefont {Chapman}}, \bibinfo {author}
  {\bibfnamefont {J.}~\bibnamefont {Claes}}, \bibinfo {author} {\bibfnamefont
  {S.~J.}\ \bibnamefont {De~Graaf}}, \bibinfo {author} {\bibfnamefont
  {J.~W.~O.}\ \bibnamefont {Garmon}}, \bibinfo {author} {\bibfnamefont {W.~D.}\
  \bibnamefont {Kalfus}}, \bibinfo {author} {\bibfnamefont {Y.}~\bibnamefont
  {Lu}}, \bibinfo {author} {\bibfnamefont {A.}~\bibnamefont {Maiti}}, \bibinfo
  {author} {\bibfnamefont {K.}~\bibnamefont {Sahay}}, \bibinfo {author}
  {\bibfnamefont {N.}~\bibnamefont {Thakur}}, \bibinfo {author} {\bibfnamefont
  {T.}~\bibnamefont {Tsunoda}}, \bibinfo {author} {\bibfnamefont {S.~H.}\
  \bibnamefont {Xue}}, \bibinfo {author} {\bibfnamefont {L.}~\bibnamefont
  {Frunzio}}, \bibinfo {author} {\bibfnamefont {S.~M.}\ \bibnamefont {Girvin}},
  \bibinfo {author} {\bibfnamefont {S.}~\bibnamefont {Puri}},\ and\ \bibinfo
  {author} {\bibfnamefont {R.~J.}\ \bibnamefont {Schoelkopf}},\ }\bibfield
  {title} {\bibinfo {title} {Dual-rail encoding with superconducting
  cavities},\ }\href {https://doi.org/10.1073/pnas.2221736120} {\bibfield
  {journal} {\bibinfo  {journal} {Proceedings of the National Academy of
  Sciences}\ }\textbf {\bibinfo {volume} {120}},\ \bibinfo {pages}
  {e2221736120} (\bibinfo {year} {2023})}\BibitemShut {NoStop}%
\bibitem [{\citenamefont {Grassl}\ \emph {et~al.}(1997)\citenamefont {Grassl},
  \citenamefont {Beth},\ and\ \citenamefont {Pellizzari}}]{grassl_codes_1997}%
  \BibitemOpen
  \bibfield  {author} {\bibinfo {author} {\bibfnamefont {M.}~\bibnamefont
  {Grassl}}, \bibinfo {author} {\bibfnamefont {T.}~\bibnamefont {Beth}},\ and\
  \bibinfo {author} {\bibfnamefont {T.}~\bibnamefont {Pellizzari}},\ }\bibfield
   {title} {\bibinfo {title} {Codes for the quantum erasure channel},\ }\href
  {https://doi.org/10.1103/PhysRevA.56.33} {\bibfield  {journal} {\bibinfo
  {journal} {Physical Review A}\ }\textbf {\bibinfo {volume} {56}},\ \bibinfo
  {pages} {33} (\bibinfo {year} {1997})}\BibitemShut {NoStop}%
\bibitem [{\citenamefont {Knill}(2004)}]{knill_scalable_2004}%
  \BibitemOpen
  \bibfield  {author} {\bibinfo {author} {\bibfnamefont {E.}~\bibnamefont
  {Knill}},\ }\href@noop {} {\bibinfo {title} {Scalable quantum computation in
  the presence of large detected-error rates}},\ \bibinfo {howpublished}
  {arXiv:quant-ph/0312190} (\bibinfo {year} {2004})\BibitemShut {NoStop}%
\bibitem [{\citenamefont {Sahay}\ \emph {et~al.}(2023)\citenamefont {Sahay},
  \citenamefont {Jin}, \citenamefont {Claes}, \citenamefont {Thompson},\ and\
  \citenamefont {Puri}}]{sahay_high_2023}%
  \BibitemOpen
  \bibfield  {author} {\bibinfo {author} {\bibfnamefont {K.}~\bibnamefont
  {Sahay}}, \bibinfo {author} {\bibfnamefont {J.}~\bibnamefont {Jin}}, \bibinfo
  {author} {\bibfnamefont {J.}~\bibnamefont {Claes}}, \bibinfo {author}
  {\bibfnamefont {J.~D.}\ \bibnamefont {Thompson}},\ and\ \bibinfo {author}
  {\bibfnamefont {S.}~\bibnamefont {Puri}},\ }\bibfield  {title} {\bibinfo
  {title} {High-{Threshold} {Codes} for {Neutral}-{Atom} {Qubits} with {Biased}
  {Erasure} {Errors}},\ }\href {https://doi.org/10.1103/PhysRevX.13.041013}
  {\bibfield  {journal} {\bibinfo  {journal} {Physical Review X}\ }\textbf
  {\bibinfo {volume} {13}},\ \bibinfo {pages} {041013} (\bibinfo {year}
  {2023})}\BibitemShut {NoStop}%
\bibitem [{\citenamefont {Campbell}(2020)}]{campbell2020certified}%
  \BibitemOpen
  \bibfield  {author} {\bibinfo {author} {\bibfnamefont {W.~C.}\ \bibnamefont
  {Campbell}},\ }\bibfield  {title} {\bibinfo {title} {Certified quantum
  gates},\ }\href {https://doi.org/10.1103/PhysRevA.102.022426} {\bibfield
  {journal} {\bibinfo  {journal} {Physical Review A}\ }\textbf {\bibinfo
  {volume} {102}},\ \bibinfo {pages} {022426} (\bibinfo {year}
  {2020})}\BibitemShut {NoStop}%
\bibitem [{\citenamefont {Kang}\ \emph {et~al.}(2023)\citenamefont {Kang},
  \citenamefont {Campbell},\ and\ \citenamefont {Brown}}]{kang_quantum_2023}%
  \BibitemOpen
  \bibfield  {author} {\bibinfo {author} {\bibfnamefont {M.}~\bibnamefont
  {Kang}}, \bibinfo {author} {\bibfnamefont {W.~C.}\ \bibnamefont {Campbell}},\
  and\ \bibinfo {author} {\bibfnamefont {K.~R.}\ \bibnamefont {Brown}},\
  }\bibfield  {title} {\bibinfo {title} {Quantum {Error} {Correction} with
  {Metastable} {States} of {Trapped} {Ions} {Using} {Erasure} {Conversion}},\
  }\href {https://doi.org/10.1103/PRXQuantum.4.020358} {\bibfield  {journal}
  {\bibinfo  {journal} {PRX Quantum}\ }\textbf {\bibinfo {volume} {4}},\
  \bibinfo {pages} {020358} (\bibinfo {year} {2023})}\BibitemShut {NoStop}%
\bibitem [{\citenamefont {Ma}\ \emph {et~al.}(2023)\citenamefont {Ma},
  \citenamefont {Liu}, \citenamefont {Peng}, \citenamefont {Zhang},
  \citenamefont {Jandura}, \citenamefont {Claes}, \citenamefont {Burgers},
  \citenamefont {Pupillo}, \citenamefont {Puri},\ and\ \citenamefont
  {Thompson}}]{ma2023high}%
  \BibitemOpen
  \bibfield  {author} {\bibinfo {author} {\bibfnamefont {S.}~\bibnamefont
  {Ma}}, \bibinfo {author} {\bibfnamefont {G.}~\bibnamefont {Liu}}, \bibinfo
  {author} {\bibfnamefont {P.}~\bibnamefont {Peng}}, \bibinfo {author}
  {\bibfnamefont {B.}~\bibnamefont {Zhang}}, \bibinfo {author} {\bibfnamefont
  {S.}~\bibnamefont {Jandura}}, \bibinfo {author} {\bibfnamefont
  {J.}~\bibnamefont {Claes}}, \bibinfo {author} {\bibfnamefont {A.~P.}\
  \bibnamefont {Burgers}}, \bibinfo {author} {\bibfnamefont {G.}~\bibnamefont
  {Pupillo}}, \bibinfo {author} {\bibfnamefont {S.}~\bibnamefont {Puri}},\ and\
  \bibinfo {author} {\bibfnamefont {J.~D.}\ \bibnamefont {Thompson}},\
  }\bibfield  {title} {\bibinfo {title} {High-fidelity gates and mid-circuit
  erasure conversion in an atomic qubit},\ }\href
  {https://doi.org/10.1038/s41586-023-06438-1} {\bibfield  {journal} {\bibinfo
  {journal} {Nature}\ }\textbf {\bibinfo {volume} {622}},\ \bibinfo {pages}
  {279} (\bibinfo {year} {2023})}\BibitemShut {NoStop}%
\bibitem [{\citenamefont {Scholl}\ \emph {et~al.}(2023)\citenamefont {Scholl},
  \citenamefont {Shaw}, \citenamefont {Tsai}, \citenamefont {Finkelstein},
  \citenamefont {Choi},\ and\ \citenamefont {Endres}}]{scholl2023erasure}%
  \BibitemOpen
  \bibfield  {author} {\bibinfo {author} {\bibfnamefont {P.}~\bibnamefont
  {Scholl}}, \bibinfo {author} {\bibfnamefont {A.~L.}\ \bibnamefont {Shaw}},
  \bibinfo {author} {\bibfnamefont {R.~B.-S.}\ \bibnamefont {Tsai}}, \bibinfo
  {author} {\bibfnamefont {R.}~\bibnamefont {Finkelstein}}, \bibinfo {author}
  {\bibfnamefont {J.}~\bibnamefont {Choi}},\ and\ \bibinfo {author}
  {\bibfnamefont {M.}~\bibnamefont {Endres}},\ }\bibfield  {title} {\bibinfo
  {title} {Erasure conversion in a high-fidelity {Rydberg} quantum simulator},\
  }\href {https://doi.org/10.1038/s41586-023-06516-4} {\bibfield  {journal}
  {\bibinfo  {journal} {Nature}\ }\textbf {\bibinfo {volume} {622}},\ \bibinfo
  {pages} {273} (\bibinfo {year} {2023})}\BibitemShut {NoStop}%
\bibitem [{\citenamefont {Chuang}\ and\ \citenamefont
  {Yamamoto}(1995)}]{chuang_simple_1995}%
  \BibitemOpen
  \bibfield  {author} {\bibinfo {author} {\bibfnamefont {I.~L.}\ \bibnamefont
  {Chuang}}\ and\ \bibinfo {author} {\bibfnamefont {Y.}~\bibnamefont
  {Yamamoto}},\ }\bibfield  {title} {\bibinfo {title} {Simple quantum
  computer},\ }\href {https://doi.org/10.1103/PhysRevA.52.3489} {\bibfield
  {journal} {\bibinfo  {journal} {Physical Review A}\ }\textbf {\bibinfo
  {volume} {52}},\ \bibinfo {pages} {3489} (\bibinfo {year}
  {1995})}\BibitemShut {NoStop}%
\bibitem [{\citenamefont {Duan}\ \emph {et~al.}(2010)\citenamefont {Duan},
  \citenamefont {Grassl}, \citenamefont {Ji},\ and\ \citenamefont
  {Zeng}}]{duan_multi-error-correcting_2010}%
  \BibitemOpen
  \bibfield  {author} {\bibinfo {author} {\bibfnamefont {R.}~\bibnamefont
  {Duan}}, \bibinfo {author} {\bibfnamefont {M.}~\bibnamefont {Grassl}},
  \bibinfo {author} {\bibfnamefont {Z.}~\bibnamefont {Ji}},\ and\ \bibinfo
  {author} {\bibfnamefont {B.}~\bibnamefont {Zeng}},\ }\bibfield  {title}
  {\bibinfo {title} {Multi-error-correcting amplitude damping codes},\ }in\
  \href {https://doi.org/10.1109/ISIT.2010.5513648} {\emph {\bibinfo
  {booktitle} {2010 {IEEE} {International} {Symposium} on {Information}
  {Theory}}}}\ (\bibinfo  {publisher} {IEEE},\ \bibinfo {address} {Austin, TX,
  USA},\ \bibinfo {year} {2010})\ pp.\ \bibinfo {pages}
  {2672--2676}\BibitemShut {NoStop}%
\bibitem [{\citenamefont {Zakka-Bajjani}\ \emph {et~al.}(2011)\citenamefont
  {Zakka-Bajjani}, \citenamefont {Nguyen}, \citenamefont {Lee}, \citenamefont
  {Vale}, \citenamefont {Simmonds},\ and\ \citenamefont
  {Aumentado}}]{zakka-bajjani_quantum_2011}%
  \BibitemOpen
  \bibfield  {author} {\bibinfo {author} {\bibfnamefont {E.}~\bibnamefont
  {Zakka-Bajjani}}, \bibinfo {author} {\bibfnamefont {F.}~\bibnamefont
  {Nguyen}}, \bibinfo {author} {\bibfnamefont {M.}~\bibnamefont {Lee}},
  \bibinfo {author} {\bibfnamefont {L.~R.}\ \bibnamefont {Vale}}, \bibinfo
  {author} {\bibfnamefont {R.~W.}\ \bibnamefont {Simmonds}},\ and\ \bibinfo
  {author} {\bibfnamefont {J.}~\bibnamefont {Aumentado}},\ }\bibfield  {title}
  {\bibinfo {title} {Quantum superposition of a single microwave photon in two
  different `colour' states},\ }\href {https://doi.org/10.1038/nphys2035}
  {\bibfield  {journal} {\bibinfo  {journal} {Nature Physics}\ }\textbf
  {\bibinfo {volume} {7}},\ \bibinfo {pages} {599} (\bibinfo {year}
  {2011})}\BibitemShut {NoStop}%
\bibitem [{\citenamefont {Shim}\ and\ \citenamefont
  {Tahan}(2016)}]{shim_semiconductor-inspired_2016}%
  \BibitemOpen
  \bibfield  {author} {\bibinfo {author} {\bibfnamefont {Y.-P.}\ \bibnamefont
  {Shim}}\ and\ \bibinfo {author} {\bibfnamefont {C.}~\bibnamefont {Tahan}},\
  }\bibfield  {title} {\bibinfo {title} {Semiconductor-inspired design
  principles for superconducting quantum computing},\ }\href
  {https://doi.org/10.1038/ncomms11059} {\bibfield  {journal} {\bibinfo
  {journal} {Nature Communications}\ }\textbf {\bibinfo {volume} {7}},\
  \bibinfo {pages} {11059} (\bibinfo {year} {2016})}\BibitemShut {NoStop}%
\bibitem [{\citenamefont {Campbell}\ \emph {et~al.}(2020)\citenamefont
  {Campbell}, \citenamefont {Shim}, \citenamefont {Kannan}, \citenamefont
  {Winik}, \citenamefont {Kim}, \citenamefont {Melville}, \citenamefont
  {Niedzielski}, \citenamefont {Yoder}, \citenamefont {Tahan}, \citenamefont
  {Gustavsson},\ and\ \citenamefont {Oliver}}]{campbell_universal_2020}%
  \BibitemOpen
  \bibfield  {author} {\bibinfo {author} {\bibfnamefont {D.~L.}\ \bibnamefont
  {Campbell}}, \bibinfo {author} {\bibfnamefont {Y.-P.}\ \bibnamefont {Shim}},
  \bibinfo {author} {\bibfnamefont {B.}~\bibnamefont {Kannan}}, \bibinfo
  {author} {\bibfnamefont {R.}~\bibnamefont {Winik}}, \bibinfo {author}
  {\bibfnamefont {D.~K.}\ \bibnamefont {Kim}}, \bibinfo {author} {\bibfnamefont
  {A.}~\bibnamefont {Melville}}, \bibinfo {author} {\bibfnamefont {B.~M.}\
  \bibnamefont {Niedzielski}}, \bibinfo {author} {\bibfnamefont {J.~L.}\
  \bibnamefont {Yoder}}, \bibinfo {author} {\bibfnamefont {C.}~\bibnamefont
  {Tahan}}, \bibinfo {author} {\bibfnamefont {S.}~\bibnamefont {Gustavsson}},\
  and\ \bibinfo {author} {\bibfnamefont {W.~D.}\ \bibnamefont {Oliver}},\
  }\bibfield  {title} {\bibinfo {title} {Universal {Nonadiabatic} {Control} of
  {Small}-{Gap} {Superconducting} {Qubits}},\ }\href
  {https://doi.org/10.1103/PhysRevX.10.041051} {\bibfield  {journal} {\bibinfo
  {journal} {Physical Review X}\ }\textbf {\bibinfo {volume} {10}},\ \bibinfo
  {pages} {041051} (\bibinfo {year} {2020})}\BibitemShut {NoStop}%
\bibitem [{\citenamefont {Timoney}\ \emph {et~al.}(2011)\citenamefont
  {Timoney}, \citenamefont {Baumgart}, \citenamefont {Johanning}, \citenamefont
  {Varón}, \citenamefont {Plenio}, \citenamefont {Retzker},\ and\
  \citenamefont {Wunderlich}}]{timoney_quantum_2011}%
  \BibitemOpen
  \bibfield  {author} {\bibinfo {author} {\bibfnamefont {N.}~\bibnamefont
  {Timoney}}, \bibinfo {author} {\bibfnamefont {I.}~\bibnamefont {Baumgart}},
  \bibinfo {author} {\bibfnamefont {M.}~\bibnamefont {Johanning}}, \bibinfo
  {author} {\bibfnamefont {A.~F.}\ \bibnamefont {Varón}}, \bibinfo {author}
  {\bibfnamefont {M.~B.}\ \bibnamefont {Plenio}}, \bibinfo {author}
  {\bibfnamefont {A.}~\bibnamefont {Retzker}},\ and\ \bibinfo {author}
  {\bibfnamefont {C.}~\bibnamefont {Wunderlich}},\ }\bibfield  {title}
  {\bibinfo {title} {Quantum gates and memory using microwave-dressed states},\
  }\href {https://doi.org/10.1038/nature10319} {\bibfield  {journal} {\bibinfo
  {journal} {Nature}\ }\textbf {\bibinfo {volume} {476}},\ \bibinfo {pages}
  {185} (\bibinfo {year} {2011})}\BibitemShut {NoStop}%
\bibitem [{\citenamefont {Miao}\ \emph {et~al.}(2020)\citenamefont {Miao},
  \citenamefont {Blanton}, \citenamefont {Anderson}, \citenamefont {Bourassa},
  \citenamefont {Crook}, \citenamefont {Wolfowicz}, \citenamefont {Abe},
  \citenamefont {Ohshima},\ and\ \citenamefont
  {Awschalom}}]{miao_universal_2020}%
  \BibitemOpen
  \bibfield  {author} {\bibinfo {author} {\bibfnamefont {K.~C.}\ \bibnamefont
  {Miao}}, \bibinfo {author} {\bibfnamefont {J.~P.}\ \bibnamefont {Blanton}},
  \bibinfo {author} {\bibfnamefont {C.~P.}\ \bibnamefont {Anderson}}, \bibinfo
  {author} {\bibfnamefont {A.}~\bibnamefont {Bourassa}}, \bibinfo {author}
  {\bibfnamefont {A.~L.}\ \bibnamefont {Crook}}, \bibinfo {author}
  {\bibfnamefont {G.}~\bibnamefont {Wolfowicz}}, \bibinfo {author}
  {\bibfnamefont {H.}~\bibnamefont {Abe}}, \bibinfo {author} {\bibfnamefont
  {T.}~\bibnamefont {Ohshima}},\ and\ \bibinfo {author} {\bibfnamefont {D.~D.}\
  \bibnamefont {Awschalom}},\ }\bibfield  {title} {\bibinfo {title} {Universal
  coherence protection in a solid-state spin qubit},\ }\href
  {https://doi.org/10.1126/science.abc5186} {\bibfield  {journal} {\bibinfo
  {journal} {Science}\ }\textbf {\bibinfo {volume} {369}},\ \bibinfo {pages}
  {1493} (\bibinfo {year} {2020})}\BibitemShut {NoStop}%
\bibitem [{\citenamefont {Huang}\ \emph {et~al.}(2021)\citenamefont {Huang},
  \citenamefont {Mundada}, \citenamefont {Gyenis}, \citenamefont {Schuster},
  \citenamefont {Houck},\ and\ \citenamefont {Koch}}]{huang_engineering_2021}%
  \BibitemOpen
  \bibfield  {author} {\bibinfo {author} {\bibfnamefont {Z.}~\bibnamefont
  {Huang}}, \bibinfo {author} {\bibfnamefont {P.~S.}\ \bibnamefont {Mundada}},
  \bibinfo {author} {\bibfnamefont {A.}~\bibnamefont {Gyenis}}, \bibinfo
  {author} {\bibfnamefont {D.~I.}\ \bibnamefont {Schuster}}, \bibinfo {author}
  {\bibfnamefont {A.~A.}\ \bibnamefont {Houck}},\ and\ \bibinfo {author}
  {\bibfnamefont {J.}~\bibnamefont {Koch}},\ }\bibfield  {title} {\bibinfo
  {title} {Engineering {Dynamical} {Sweet} {Spots} to {Protect} {Qubits} from
  $1/f$ {Noise}},\ }\href {https://doi.org/10.1103/PhysRevApplied.15.034065}
  {\bibfield  {journal} {\bibinfo  {journal} {Physical Review Applied}\
  }\textbf {\bibinfo {volume} {15}},\ \bibinfo {pages} {034065} (\bibinfo
  {year} {2021})}\BibitemShut {NoStop}%
\bibitem [{\citenamefont {Gullion}\ \emph {et~al.}(1990)\citenamefont
  {Gullion}, \citenamefont {Baker},\ and\ \citenamefont
  {Conradi}}]{gullion_new_1990}%
  \BibitemOpen
  \bibfield  {author} {\bibinfo {author} {\bibfnamefont {T.}~\bibnamefont
  {Gullion}}, \bibinfo {author} {\bibfnamefont {D.~B.}\ \bibnamefont {Baker}},\
  and\ \bibinfo {author} {\bibfnamefont {M.~S.}\ \bibnamefont {Conradi}},\
  }\bibfield  {title} {\bibinfo {title} {New, compensated {Carr}-{Purcell}
  sequences},\ }\href {https://doi.org/10.1016/0022-2364(90)90331-3} {\bibfield
   {journal} {\bibinfo  {journal} {Journal of Magnetic Resonance (1969)}\
  }\textbf {\bibinfo {volume} {89}},\ \bibinfo {pages} {479} (\bibinfo {year}
  {1990})}\BibitemShut {NoStop}%
\bibitem [{\citenamefont {McKay}\ \emph {et~al.}(2017)\citenamefont {McKay},
  \citenamefont {Wood}, \citenamefont {Sheldon}, \citenamefont {Chow},\ and\
  \citenamefont {Gambetta}}]{mckay_efficient_2017}%
  \BibitemOpen
  \bibfield  {author} {\bibinfo {author} {\bibfnamefont {D.~C.}\ \bibnamefont
  {McKay}}, \bibinfo {author} {\bibfnamefont {C.~J.}\ \bibnamefont {Wood}},
  \bibinfo {author} {\bibfnamefont {S.}~\bibnamefont {Sheldon}}, \bibinfo
  {author} {\bibfnamefont {J.~M.}\ \bibnamefont {Chow}},\ and\ \bibinfo
  {author} {\bibfnamefont {J.~M.}\ \bibnamefont {Gambetta}},\ }\bibfield
  {title} {\bibinfo {title} {Efficient {Z} gates for quantum computing},\
  }\href {https://doi.org/10.1103/PhysRevA.96.022330} {\bibfield  {journal}
  {\bibinfo  {journal} {Physical Review A}\ }\textbf {\bibinfo {volume} {96}},\
  \bibinfo {pages} {022330} (\bibinfo {year} {2017})}\BibitemShut {NoStop}%
\bibitem [{\citenamefont {Livingston}\ \emph {et~al.}(2022)\citenamefont
  {Livingston}, \citenamefont {Blok}, \citenamefont {Flurin}, \citenamefont
  {Dressel}, \citenamefont {Jordan},\ and\ \citenamefont
  {Siddiqi}}]{livingston_experimental_2022}%
  \BibitemOpen
  \bibfield  {author} {\bibinfo {author} {\bibfnamefont {W.~P.}\ \bibnamefont
  {Livingston}}, \bibinfo {author} {\bibfnamefont {M.~S.}\ \bibnamefont
  {Blok}}, \bibinfo {author} {\bibfnamefont {E.}~\bibnamefont {Flurin}},
  \bibinfo {author} {\bibfnamefont {J.}~\bibnamefont {Dressel}}, \bibinfo
  {author} {\bibfnamefont {A.~N.}\ \bibnamefont {Jordan}},\ and\ \bibinfo
  {author} {\bibfnamefont {I.}~\bibnamefont {Siddiqi}},\ }\bibfield  {title}
  {\bibinfo {title} {Experimental demonstration of continuous quantum error
  correction},\ }\href {https://doi.org/10.1038/s41467-022-29906-0} {\bibfield
  {journal} {\bibinfo  {journal} {Nature Communications}\ }\textbf {\bibinfo
  {volume} {13}},\ \bibinfo {pages} {2307} (\bibinfo {year}
  {2022})}\BibitemShut {NoStop}%
\bibitem [{\citenamefont {Heinsoo}\ \emph {et~al.}(2018)\citenamefont
  {Heinsoo}, \citenamefont {Andersen}, \citenamefont {Remm}, \citenamefont
  {Krinner}, \citenamefont {Walter}, \citenamefont {Salathé}, \citenamefont
  {Gasparinetti}, \citenamefont {Besse}, \citenamefont {Potočnik},
  \citenamefont {Wallraff},\ and\ \citenamefont
  {Eichler}}]{heinsoo_rapid_2018}%
  \BibitemOpen
  \bibfield  {author} {\bibinfo {author} {\bibfnamefont {J.}~\bibnamefont
  {Heinsoo}}, \bibinfo {author} {\bibfnamefont {C.~K.}\ \bibnamefont
  {Andersen}}, \bibinfo {author} {\bibfnamefont {A.}~\bibnamefont {Remm}},
  \bibinfo {author} {\bibfnamefont {S.}~\bibnamefont {Krinner}}, \bibinfo
  {author} {\bibfnamefont {T.}~\bibnamefont {Walter}}, \bibinfo {author}
  {\bibfnamefont {Y.}~\bibnamefont {Salathé}}, \bibinfo {author}
  {\bibfnamefont {S.}~\bibnamefont {Gasparinetti}}, \bibinfo {author}
  {\bibfnamefont {J.-C.}\ \bibnamefont {Besse}}, \bibinfo {author}
  {\bibfnamefont {A.}~\bibnamefont {Potočnik}}, \bibinfo {author}
  {\bibfnamefont {A.}~\bibnamefont {Wallraff}},\ and\ \bibinfo {author}
  {\bibfnamefont {C.}~\bibnamefont {Eichler}},\ }\bibfield  {title} {\bibinfo
  {title} {Rapid {High}-fidelity {Multiplexed} {Readout} of {Superconducting}
  {Qubits}},\ }\href {https://doi.org/10.1103/PhysRevApplied.10.034040}
  {\bibfield  {journal} {\bibinfo  {journal} {Physical Review Applied}\
  }\textbf {\bibinfo {volume} {10}},\ \bibinfo {pages} {034040} (\bibinfo
  {year} {2018})}\BibitemShut {NoStop}%
\bibitem [{\citenamefont {Ristè}\ \emph {et~al.}(2012)\citenamefont {Ristè},
  \citenamefont {Bultink}, \citenamefont {Lehnert},\ and\ \citenamefont
  {DiCarlo}}]{riste_feedback_2012}%
  \BibitemOpen
  \bibfield  {author} {\bibinfo {author} {\bibfnamefont {D.}~\bibnamefont
  {Ristè}}, \bibinfo {author} {\bibfnamefont {C.~C.}\ \bibnamefont {Bultink}},
  \bibinfo {author} {\bibfnamefont {K.~W.}\ \bibnamefont {Lehnert}},\ and\
  \bibinfo {author} {\bibfnamefont {L.}~\bibnamefont {DiCarlo}},\ }\bibfield
  {title} {\bibinfo {title} {Feedback {Control} of a {Solid}-{State} {Qubit}
  {Using} {High}-{Fidelity} {Projective} {Measurement}},\ }\href
  {https://doi.org/10.1103/PhysRevLett.109.240502} {\bibfield  {journal}
  {\bibinfo  {journal} {Physical Review Letters}\ }\textbf {\bibinfo {volume}
  {109}},\ \bibinfo {pages} {240502} (\bibinfo {year} {2012})}\BibitemShut
  {NoStop}%
\bibitem [{\citenamefont {McEwen}\ \emph {et~al.}(2021)\citenamefont {McEwen},
  \citenamefont {Kafri}, \citenamefont {Chen}, \citenamefont {Atalaya},
  \citenamefont {Satzinger}, \citenamefont {Quintana}, \citenamefont {Klimov},
  \citenamefont {Sank}, \citenamefont {Gidney}, \citenamefont {Fowler},
  \citenamefont {Arute}, \citenamefont {Arya}, \citenamefont {Buckley},
  \citenamefont {Burkett}, \citenamefont {Bushnell} \emph
  {et~al.}}]{mcewen_removing_2021}%
  \BibitemOpen
  \bibfield  {author} {\bibinfo {author} {\bibfnamefont {M.}~\bibnamefont
  {McEwen}}, \bibinfo {author} {\bibfnamefont {D.}~\bibnamefont {Kafri}},
  \bibinfo {author} {\bibfnamefont {Z.}~\bibnamefont {Chen}}, \bibinfo {author}
  {\bibfnamefont {J.}~\bibnamefont {Atalaya}}, \bibinfo {author} {\bibfnamefont
  {K.~J.}\ \bibnamefont {Satzinger}}, \bibinfo {author} {\bibfnamefont
  {C.}~\bibnamefont {Quintana}}, \bibinfo {author} {\bibfnamefont {P.~V.}\
  \bibnamefont {Klimov}}, \bibinfo {author} {\bibfnamefont {D.}~\bibnamefont
  {Sank}}, \bibinfo {author} {\bibfnamefont {C.}~\bibnamefont {Gidney}},
  \bibinfo {author} {\bibfnamefont {A.~G.}\ \bibnamefont {Fowler}}, \bibinfo
  {author} {\bibfnamefont {F.}~\bibnamefont {Arute}}, \bibinfo {author}
  {\bibfnamefont {K.}~\bibnamefont {Arya}}, \bibinfo {author} {\bibfnamefont
  {B.}~\bibnamefont {Buckley}}, \bibinfo {author} {\bibfnamefont
  {B.}~\bibnamefont {Burkett}}, \bibinfo {author} {\bibfnamefont
  {N.}~\bibnamefont {Bushnell}}, \emph {et~al.},\ }\bibfield  {title} {\bibinfo
  {title} {Removing leakage-induced correlated errors in superconducting
  quantum error correction},\ }\href
  {https://doi.org/10.1038/s41467-021-21982-y} {\bibfield  {journal} {\bibinfo
  {journal} {Nature Communications}\ }\textbf {\bibinfo {volume} {12}},\
  \bibinfo {pages} {1761} (\bibinfo {year} {2021})}\BibitemShut {NoStop}%
\bibitem [{\citenamefont {Marques}\ \emph {et~al.}(2023)\citenamefont
  {Marques}, \citenamefont {Ali}, \citenamefont {Varbanov}, \citenamefont
  {Finkel}, \citenamefont {Veen}, \citenamefont {Van Der~Meer}, \citenamefont
  {Valles-Sanclemente}, \citenamefont {Muthusubramanian}, \citenamefont
  {Beekman}, \citenamefont {Haider}, \citenamefont {Terhal},\ and\
  \citenamefont {DiCarlo}}]{marques_all-microwave_2023}%
  \BibitemOpen
  \bibfield  {author} {\bibinfo {author} {\bibfnamefont {J.}~\bibnamefont
  {Marques}}, \bibinfo {author} {\bibfnamefont {H.}~\bibnamefont {Ali}},
  \bibinfo {author} {\bibfnamefont {B.}~\bibnamefont {Varbanov}}, \bibinfo
  {author} {\bibfnamefont {M.}~\bibnamefont {Finkel}}, \bibinfo {author}
  {\bibfnamefont {H.}~\bibnamefont {Veen}}, \bibinfo {author} {\bibfnamefont
  {S.}~\bibnamefont {Van Der~Meer}}, \bibinfo {author} {\bibfnamefont
  {S.}~\bibnamefont {Valles-Sanclemente}}, \bibinfo {author} {\bibfnamefont
  {N.}~\bibnamefont {Muthusubramanian}}, \bibinfo {author} {\bibfnamefont
  {M.}~\bibnamefont {Beekman}}, \bibinfo {author} {\bibfnamefont
  {N.}~\bibnamefont {Haider}}, \bibinfo {author} {\bibfnamefont
  {B.}~\bibnamefont {Terhal}},\ and\ \bibinfo {author} {\bibfnamefont
  {L.}~\bibnamefont {DiCarlo}},\ }\bibfield  {title} {\bibinfo {title}
  {All-{Microwave} {Leakage} {Reduction} {Units} for {Quantum} {Error}
  {Correction} with {Superconducting} {Transmon} {Qubits}},\ }\href
  {https://doi.org/10.1103/PhysRevLett.130.250602} {\bibfield  {journal}
  {\bibinfo  {journal} {Physical Review Letters}\ }\textbf {\bibinfo {volume}
  {130}},\ \bibinfo {pages} {250602} (\bibinfo {year} {2023})}\BibitemShut
  {NoStop}%
\bibitem [{\citenamefont {Khodjasteh}\ and\ \citenamefont
  {Viola}(2009)}]{khodjasteh_dynamically_2009}%
  \BibitemOpen
  \bibfield  {author} {\bibinfo {author} {\bibfnamefont {K.}~\bibnamefont
  {Khodjasteh}}\ and\ \bibinfo {author} {\bibfnamefont {L.}~\bibnamefont
  {Viola}},\ }\bibfield  {title} {\bibinfo {title} {Dynamically
  {Error}-{Corrected} {Gates} for {Universal} {Quantum} {Computation}},\ }\href
  {https://doi.org/10.1103/PhysRevLett.102.080501} {\bibfield  {journal}
  {\bibinfo  {journal} {Physical Review Letters}\ }\textbf {\bibinfo {volume}
  {102}},\ \bibinfo {pages} {080501} (\bibinfo {year} {2009})}\BibitemShut
  {NoStop}%
\bibitem [{\citenamefont {McArdle}\ \emph {et~al.}(2019)\citenamefont
  {McArdle}, \citenamefont {Yuan},\ and\ \citenamefont
  {Benjamin}}]{mcardle2019error}%
  \BibitemOpen
  \bibfield  {author} {\bibinfo {author} {\bibfnamefont {S.}~\bibnamefont
  {McArdle}}, \bibinfo {author} {\bibfnamefont {X.}~\bibnamefont {Yuan}},\ and\
  \bibinfo {author} {\bibfnamefont {S.}~\bibnamefont {Benjamin}},\ }\bibfield
  {title} {\bibinfo {title} {Error-{Mitigated} {Digital} {Quantum}
  {Simulation}},\ }\href {https://doi.org/10.1103/PhysRevLett.122.180501}
  {\bibfield  {journal} {\bibinfo  {journal} {Physical Review Letters}\
  }\textbf {\bibinfo {volume} {122}},\ \bibinfo {pages} {180501} (\bibinfo
  {year} {2019})}\BibitemShut {NoStop}%
\bibitem [{\citenamefont {Yang}\ \emph {et~al.}(2023)\citenamefont {Yang},
  \citenamefont {Shen}, \citenamefont {Cao},\ and\ \citenamefont
  {Wang}}]{yang_post-selection_2023}%
  \BibitemOpen
  \bibfield  {author} {\bibinfo {author} {\bibfnamefont {T.-Y.}\ \bibnamefont
  {Yang}}, \bibinfo {author} {\bibfnamefont {Y.-X.}\ \bibnamefont {Shen}},
  \bibinfo {author} {\bibfnamefont {Z.-K.}\ \bibnamefont {Cao}},\ and\ \bibinfo
  {author} {\bibfnamefont {X.-B.}\ \bibnamefont {Wang}},\ }\bibfield  {title}
  {\bibinfo {title} {Post-selection in noisy {Gaussian} boson sampling: part is
  better than whole},\ }\href {https://doi.org/10.1088/2058-9565/acf06c}
  {\bibfield  {journal} {\bibinfo  {journal} {Quantum Science and Technology}\
  }\textbf {\bibinfo {volume} {8}},\ \bibinfo {pages} {045020} (\bibinfo {year}
  {2023})}\BibitemShut {NoStop}%
\bibitem [{\citenamefont {Simon}(1997)}]{simon1997power}%
  \BibitemOpen
  \bibfield  {author} {\bibinfo {author} {\bibfnamefont {D.~R.}\ \bibnamefont
  {Simon}},\ }\bibfield  {title} {\bibinfo {title} {On the {Power} of {Quantum}
  {Computation}},\ }\href {https://doi.org/10.1137/S0097539796298637}
  {\bibfield  {journal} {\bibinfo  {journal} {SIAM Journal on Computing}\
  }\textbf {\bibinfo {volume} {26}},\ \bibinfo {pages} {1474} (\bibinfo {year}
  {1997})}\BibitemShut {NoStop}%
\bibitem [{\citenamefont {Botelho}\ \emph {et~al.}(2022)\citenamefont
  {Botelho}, \citenamefont {Glos}, \citenamefont {Kundu}, \citenamefont
  {Miszczak}, \citenamefont {Salehi},\ and\ \citenamefont
  {Zimbor\'as}}]{botelho2022error}%
  \BibitemOpen
  \bibfield  {author} {\bibinfo {author} {\bibfnamefont {L.}~\bibnamefont
  {Botelho}}, \bibinfo {author} {\bibfnamefont {A.}~\bibnamefont {Glos}},
  \bibinfo {author} {\bibfnamefont {A.}~\bibnamefont {Kundu}}, \bibinfo
  {author} {\bibfnamefont {J.~A.}\ \bibnamefont {Miszczak}}, \bibinfo {author}
  {\bibfnamefont {O.}~\bibnamefont {Salehi}},\ and\ \bibinfo {author}
  {\bibfnamefont {Z.}~\bibnamefont {Zimbor\'as}},\ }\bibfield  {title}
  {\bibinfo {title} {Error mitigation for variational quantum algorithms
  through mid-circuit measurements},\ }\href
  {https://doi.org/10.1103/PhysRevA.105.022441} {\bibfield  {journal} {\bibinfo
   {journal} {Physical Review A}\ }\textbf {\bibinfo {volume} {105}},\ \bibinfo
  {pages} {022441} (\bibinfo {year} {2022})}\BibitemShut {NoStop}%
\bibitem [{\citenamefont {Chou}\ \emph {et~al.}(2023)\citenamefont {Chou},
  \citenamefont {Shemma}, \citenamefont {McCarrick}, \citenamefont {Chien},
  \citenamefont {Teoh}, \citenamefont {Winkel}, \citenamefont {Anderson},
  \citenamefont {Chen}, \citenamefont {Curtis}, \citenamefont {de~Graaf},
  \citenamefont {Garmon}, \citenamefont {Gudlewski}, \citenamefont {Kalfus},
  \citenamefont {Keen} \emph {et~al.}}]{chou2023demonstrating}%
  \BibitemOpen
  \bibfield  {author} {\bibinfo {author} {\bibfnamefont {K.~S.}\ \bibnamefont
  {Chou}}, \bibinfo {author} {\bibfnamefont {T.}~\bibnamefont {Shemma}},
  \bibinfo {author} {\bibfnamefont {H.}~\bibnamefont {McCarrick}}, \bibinfo
  {author} {\bibfnamefont {T.-C.}\ \bibnamefont {Chien}}, \bibinfo {author}
  {\bibfnamefont {J.~D.}\ \bibnamefont {Teoh}}, \bibinfo {author}
  {\bibfnamefont {P.}~\bibnamefont {Winkel}}, \bibinfo {author} {\bibfnamefont
  {A.}~\bibnamefont {Anderson}}, \bibinfo {author} {\bibfnamefont
  {J.}~\bibnamefont {Chen}}, \bibinfo {author} {\bibfnamefont {J.}~\bibnamefont
  {Curtis}}, \bibinfo {author} {\bibfnamefont {S.~J.}\ \bibnamefont
  {de~Graaf}}, \bibinfo {author} {\bibfnamefont {J.~W.~O.}\ \bibnamefont
  {Garmon}}, \bibinfo {author} {\bibfnamefont {B.}~\bibnamefont {Gudlewski}},
  \bibinfo {author} {\bibfnamefont {W.~D.}\ \bibnamefont {Kalfus}}, \bibinfo
  {author} {\bibfnamefont {T.}~\bibnamefont {Keen}}, \emph {et~al.},\
  }\href@noop {} {\bibinfo {title} {Demonstrating a superconducting dual-rail
  cavity qubit with erasure-detected logical measurements}},\ \bibinfo
  {howpublished} {arXiv:2307.03169} (\bibinfo {year} {2023})\BibitemShut
  {NoStop}%
\bibitem [{\citenamefont {Clerk}\ \emph {et~al.}(2010)\citenamefont {Clerk},
  \citenamefont {Devoret}, \citenamefont {Girvin}, \citenamefont {Marquardt},\
  and\ \citenamefont {Schoelkopf}}]{clerk_introduction_2010}%
  \BibitemOpen
  \bibfield  {author} {\bibinfo {author} {\bibfnamefont {A.~A.}\ \bibnamefont
  {Clerk}}, \bibinfo {author} {\bibfnamefont {M.~H.}\ \bibnamefont {Devoret}},
  \bibinfo {author} {\bibfnamefont {S.~M.}\ \bibnamefont {Girvin}}, \bibinfo
  {author} {\bibfnamefont {F.}~\bibnamefont {Marquardt}},\ and\ \bibinfo
  {author} {\bibfnamefont {R.~J.}\ \bibnamefont {Schoelkopf}},\ }\bibfield
  {title} {\bibinfo {title} {Introduction to quantum noise, measurement, and
  amplification},\ }\href {https://doi.org/10.1103/RevModPhys.82.1155}
  {\bibfield  {journal} {\bibinfo  {journal} {Reviews of Modern Physics}\
  }\textbf {\bibinfo {volume} {82}},\ \bibinfo {pages} {1155} (\bibinfo {year}
  {2010})}\BibitemShut {NoStop}%
\bibitem [{\citenamefont {Krinner}\ \emph {et~al.}(2022)\citenamefont
  {Krinner}, \citenamefont {Lacroix}, \citenamefont {Remm}, \citenamefont
  {Di~Paolo}, \citenamefont {Genois}, \citenamefont {Leroux}, \citenamefont
  {Hellings}, \citenamefont {Lazar}, \citenamefont {Swiadek}, \citenamefont
  {Herrmann}, \citenamefont {Norris}, \citenamefont {Andersen}, \citenamefont
  {Müller}, \citenamefont {Blais}, \citenamefont {Eichler},\ and\
  \citenamefont {Wallraff}}]{krinner_realizing_2022}%
  \BibitemOpen
  \bibfield  {author} {\bibinfo {author} {\bibfnamefont {S.}~\bibnamefont
  {Krinner}}, \bibinfo {author} {\bibfnamefont {N.}~\bibnamefont {Lacroix}},
  \bibinfo {author} {\bibfnamefont {A.}~\bibnamefont {Remm}}, \bibinfo {author}
  {\bibfnamefont {A.}~\bibnamefont {Di~Paolo}}, \bibinfo {author}
  {\bibfnamefont {E.}~\bibnamefont {Genois}}, \bibinfo {author} {\bibfnamefont
  {C.}~\bibnamefont {Leroux}}, \bibinfo {author} {\bibfnamefont
  {C.}~\bibnamefont {Hellings}}, \bibinfo {author} {\bibfnamefont
  {S.}~\bibnamefont {Lazar}}, \bibinfo {author} {\bibfnamefont
  {F.}~\bibnamefont {Swiadek}}, \bibinfo {author} {\bibfnamefont
  {J.}~\bibnamefont {Herrmann}}, \bibinfo {author} {\bibfnamefont {G.~J.}\
  \bibnamefont {Norris}}, \bibinfo {author} {\bibfnamefont {C.~K.}\
  \bibnamefont {Andersen}}, \bibinfo {author} {\bibfnamefont {M.}~\bibnamefont
  {Müller}}, \bibinfo {author} {\bibfnamefont {A.}~\bibnamefont {Blais}},
  \bibinfo {author} {\bibfnamefont {C.}~\bibnamefont {Eichler}},\ and\ \bibinfo
  {author} {\bibfnamefont {A.}~\bibnamefont {Wallraff}},\ }\bibfield  {title}
  {\bibinfo {title} {Realizing repeated quantum error correction in a
  distance-three surface code},\ }\href
  {https://doi.org/10.1038/s41586-022-04566-8} {\bibfield  {journal} {\bibinfo
  {journal} {Nature}\ }\textbf {\bibinfo {volume} {605}},\ \bibinfo {pages}
  {669} (\bibinfo {year} {2022})}\BibitemShut {NoStop}%
\bibitem [{\citenamefont {Macklin}\ \emph {et~al.}(2015)\citenamefont
  {Macklin}, \citenamefont {O’Brien}, \citenamefont {Hover}, \citenamefont
  {Schwartz}, \citenamefont {Bolkhovsky}, \citenamefont {Zhang}, \citenamefont
  {Oliver},\ and\ \citenamefont {Siddiqi}}]{macklin_nearquantum-limited_2015}%
  \BibitemOpen
  \bibfield  {author} {\bibinfo {author} {\bibfnamefont {C.}~\bibnamefont
  {Macklin}}, \bibinfo {author} {\bibfnamefont {K.}~\bibnamefont {O’Brien}},
  \bibinfo {author} {\bibfnamefont {D.}~\bibnamefont {Hover}}, \bibinfo
  {author} {\bibfnamefont {M.~E.}\ \bibnamefont {Schwartz}}, \bibinfo {author}
  {\bibfnamefont {V.}~\bibnamefont {Bolkhovsky}}, \bibinfo {author}
  {\bibfnamefont {X.}~\bibnamefont {Zhang}}, \bibinfo {author} {\bibfnamefont
  {W.~D.}\ \bibnamefont {Oliver}},\ and\ \bibinfo {author} {\bibfnamefont
  {I.}~\bibnamefont {Siddiqi}},\ }\bibfield  {title} {\bibinfo {title} {A
  near–quantum-limited {Josephson} traveling-wave parametric amplifier},\
  }\href {https://doi.org/10.1126/science.aaa8525} {\bibfield  {journal}
  {\bibinfo  {journal} {Science}\ }\textbf {\bibinfo {volume} {350}},\ \bibinfo
  {pages} {307} (\bibinfo {year} {2015})}\BibitemShut {NoStop}%
\bibitem [{\citenamefont {Ryan}\ \emph {et~al.}(2015)\citenamefont {Ryan},
  \citenamefont {Johnson}, \citenamefont {Gambetta}, \citenamefont {Chow},
  \citenamefont {Da~Silva}, \citenamefont {Dial},\ and\ \citenamefont
  {Ohki}}]{ryan_tomography_2015}%
  \BibitemOpen
  \bibfield  {author} {\bibinfo {author} {\bibfnamefont {C.~A.}\ \bibnamefont
  {Ryan}}, \bibinfo {author} {\bibfnamefont {B.~R.}\ \bibnamefont {Johnson}},
  \bibinfo {author} {\bibfnamefont {J.~M.}\ \bibnamefont {Gambetta}}, \bibinfo
  {author} {\bibfnamefont {J.~M.}\ \bibnamefont {Chow}}, \bibinfo {author}
  {\bibfnamefont {M.~P.}\ \bibnamefont {Da~Silva}}, \bibinfo {author}
  {\bibfnamefont {O.~E.}\ \bibnamefont {Dial}},\ and\ \bibinfo {author}
  {\bibfnamefont {T.~A.}\ \bibnamefont {Ohki}},\ }\bibfield  {title} {\bibinfo
  {title} {Tomography via correlation of noisy measurement records},\ }\href
  {https://doi.org/10.1103/PhysRevA.91.022118} {\bibfield  {journal} {\bibinfo
  {journal} {Physical Review A}\ }\textbf {\bibinfo {volume} {91}},\ \bibinfo
  {pages} {022118} (\bibinfo {year} {2015})}\BibitemShut {NoStop}%
\bibitem [{\citenamefont {Sank}\ \emph {et~al.}(2016)\citenamefont {Sank},
  \citenamefont {Chen}, \citenamefont {Khezri}, \citenamefont {Kelly},
  \citenamefont {Barends}, \citenamefont {Campbell}, \citenamefont {Chen},
  \citenamefont {Chiaro}, \citenamefont {Dunsworth}, \citenamefont {Fowler},
  \citenamefont {Jeffrey}, \citenamefont {Lucero}, \citenamefont {Megrant},
  \citenamefont {Mutus}, \citenamefont {Neeley} \emph
  {et~al.}}]{sank_measurement-induced_2016}%
  \BibitemOpen
  \bibfield  {author} {\bibinfo {author} {\bibfnamefont {D.}~\bibnamefont
  {Sank}}, \bibinfo {author} {\bibfnamefont {Z.}~\bibnamefont {Chen}}, \bibinfo
  {author} {\bibfnamefont {M.}~\bibnamefont {Khezri}}, \bibinfo {author}
  {\bibfnamefont {J.}~\bibnamefont {Kelly}}, \bibinfo {author} {\bibfnamefont
  {R.}~\bibnamefont {Barends}}, \bibinfo {author} {\bibfnamefont
  {B.}~\bibnamefont {Campbell}}, \bibinfo {author} {\bibfnamefont
  {Y.}~\bibnamefont {Chen}}, \bibinfo {author} {\bibfnamefont {B.}~\bibnamefont
  {Chiaro}}, \bibinfo {author} {\bibfnamefont {A.}~\bibnamefont {Dunsworth}},
  \bibinfo {author} {\bibfnamefont {A.}~\bibnamefont {Fowler}}, \bibinfo
  {author} {\bibfnamefont {E.}~\bibnamefont {Jeffrey}}, \bibinfo {author}
  {\bibfnamefont {E.}~\bibnamefont {Lucero}}, \bibinfo {author} {\bibfnamefont
  {A.}~\bibnamefont {Megrant}}, \bibinfo {author} {\bibfnamefont
  {J.}~\bibnamefont {Mutus}}, \bibinfo {author} {\bibfnamefont
  {M.}~\bibnamefont {Neeley}}, \emph {et~al.},\ }\bibfield  {title} {\bibinfo
  {title} {Measurement-{Induced} {State} {Transitions} in a {Superconducting}
  {Qubit}: {Beyond} the {Rotating} {Wave} {Approximation}},\ }\href
  {https://doi.org/10.1103/PhysRevLett.117.190503} {\bibfield  {journal}
  {\bibinfo  {journal} {Physical Review Letters}\ }\textbf {\bibinfo {volume}
  {117}},\ \bibinfo {pages} {190503} (\bibinfo {year} {2016})}\BibitemShut
  {NoStop}%
\bibitem [{\citenamefont {Khezri}\ \emph {et~al.}(2023)\citenamefont {Khezri},
  \citenamefont {Opremcak}, \citenamefont {Chen}, \citenamefont {Miao},
  \citenamefont {McEwen}, \citenamefont {Bengtsson}, \citenamefont {White},
  \citenamefont {Naaman}, \citenamefont {Sank}, \citenamefont {Korotkov},
  \citenamefont {Chen},\ and\ \citenamefont
  {Smelyanskiy}}]{khezri_measurement-induced_2023}%
  \BibitemOpen
  \bibfield  {author} {\bibinfo {author} {\bibfnamefont {M.}~\bibnamefont
  {Khezri}}, \bibinfo {author} {\bibfnamefont {A.}~\bibnamefont {Opremcak}},
  \bibinfo {author} {\bibfnamefont {Z.}~\bibnamefont {Chen}}, \bibinfo {author}
  {\bibfnamefont {K.~C.}\ \bibnamefont {Miao}}, \bibinfo {author}
  {\bibfnamefont {M.}~\bibnamefont {McEwen}}, \bibinfo {author} {\bibfnamefont
  {A.}~\bibnamefont {Bengtsson}}, \bibinfo {author} {\bibfnamefont
  {T.}~\bibnamefont {White}}, \bibinfo {author} {\bibfnamefont
  {O.}~\bibnamefont {Naaman}}, \bibinfo {author} {\bibfnamefont
  {D.}~\bibnamefont {Sank}}, \bibinfo {author} {\bibfnamefont {A.~N.}\
  \bibnamefont {Korotkov}}, \bibinfo {author} {\bibfnamefont {Y.}~\bibnamefont
  {Chen}},\ and\ \bibinfo {author} {\bibfnamefont {V.}~\bibnamefont
  {Smelyanskiy}},\ }\bibfield  {title} {\bibinfo {title} {Measurement-induced
  state transitions in a superconducting qubit: {Within} the rotating-wave
  approximation},\ }\href {https://doi.org/10.1103/PhysRevApplied.20.054008}
  {\bibfield  {journal} {\bibinfo  {journal} {Physical Review Applied}\
  }\textbf {\bibinfo {volume} {20}},\ \bibinfo {pages} {054008} (\bibinfo
  {year} {2023})}\BibitemShut {NoStop}%
\bibitem [{\citenamefont {Rudinger}\ \emph {et~al.}(2017)\citenamefont
  {Rudinger}, \citenamefont {Kimmel}, \citenamefont {Lobser},\ and\
  \citenamefont {Maunz}}]{rudinger_experimental_2017}%
  \BibitemOpen
  \bibfield  {author} {\bibinfo {author} {\bibfnamefont {K.}~\bibnamefont
  {Rudinger}}, \bibinfo {author} {\bibfnamefont {S.}~\bibnamefont {Kimmel}},
  \bibinfo {author} {\bibfnamefont {D.}~\bibnamefont {Lobser}},\ and\ \bibinfo
  {author} {\bibfnamefont {P.}~\bibnamefont {Maunz}},\ }\bibfield  {title}
  {\bibinfo {title} {Experimental {Demonstration} of a {Cheap} and {Accurate}
  {Phase} {Estimation}},\ }\href
  {https://doi.org/10.1103/PhysRevLett.118.190502} {\bibfield  {journal}
  {\bibinfo  {journal} {Physical Review Letters}\ }\textbf {\bibinfo {volume}
  {118}},\ \bibinfo {pages} {190502} (\bibinfo {year} {2017})}\BibitemShut
  {NoStop}%
\bibitem [{\citenamefont {Lucero}\ \emph {et~al.}(2010)\citenamefont {Lucero},
  \citenamefont {Kelly}, \citenamefont {Bialczak}, \citenamefont {Lenander},
  \citenamefont {Mariantoni}, \citenamefont {Neeley}, \citenamefont
  {O’Connell}, \citenamefont {Sank}, \citenamefont {Wang}, \citenamefont
  {Weides}, \citenamefont {Wenner}, \citenamefont {Yamamoto}, \citenamefont
  {Cleland},\ and\ \citenamefont {Martinis}}]{lucero_reduced_2010}%
  \BibitemOpen
  \bibfield  {author} {\bibinfo {author} {\bibfnamefont {E.}~\bibnamefont
  {Lucero}}, \bibinfo {author} {\bibfnamefont {J.}~\bibnamefont {Kelly}},
  \bibinfo {author} {\bibfnamefont {R.~C.}\ \bibnamefont {Bialczak}}, \bibinfo
  {author} {\bibfnamefont {M.}~\bibnamefont {Lenander}}, \bibinfo {author}
  {\bibfnamefont {M.}~\bibnamefont {Mariantoni}}, \bibinfo {author}
  {\bibfnamefont {M.}~\bibnamefont {Neeley}}, \bibinfo {author} {\bibfnamefont
  {A.~D.}\ \bibnamefont {O’Connell}}, \bibinfo {author} {\bibfnamefont
  {D.}~\bibnamefont {Sank}}, \bibinfo {author} {\bibfnamefont {H.}~\bibnamefont
  {Wang}}, \bibinfo {author} {\bibfnamefont {M.}~\bibnamefont {Weides}},
  \bibinfo {author} {\bibfnamefont {J.}~\bibnamefont {Wenner}}, \bibinfo
  {author} {\bibfnamefont {T.}~\bibnamefont {Yamamoto}}, \bibinfo {author}
  {\bibfnamefont {A.~N.}\ \bibnamefont {Cleland}},\ and\ \bibinfo {author}
  {\bibfnamefont {J.~M.}\ \bibnamefont {Martinis}},\ }\bibfield  {title}
  {\bibinfo {title} {Reduced phase error through optimized control of a
  superconducting qubit},\ }\href {https://doi.org/10.1103/PhysRevA.82.042339}
  {\bibfield  {journal} {\bibinfo  {journal} {Physical Review A}\ }\textbf
  {\bibinfo {volume} {82}},\ \bibinfo {pages} {042339} (\bibinfo {year}
  {2010})}\BibitemShut {NoStop}%
\bibitem [{\citenamefont {Knill}\ \emph {et~al.}(2000)\citenamefont {Knill},
  \citenamefont {Laflamme}, \citenamefont {Martinez},\ and\ \citenamefont
  {Tseng}}]{knill_algorithmic_2000}%
  \BibitemOpen
  \bibfield  {author} {\bibinfo {author} {\bibfnamefont {E.}~\bibnamefont
  {Knill}}, \bibinfo {author} {\bibfnamefont {R.}~\bibnamefont {Laflamme}},
  \bibinfo {author} {\bibfnamefont {R.}~\bibnamefont {Martinez}},\ and\
  \bibinfo {author} {\bibfnamefont {C.-H.}\ \bibnamefont {Tseng}},\ }\bibfield
  {title} {\bibinfo {title} {An algorithmic benchmark for quantum information
  processing},\ }\href {https://doi.org/10.1038/35006012} {\bibfield  {journal}
  {\bibinfo  {journal} {Nature}\ }\textbf {\bibinfo {volume} {404}},\ \bibinfo
  {pages} {368} (\bibinfo {year} {2000})}\BibitemShut {NoStop}%
\bibitem [{\citenamefont {Weiss}\ \emph {et~al.}(2022)\citenamefont {Weiss},
  \citenamefont {Zhang}, \citenamefont {Ding}, \citenamefont {Ma},
  \citenamefont {Schuster},\ and\ \citenamefont {Koch}}]{weiss_fast_2022}%
  \BibitemOpen
  \bibfield  {author} {\bibinfo {author} {\bibfnamefont {D.}~\bibnamefont
  {Weiss}}, \bibinfo {author} {\bibfnamefont {H.}~\bibnamefont {Zhang}},
  \bibinfo {author} {\bibfnamefont {C.}~\bibnamefont {Ding}}, \bibinfo {author}
  {\bibfnamefont {Y.}~\bibnamefont {Ma}}, \bibinfo {author} {\bibfnamefont
  {D.~I.}\ \bibnamefont {Schuster}},\ and\ \bibinfo {author} {\bibfnamefont
  {J.}~\bibnamefont {Koch}},\ }\bibfield  {title} {\bibinfo {title} {Fast
  {High}-{Fidelity} {Gates} for {Galvanically}-{Coupled} {Fluxonium} {Qubits}
  {Using} {Strong} {Flux} {Modulation}},\ }\href
  {https://doi.org/10.1103/PRXQuantum.3.040336} {\bibfield  {journal} {\bibinfo
   {journal} {PRX Quantum}\ }\textbf {\bibinfo {volume} {3}},\ \bibinfo {pages}
  {040336} (\bibinfo {year} {2022})}\BibitemShut {NoStop}%
\bibitem [{\citenamefont {Bylander}\ \emph {et~al.}(2011)\citenamefont
  {Bylander}, \citenamefont {Gustavsson}, \citenamefont {Yan}, \citenamefont
  {Yoshihara}, \citenamefont {Harrabi}, \citenamefont {Fitch}, \citenamefont
  {Cory}, \citenamefont {Nakamura}, \citenamefont {Tsai},\ and\ \citenamefont
  {Oliver}}]{bylander_noise_2011}%
  \BibitemOpen
  \bibfield  {author} {\bibinfo {author} {\bibfnamefont {J.}~\bibnamefont
  {Bylander}}, \bibinfo {author} {\bibfnamefont {S.}~\bibnamefont
  {Gustavsson}}, \bibinfo {author} {\bibfnamefont {F.}~\bibnamefont {Yan}},
  \bibinfo {author} {\bibfnamefont {F.}~\bibnamefont {Yoshihara}}, \bibinfo
  {author} {\bibfnamefont {K.}~\bibnamefont {Harrabi}}, \bibinfo {author}
  {\bibfnamefont {G.}~\bibnamefont {Fitch}}, \bibinfo {author} {\bibfnamefont
  {D.~G.}\ \bibnamefont {Cory}}, \bibinfo {author} {\bibfnamefont
  {Y.}~\bibnamefont {Nakamura}}, \bibinfo {author} {\bibfnamefont {J.-S.}\
  \bibnamefont {Tsai}},\ and\ \bibinfo {author} {\bibfnamefont {W.~D.}\
  \bibnamefont {Oliver}},\ }\bibfield  {title} {\bibinfo {title} {Noise
  spectroscopy through dynamical decoupling with a superconducting flux
  qubit},\ }\href {https://doi.org/10.1038/nphys1994} {\bibfield  {journal}
  {\bibinfo  {journal} {Nature Physics}\ }\textbf {\bibinfo {volume} {7}},\
  \bibinfo {pages} {565} (\bibinfo {year} {2011})}\BibitemShut {NoStop}%
\bibitem [{noi()}]{noise_spectrum_footnote}%
  \BibitemOpen
  \href@noop {} {}\bibinfo {note} {The noise spectrum $S(\omega)$ drives
  bit-flips within the dual-rail subspace with rates given by Fermi's golden
  rule as $\Gamma_{0 \to 1} = \Gamma_{1 \to 0} = \frac{1}{4} S(2g)$, where the
  $1/4$ accounts for the matrix element between logical states
  \cite{clerk_introduction_2010}. Adding both bit-flip rates leads to a
  combined $T_1$ limit of $T_1^{DR} = 2 / S(2g)$.}\BibitemShut {Stop}%
\bibitem [{\citenamefont {Krantz}\ \emph {et~al.}(2019)\citenamefont {Krantz},
  \citenamefont {Kjaergaard}, \citenamefont {Yan}, \citenamefont {Orlando},
  \citenamefont {Gustavsson},\ and\ \citenamefont
  {Oliver}}]{krantz_quantum_2019}%
  \BibitemOpen
  \bibfield  {author} {\bibinfo {author} {\bibfnamefont {P.}~\bibnamefont
  {Krantz}}, \bibinfo {author} {\bibfnamefont {M.}~\bibnamefont {Kjaergaard}},
  \bibinfo {author} {\bibfnamefont {F.}~\bibnamefont {Yan}}, \bibinfo {author}
  {\bibfnamefont {T.~P.}\ \bibnamefont {Orlando}}, \bibinfo {author}
  {\bibfnamefont {S.}~\bibnamefont {Gustavsson}},\ and\ \bibinfo {author}
  {\bibfnamefont {W.~D.}\ \bibnamefont {Oliver}},\ }\bibfield  {title}
  {\bibinfo {title} {A quantum engineer's guide to superconducting qubits},\
  }\href {https://doi.org/10.1063/1.5089550} {\bibfield  {journal} {\bibinfo
  {journal} {Applied Physics Reviews}\ }\textbf {\bibinfo {volume} {6}},\
  \bibinfo {pages} {021318} (\bibinfo {year} {2019})}\BibitemShut {NoStop}%
\bibitem [{\citenamefont {Jurcevic}\ \emph {et~al.}(2021)\citenamefont
  {Jurcevic}, \citenamefont {Javadi-Abhari}, \citenamefont {Bishop},
  \citenamefont {Lauer}, \citenamefont {Bogorin}, \citenamefont {Brink},
  \citenamefont {Capelluto}, \citenamefont {Günlük}, \citenamefont {Itoko},
  \citenamefont {Kanazawa}, \citenamefont {Kandala}, \citenamefont {Keefe},
  \citenamefont {Krsulich}, \citenamefont {Landers}, \citenamefont
  {Lewandowski} \emph {et~al.}}]{jurcevic_demonstration_2021}%
  \BibitemOpen
  \bibfield  {author} {\bibinfo {author} {\bibfnamefont {P.}~\bibnamefont
  {Jurcevic}}, \bibinfo {author} {\bibfnamefont {A.}~\bibnamefont
  {Javadi-Abhari}}, \bibinfo {author} {\bibfnamefont {L.~S.}\ \bibnamefont
  {Bishop}}, \bibinfo {author} {\bibfnamefont {I.}~\bibnamefont {Lauer}},
  \bibinfo {author} {\bibfnamefont {D.~F.}\ \bibnamefont {Bogorin}}, \bibinfo
  {author} {\bibfnamefont {M.}~\bibnamefont {Brink}}, \bibinfo {author}
  {\bibfnamefont {L.}~\bibnamefont {Capelluto}}, \bibinfo {author}
  {\bibfnamefont {O.}~\bibnamefont {Günlük}}, \bibinfo {author}
  {\bibfnamefont {T.}~\bibnamefont {Itoko}}, \bibinfo {author} {\bibfnamefont
  {N.}~\bibnamefont {Kanazawa}}, \bibinfo {author} {\bibfnamefont
  {A.}~\bibnamefont {Kandala}}, \bibinfo {author} {\bibfnamefont {G.~A.}\
  \bibnamefont {Keefe}}, \bibinfo {author} {\bibfnamefont {K.}~\bibnamefont
  {Krsulich}}, \bibinfo {author} {\bibfnamefont {W.}~\bibnamefont {Landers}},
  \bibinfo {author} {\bibfnamefont {E.~P.}\ \bibnamefont {Lewandowski}}, \emph
  {et~al.},\ }\bibfield  {title} {\bibinfo {title} {Demonstration of quantum
  volume 64 on a superconducting quantum computing system},\ }\href
  {https://doi.org/10.1088/2058-9565/abe519} {\bibfield  {journal} {\bibinfo
  {journal} {Quantum Science and Technology}\ }\textbf {\bibinfo {volume}
  {6}},\ \bibinfo {pages} {025020} (\bibinfo {year} {2021})}\BibitemShut
  {NoStop}%
\end{thebibliography}
\end{document}